\newcommand{\blind}{1}

\documentclass[authoryear,12pt]{article}
\usepackage[a4paper, total={6.5in, 10in}]{geometry}
\usepackage{setspace}
\usepackage{natbib}
\usepackage[export]{adjustbox}
\usepackage{import}
\usepackage{xifthen}
\usepackage{pdfpages}
\usepackage{transparent}
\usepackage{fontenc, inputenc, subcaption, graphicx, sectsty, titlesec, titling, caption, fancyhdr}
\usepackage{amsthm,amsmath,amssymb,nccmath,xcolor}
\usepackage{paralist,multirow,arydshln}
\usepackage[noend]{algpseudocode}
\usepackage[ruled, vlined]{algorithm2e}
\usepackage{float}
\usepackage[colorlinks=true,linkcolor=black, citecolor=blue, urlcolor=purple]{hyperref}
\usepackage{comment}

\makeatletter
\newcommand*{\addFileDependency}[1]{
  \typeout{(#1)}
  \@addtofilelist{#1}
  \IfFileExists{#1}{}{\typeout{No file #1.}}
}
\makeatother

\newcommand{\trans}{^\top} 
\newcommand{\del}[1]{\frac{\partial}{\partial #1}} 
\newcommand{\deltwo}[1]{\mfrac{\partial^2}{\partial #1^2}}
\newcommand{\deltwom}[2]{\mfrac{\partial^2}{\partial #1 \partial #2}}

\newcommand{\Cov}{\mathrm{Cov}}
\newcommand{\tK}{\widetilde{K}}
\newcommand{\given}{\, | \;}

\def\bbeta{\mbox{\boldmath${\beta}$}}
\def\bgamma{\mbox{\boldmath$\gamma$}}
\def\bGamma{\mbox{\boldmath$\Gamma$}}
\def\bdelta{\mbox{\boldmath${\delta}$}}
\def\btheta{\mbox{\boldmath$\theta$}}
\def\bmu{\mbox{\boldmath$\mu$}}

\def\bleta{\mbox{\boldmath$\medmath{\eta}$}}

\def\A{\mbox{\boldmath$A$}}

\def\a{\mbox{$\bf a$}}

\def\C{\mbox{$\bf C$}}
\def\c{\mbox{$\bf c$}}

\def\E{\mbox{$\bf E$}}
\def\e{\mbox{$\bf e$}}
\def\F{\mbox{$\bf F$}}

\def\K{\mbox{$\bf K$}}
\def\k{\mbox{$\bf k$}}

\def\L{\mbox{$ \bf L$}}

\def\m{\mbox{$\bf m$}}
\def\N{\mbox{$\bf N$}}
\def\n{\mbox{$\bf n$}}

\def\R{\mbox{$\bf R$}}

\def\s{\mbox{$\bf s$}}
\def\S{\mbox{$\bf S$}}
\def\t{\mbox{$\bf t$}}

\def\U{\mbox{$\bf U$}}
\def\u{\mbox{$\bf u$}}

\def\v{\mbox{$\bf v$}}

\def\w{\mbox{$\bf w$}}
\def\X{\mbox{$\bf X$}}
\def\x{\mbox{$\bf x$}}
\def\Y{\mbox{$\bf Y$}}

\def\Z{\mbox{$\bf Z$}}
\def\z{\mbox{$\bf z$}}
\def\1{\mbox{$\bf 1$}}
\def\Rone{\mbox{$\mathbb{R}$}}
\def\Rtwo{\mbox{$\mathbb{R}^2$}}
\def\Rthree{\mbox{$\mathbb{R}^3$}}
\def\Rd{\mbox{$\mathbb{R}^{d}$}}

\newtheorem{myres}{Result}

\begin{document}
	\def\spacingset#1{\renewcommand{\baselinestretch}%
		{#1}\small\normalsize} \spacingset{1}
	
	\if1\blind
	{
		\title{\bf Bayesian Modeling with Spatial Curvature Processes}
		\author{Aritra Halder$^{a}$, Sudipto Banerjee$^{b}$ and Dipak K. Dey$^{c}$ \\
		    $^{a}$Department of Biostatistics,\\
		    Drexel University, Philadelphia, PA, USA.\\
			$^{b}$Department of Biostatistics,\\
			University of California,
			Los Angeles, CA, USA.\\
			$^{c}$Department of Statistics,\\
			University of Connecticut,
			Storrs, CT, USA.}
		\maketitle
	} \fi

	\if0\blind
	{
		\bigskip
		\bigskip
		\bigskip
		\begin{center}
			{\LARGE\bf Bayesian Modeling with Spatial Curvature Processes}
		\end{center}
		\medskip
	} \fi

	\begin{abstract}

	\noindent Spatial process models are widely used for modeling point-referenced variables arising from diverse scientific domains. Analyzing the resulting random surface provides deeper insights into the nature of latent dependence within the studied response. We develop Bayesian modeling and inference for rapid changes on the response surface to assess directional curvature along a given trajectory. Such trajectories or curves of rapid change, often referred to as \emph{wombling} boundaries, occur in geographic space in the form of rivers in a flood plain, roads, mountains or plateaus or other topographic features leading to high gradients on the response surface. We demonstrate fully model based Bayesian inference on directional curvature processes to analyze differential behavior in responses along wombling boundaries. We illustrate our methodology with a number of simulated experiments followed by multiple applications featuring the Boston Housing data; Meuse river data; and temperature data from the Northeastern United States.
\end{abstract}


\spacingset{1.9} 
\section{Introduction}
Spatial data science %
manifests in a variety of domains including environmental and geographical information systems (GIS) \citep[][]{webster2007geostatistics, burrough2015principles, schabenberger2017statistical, plant2018spatial}, digital cartography and terrain modeling  \citep[][]{law2000simulation, santner2003design, jones2014geographical, vaughan2018mapping}, imaging \citep[][]{winkler2003image, chiu2013stochastic, dryden2016statistical}, spatial econometrics and land use \citep{lesage2009introduction}, public health and epidemiology \citep[][]{elliot2000spatial, waller2004applied, lawson2013statistical} and 
public policy \citep[][]{haining1993spatial, wise2007gis}. Spatial data analysis seeks to estimate an underlying spatial surface representing the process generating the data. Specific inferential interest resides with local features of the surface including rates of change of the process at points and along ``spatial boundaries'' to understand the behavior of the underlying process and identify lurking explanatory variables or risk factors. This exercise is often referred to as ``wombling'', named after a seminal paper by \cite{womble1951differential}; \citep[also see][]{gleyze2001wombling}. 
For regionally aggregated data, it identifies boundaries delineating neighboring regions and has been used to study health disparities \citep[][]{lu2005bayesian, li2015bayesian, gao2022wombling} and ecological boundaries \citep{fitzpatrick2010ecological}. For point-referenced data, where variables are mapped at locations within an Euclidean coordinate frame 
with a sufficiently smooth spatial surface, it refers to estimating spatial gradients and identifying boundaries representing large gradients \citep[][]{banerjee2003directional, banerjee2006bayesian, qu2021boundary}. 

{Our current contribution develops Bayesian inference for spatial curvature along curves on Euclidean domains. Modeling curvature will require smoothness considerations of the process} \citep[][]{adler1981geometry, kent1989continuity, stein1999interpolation, banerjee2003smoothness}.  Observations over a finite set of locations from these processes cannot visually inform about smoothness. Therefore, smoothness of the process is specified from mechanistic considerations which can be introduced through prior specifications as needed
. 
{While Bayesian inference for first order derivatives and directional gradients have received considerable attention \citep[see, e.g.,][for inferential developments involving spatial gradients from diverse modeling and application perspectives]{morris1993bayesian, banerjee2003directional, majumdar2006gradients, liang2009bayesian, heaton2014wombling, terres2015using, wang2016estimating, terres2016spatial, wang2018process, qu2021boundary} 
such processes inform about directional change, but do not enable inference on} 
curvature (departure from flatness) of the spatial surface.

Analyzing surface roughness from sampling considerations can be traced at least as far back as \cite{greenwood1984unified}.  
{We offer full inference with uncertainty quantification about spatial curvature at a point and average curvature along a curve from observed data after accounting for explanatory variables.} Considering second-order finite differences we establish a valid spatial curvature process as a limit of such finite difference processes. When formulating directional curvature, we favor the normal direction corresponding to a chosen curve and devise a ``wombling'' measure to track curvature of the surface along the curve. We derive and exploit analytical expressions of higher order processes to avoid numerical finite differences. The Bayesian inferential framework delivers exact posterior inference for the above constructs on the response as well as latent (or residual) processes
. 

Section~\ref{sec::sgcp} develops the directional curvature processes through a differential operator. 
Section \ref{sec::bcw} develops the vector analytic framework for curvilinear wombling using curvature processes. Section~\ref{sec::model} builds a hierarchical model to exploit the preceding distribution theory and conduct curvature analysis on the response and the latent process. Section~\ref{sec::syn-exp} presents detailed simulation experiments for assessing directional gradients and curvatures. Section \ref{sec::app} considers applications 
to three different data sets: Boston housing data, Meuse river data, and Northeastern US Temperatures (the third data is presented in the Supplement). 

\section{Spatial Curvature Processes} \label{sec::sgcp} 
Let $\{Y(\s): \s \in {\cal S} \subset \Rd\}$ be a univariate weakly stationary random field with zero mean, finite second moment and a positive definite covariance function $K(\s,\s')=\Cov\left(Y(\s),Y(\s')\right)$ for locations $\s,\s' \in \Rd$. 
In particular, under {\em isotropy} we assume $K(\s,\s')=\widetilde{K}\left(||\s-\s'||\right)$, where $||\s-\s'||$ is the Euclidean distance between the locations $\s, \s'$ \citep{matern2013spatial}.
{Building upon notions of mean square smoothness \citep[see, e.g.,][]{stein1999interpolation} at an arbitrary location $\s_0$ in $\mathbb{R}^d$, we focus upon second order differentiability, $\displaystyle Y(\s_0+h\u)=Y(\s_0)+h\u\trans \nabla Y(\s_0) + h^2\u\trans\nabla^2 Y(\s_0)\u/2+r_2(\s_0,h^2||\u||)$, where $r_2(\s_0,h^2||\u||)/h^2\to 0$ as $h\to 0$ in the $L_2$ sense} and $\nabla$ and $\nabla^2$ are the gradient and Hessian operators, respectively. 

{For the scalar $h$ and unit vectors $\u$, $\v$, we define 
$\displaystyle Y^{(2)}_{\u,\v,h}(\s_0)=(Y(\s_0+h(\u+\v))-Y(\s_0+h\u)-Y(\s_0+h\v)+Y(\s_0))/h^{2}$
to be the 
second order finite difference processes 
in the directions 
$\u$, $\v$ 
at scale $h$. Being a linear function of stationary processes 
it is well-defined. Passing to limits, 
$D^{(2)}_{\u,\v}Y(\s_0)=\lim_{h\to 0}Y^{(2)}_{\u,\v,h}(\s_0)$. Provided the limit exists, 
$D^{(2)}_{\u,\u}Y(\s_0)$ is defined as the directional curvature process. If $Y(\s)$ is a mean square second order differentiable process in $\Rd$ for every $\s \in \Rd$ then 
$D^{(2)}_{\u,\v}Y(\s)=\u\trans\nabla^2Y(\s)\v$ is 
well-defined with
$D^{(2)}_{\u,\v}Y(\s) = \lim_{h\to 0}\left(h^2\u\trans\nabla^2Y(\s)\v+\widetilde{r_2}\right)/h^2=\u\trans\nabla^2Y(\s)\v$, 
where $\widetilde{r_2} = r_2(\s,h^2||\u+\v||)-r_2(\s,h^2||\u||)-r_2(\s,h^2||\v||)$.
In practice, we need only work with computing these derivatives for an orthonormal basis of $\Rd$, say the Euclidean canonical unit vectors along each axis $\{\e_1,\ldots,\e_d\}$. If $\u=\sum_{i=1}^{d}u_i\e_i$, and $\v=\sum_{i=1}^d v_i\e_i$ are arbitrary unit vectors, we can compute 
$D^{(2)}_{\u,\v}Y(\s)=\sum_{i=1}^{d}\sum_{j=1}^{d}u_iD^{(2)}_{\e_i,\e_j}Y(\s)v_j$.
The directional curvature process is linear in 
the sense that
$D^{(2)}_{-\u,-\v}Y(\s)=D^{(2)}_{\u,\v}Y(\s)$, $D^{(2)}_{\u,-\v}Y(\s)=D^{(2)}_{-\u,\v}Y(\s)=-D^{(2)}_{\u,\v}Y(\s)$. Since $D^{(2)}_{\w,\w}Y(\s)=||\w||^2D^{(2)}_{\u,\u}Y(\s)$, where $\w=||\w||\u$ and $\u$ is a unit direction, we henceforth only consider unit directions. 
First order directional gradient processes, $D^{(1)}_{\u}Y(\s)$, are reviewed in \cite{banerjee2006bayesian} and in Section~S1.1 
of the Supplement. Choosing a direction is emphasized with respect to interpreting the directional 
curvature processes. 
Directional curvature is the change in the normal to the surface $Y(\s)$ at $\s_0$ when moving along a slice of the surface in the direction $\w$. The associated algebraic sign locally classifies the nature of curvature at $\s_0$---for instance, convex or concave ellipsoids \citep[see][]{stevens1981visual}. A detailed discussion, with illustration, is available in Section~S2 
of the Supplement.}

Since $\nabla^2Y(\s)$ is a symmetric matrix, to avoid singularities arising from duplication we modify $D^{(2)}_{\u,\v}Y(\s)$ as follows. If $vech$ is the usual half-vectorization operator for symmetric matrices 
and $\mathcal{D}_d$ is the duplication matrix \citep{magnus1980elimination} of order $d^2\times d(d+1)/2$ then, $D^{(2)}_{\u,\v}Y(\s)=
\c\trans_{\u,\v} vech\left(\nabla^2 Y(\s)\right)$
where $\c_{\u,\v}\trans=(\u\otimes\v)\trans\mathcal{D}_d$ and $\otimes$ is the Kronecker product for matrices. If $\u=(u_1,u_2)\trans, \v=(v_1,v_2)\trans \in \Rtwo$, then $\c_{\u,\v}=(\u\otimes\v)\trans\mathcal{D}_2=\left(v_1u_1,v_1u_2+v_2u_1,v_2u_2\right)\trans$.
The process $vech\left(\nabla^2 Y(\s)\right)$ in $\mathbb{R}^{d(d+1)/2}$ consists of the pure and mixed second order derivatives in $\nabla^2Y(\s)$. The distributions needed for inference on directional curvature processes depend on $vech\left(\nabla^2 Y(\s)\right)$ rather than $\nabla^2Y(\s)$. We refer to $(\nabla Y(\s)\trans, vech(\nabla^2 Y(\s))\trans)\trans$ as the differential process and $\{\u\trans\nabla Y(\s)$, $\c_{\u,\u}\trans vech(\nabla^2 Y(\s))\}$ as the directional differential processes induced by $Y(\s)$ along $\u$.

{Inference for differential processes requires $(Y(\s),\nabla Y(\s)^{\trans}, vech(\nabla^2 Y(\s))^{\trans})$ to be a valid multivariate process.} Its existence is derived from the limit of corresponding finite difference approximations,
{which yields the cross-covariance matrix depending on fourth (and lower) order derivatives of $K$.} 
We %
investigate 
the parent 
and %
differential processes 
using 
a 
differential operator 
${\cal L}: \mathbb{R}^{1}\to\mathbb{R}^m$, $m=1+d+d(d+1)/2$
, 
where 
${\cal L}Y=\left(Y,\nabla Y\trans,vech(\nabla^2 Y)\trans\right)\trans$. 
The resulting process ${\cal L}Y(\s)$ is also stationary with a zero mean and a cross-covariance matrix
\begin{equation}\label{eq::cov-delta}
	V_{{\cal L}Y}(\Delta)=\begin{pmatrix}
		K(\Delta) & -(\nabla K(\Delta))\trans & vech(\nabla^2 K(\Delta))\trans\\
		\nabla K(\Delta) & -\nabla^2 K(\Delta) & \nabla^3 K(\Delta)\trans\\
		vech(\nabla^2 K(\Delta)) & -\nabla^3 K(\Delta) & \nabla^4 K(\Delta)
	\end{pmatrix}\;,
\end{equation}
where $\Delta=\s-\s'$,  $\nabla K(\Delta)$ is the $d\times 1$ gradient, $\nabla^2K(\Delta)$ is the $d\times d$ Hessian, $\nabla^3K(\Delta)$ is the $d(d+1)/2\times d$ matrix of third derivatives and $\nabla^4K(\Delta)$ is the $d(d+1)/2\times d(d+1)/2$ matrix of fourth order derivatives associated with $K(\Delta)$. 
Under isotropy, $\nabla K(\Delta)=\frac{\nabla\widetilde{K}(||\Delta||)}{||\Delta||}\Delta$, if $A_0=\left(\nabla^2\widetilde{K}(||\Delta||)-\frac{\nabla\widetilde{K}(||\Delta||)}{||\Delta||}\right)$ then, $\nabla^2 K(\Delta)=\frac{\nabla\widetilde{K}(||\Delta||)}{||\Delta||} I_d+A_0\frac{\Delta\Delta\trans}{||\Delta||^2}$, $\nabla^3K(\Delta)=A_0\bigg\{\frac{vech(I_d)\trans\otimes\Delta}{||\Delta||^2}-3\frac{vech(\Delta\Delta\trans)\trans\otimes\Delta }{||\Delta||^4}+\frac{1}{||\Delta||^2}\left(\frac{\partial vech(\Delta\Delta\trans)}{\partial\Delta}\right)\bigg\}+\nabla^3\widetilde{K}(||\Delta||)\cdot\frac{vech(\Delta\Delta\trans)\trans\otimes\Delta}{||\Delta||^3}$,
where 
{$\widetilde{K}(\Delta)$ and its derivatives are analytically computed for our covariance functions of interest in Section~S3 
of the Supplement.} 
Let $A_1=\frac{\partial\Delta\otimes vech(I_d)\trans}{\partial \Delta}$, $A_2=\frac{\partial\Delta\otimes vech(\Delta\Delta\trans)\trans}{\partial \Delta}$, $A_3=\frac{\partial}{\partial\Delta}\left(\frac{\partial vech(\Delta\Delta\trans)}{\partial\Delta}\right)$ be reordered tensors (matrices) of order $d(d+1)/2\times d(d+1)/2$ conforming to the order of corresponding elements in $vech$. Let $A_4$ be the element-wise product of $\Delta$ with $\left(\frac{\partial vech(\Delta\Delta\trans)}{\partial\Delta}\right)$ in the same order, $B_1=vech(\Delta\Delta\trans)vech(I_d)\trans$ and $B_2=vech(\Delta\Delta\trans)vech(\Delta\Delta\trans)\trans$. Then, $\nabla^4 K(\Delta)$ is,
\begin{equation}\label{eq::nabla-k}
		\begin{split}
			A_0\left\{\frac{A_1}{||\Delta||^2}-3\frac{A_2}{||\Delta||^4}+\frac{A_3}{||\Delta||^2}
			-(1+A_4)\left(\frac{2B_1}{||\Delta||^4}+\frac{B_1}{||\Delta||^3}\right)+3\left(\frac{4B_2}{||\Delta||^6}+\frac{B_2}{||\Delta||^5}\right)\right\}\\
			+\nabla^3\widetilde{K}(||\Delta||)\Bigg(\frac{B_1}{||\Delta||^3}+\frac{A_2}{||\Delta||^3}+\frac{A_4}{||\Delta||^3}-6\frac{B_2}{||\Delta||^5}\Bigg)+\nabla^4\widetilde{K}(||\Delta||)\frac{B_2}{||\Delta||^4}\;.
		\end{split}
\end{equation}
The resulting multivariate differential process, ${\cal L}Y$, is stationary but not isotropic. Evidently, for the differential operator to be well-defined under isotropy, $\nabla^4K({\bf 0})$ must exist since $var(D^{(2)}_{\u,\u}Y(\s)) = \nabla^4\widetilde{K}({\bf 0})$ \citep[analogous to results in][Section 3]{banerjee2003directional}. The directional differential operator is defined analogously as ${\cal L}_{\u}Y(\s)$ 
such that ${\cal L}_{\u}:\Rone\to\Rthree$. If $a_0=\left(1-\frac{(\u\trans\Delta)^2}{||\Delta||^2}\right)$, then analogous to (\ref{eq::nabla-k}) the covariance function of the directional curvature process, $\Cov\left(\c_{\u,\u}\trans vech(\nabla^2Y(\s)),\c_{\u,\u}\trans vech(\nabla^2Y(\s'))\right)=\frac{3}{||\Delta||^2}(5a_0-4)a_0A_0+\frac{6}{||\Delta||}(1-a_0)a_0\nabla^3\widetilde{K}(||\Delta||)+(1-a_0)^2\nabla^4\widetilde{K}(||\Delta||)$.
To characterize covariance functions that admit such processes, we turn to spectral theory. Recall that for a positive definite function $K$ defined in $\mathbb{R}$, Bochner's theorem \citep[see e.g.,][]{williams2006gaussian} establishes the existence of a finite positive spectral measure ${\cal F}$ on $\mathbb{R}$. $K$ can be expressed as the inverse Fourier transform of ${\cal F}$, $K(t)=\int_{\mathbb{R}}e^{-i\lambda t}{\cal F}(d\lambda)$. In cases where ${\cal F}$ admits a spectral density, $K(t)=\int e^{-i\lambda t}f(\lambda)\,d\lambda$. For $\nabla^4K$ to exist, a trivial extension of the result in \cite{wang2018process} requires that $f$ possess a finite fourth moment. Examples of covariance kernels that satisfy this condition are (a) the squared exponential covariance kernel with $K(t)=\exp(-t^2)$ ($\sigma^2=\phi=1$), and $f(\lambda)=1/2\sqrt{\pi}\exp(-\lambda^2/4)$ then, $\frac{1}{2\sqrt{\pi}}\int_{\mathbb{R}}\lambda^4\exp(-\lambda^2/4)\,d\lambda=3(\sqrt{2})^4=12$; and (b) the Mat\'ern class with fractal parameter, $\nu$; $f(\lambda)$ is known to belong to the $t$-family \citep[see e.g.,][]{stein1999interpolation} with $f(\lambda)=C(\phi,\nu)/(c(\phi,\nu)+\lambda^2)^{\nu+1/2}$ then, $\int_{\mathbb{R}}\lambda^4C(\phi,\nu)/(c(\phi,\nu)+\lambda^2)^{\nu+1/2}\,d\lambda<\infty$, for all $\nu>2$ (since the fourth central moment for the $t$-distribution exists if $\nu>2$). Here, we consider formulating the directional differential processes using these two classes of covariance functions (a) the squared exponential, $\widetilde{K}(||\Delta||)=\sigma^2\exp(-\phi||\Delta||^{\nu})$, $\nu=2$; and (b) members of the Mat\'ern class, $\widetilde{K}(||\Delta||)=\sigma^2(\phi||\Delta||)^\nu K_\nu(\phi||\Delta||)$, where $K_\nu$ is the modified Bessel function of order $\nu$ \citep[see e.g.,][]{abramowitz1988handbook}, and $\nu$ controls the smoothness of process realizations. We are particularly interested in $\nu=5/2$.

The multivariate process, ${\cal L}Y(\s)$, is valid under the above assumptions without any further specific parametric assumptions over what has been outlined above. To facilitate inference for ${\cal L}Y(\s)$, a probability distribution is specified for the parent process. 
We assume that $Y(\s)\sim GP(\mu(\s,\bbeta),K(\cdot;\sigma^2,\phi))$ is a stationary process specified on $\Rd$. In what follows we also assume that $K=K(\cdot;\sigma^2,\phi)$ admits four derivatives. There are some immediate implications of a Gaussian assumption on the parent process. If $Y_1(\cdot)$ and $Y_2(\cdot)$ are zero mean, independent stationary Gaussian processes on $\Rd$, then (i) the differential processes ${\cal L}Y_1$ and ${\cal L}Y_2$ are independent of each other; (ii) if $c_1, c_2 \in \Rone$ are scalars, then ${\cal L} (c_1Y_1+c_2Y_2)=c_1{\cal L}Y_1+c_2{\cal L}Y_2$ is 
{stationary} 
and (iii) any sub-vector of ${\cal L}Y$, for example $Y$ or $(Y,\nabla Y\trans)\trans$, is a stationary Gaussian processes.

If $K$ is $k$-times mean square differentiable (i.e. $\nabla^{2k}K$ exists), the proposed differential operator can be extended to include higher order derivatives of $\nabla^kY(\s)$ \citep[][]{mardia1996kriging}. 
Differential operators characterizing change in the response (and gradient) surface 
also follow valid stationary Gaussian processes. For instance, at an arbitrary location $\s_0$ the divergence operator, ${\rm div}{(Y(\s_0))} =\displaystyle{\sum_{i=1}^d{\del{\e_i}Y(\s_0)}}= \c_1\trans{\cal L}{Y(\s_0)}$, 
{where $\c_1$ is an $m\times 1$ vector with 0's in all places except for first order derivatives where it takes a value of 1,} 
and the Laplacian, defined as the divergence operator for gradients, $\Delta{ (Y(\s_0))}=\displaystyle{\sum_{i=1}^{d}(\nabla^2{Y(\s_0)})_{ii}=\sum_{i=1}^{d}{\deltwo{\e_i}Y(\s_0)}}=\c_2\trans{\cal L}{Y(\s_0)}$, 
{where $\c_2$ is a $m\times 1$ vector with 0's in all places except for pure second order derivatives where it takes a value of 1.} Furthermore, they
follow valid Gaussian processes 
{with ${\rm var}({\rm div} (Y(\s_0)))=\c_1\trans V_{\cal L}\c_1$ and ${\rm var}(\Delta( Y(\s_0)))=\c_2\trans V_{\cal L}\c_2$}.
{Let $Y(\s)$ be a Gaussian parent process with a twice-differentiable mean function $\mu(\s, \bbeta)$, i.e. $\nabla\mu(\s,\bbeta)$ and $\nabla^2\mu(\s,\bbeta)$ exist, and let $K(\cdot)$ be a covariance function with variance $\sigma^2$ and range $\phi$.} Let $\Y=(Y(\s_1),\ldots,Y(\s_L))\trans$ be the observed realization over ${\cal S}$ with mean $\bmu=(\mu(\s_1,\bbeta),\ldots,\mu(\s_L,\bbeta))\trans$ and $\Sigma_{\Y}$ be the associated $L\times L$ covariance matrix with elements $K(\s_i,\s_j)$, and $\s_0$ be an arbitrary location
. 
Let $\nabla\K_1=\left(
	\nabla K(\delta_1)\trans,\ldots,\nabla K(\delta_L)\trans\right)\trans$ and $\nabla\K_2=\left(vech(\nabla^2 K(\delta_1))\trans,\ldots,vech(\nabla^2 K(\delta_L))\trans\right)\trans$
be $L\times d$ and $L\times d(d+1)/2$ matrices, respectively, and $\delta_i=\s_i-\s_0$, $i=1,\ldots,L$. The distribution $P(\Y,\nabla Y(\s_0), vech(\nabla^2Y(\s_0)) \given \btheta)$, where $\btheta=\{\bbeta,\sigma^2,\phi\}$, is the $m_0=L+d+d(d+1)/2$-dimensional Gaussian,
\begin{equation}\label{eq::joint-dist}
{\cal N}_{m_0}\left(\begin{pmatrix} \bmu\\ \nabla\mu(\s_0)\\vech(\nabla^2\mu(\s_0))\end{pmatrix}, \begin{pmatrix} \Sigma_{\Y} & -\nabla \K_1 & \nabla\K_2\\
\nabla\K_1\trans & -\nabla^2K({\bf 0}) & \nabla^3K({\bf 0})\\
\nabla\K_2\trans & -\nabla^3K({\bf 0})\trans & \nabla^4K({\bf 0})\end{pmatrix}\right)\;,
\end{equation} 
{which is well-defined as long as the fourth order derivative of $K$ exists.} 
The posterior predictive distribution for the differential process 
{at $\s_0$ is} 
	\begin{equation}\label{eq::post-pred-dist}
		P(\nabla Y(\s_0),vech(\nabla^2Y(\s_0))\given \Y)=\int P(\nabla Y(\s_0),vech(\nabla^2Y(\s_0))\given \Y,\btheta)P(\btheta\given\Y)\,d\btheta\;.
	\end{equation}
{Posterior inference for curvature proceeds by sampling from $P(vech(\nabla^2Y(\s_0))\mid \Y) = \int P(vech(\nabla^2Y(\s_0))\mid \nabla Y(\s_0),\Y,\btheta)P(\nabla Y(\s_0)\mid \Y,\btheta)P(\btheta\mid \Y)\,d\btheta\, d\nabla Y$. We sample from (\ref{eq::post-pred-dist}) by drawing one instance of $(\nabla Y(\s_0), vech(\nabla^2Y(\s_0))$ for each sample of $\btheta$ 
obtained from $P(\btheta\given\Y)$.}
The conditional predictive distribution 
{of the differential process} is given by $\nabla Y(\s_0), vech(\nabla^2 Y(\s_0)) \given \Y,\btheta\sim {\cal N}_{m_1}\left(\bmu_1,\Sigma_1\right)$ 
where $m_1=d+d(d+1)/2$, and

{\allowdisplaybreaks
    \begin{align}
    \small
	\bmu_1&=\begin{pmatrix}\nabla \mu(\s_0)\\vech(\nabla^2\mu(\s_0))\end{pmatrix}-\begin{pmatrix} \nabla \K_1\\\nabla \K_2\end{pmatrix}\trans\Sigma_{\Y}^{-1}(\Y-\bmu)\;,\label{eq::cond-pred-1}\\
	\Sigma_1&=\begin{pmatrix} -\nabla^2K({\bf 0}) & \nabla^3K({\bf 0})\trans\\-\nabla^3K({\bf 0}) & \nabla^4K({\bf 0})\end{pmatrix}-\begin{pmatrix} \nabla \K_1\\\nabla \K_2\end{pmatrix}\trans\Sigma_{\Y}^{-1}\begin{pmatrix} -\nabla \K_1\\\nabla \K_2\end{pmatrix}\label{eq::cond-pred-2}\;.
    \end{align}
}
Analogous results follow for posterior predictive inference on the curvature process.

If $\mu(\s,\bbeta)=\mu$ is a constant, 
{as in simple ``kriging''}, then $\nabla\mu(\s)=\nabla^2\mu(\s)=0$. More generally, if $\mu(\s,\bbeta)=\x(\s)\trans\bbeta$, where $\x(\s)$ is a vector of spatially indexed covariates and $\x(\s)\trans\bbeta$ produces a twice differentiable trend surface then explicit calculation of $\nabla\mu(\s_0)$ and $\nabla^2\mu(\s_0)$ are possible. 
In case $Y(\s)=\mu(\s,\bbeta)+Z(\s)+\epsilon(\s)$, where $Z(\s)\sim GP({\bf 0},K(\cdot;\sigma^2,\phi))$ and $\epsilon(\s)\sim N(0,\tau^2)$ is a white noise process, inference on gradients for the residual spatial process, $Z(\s)$, can be performed from the posterior predictive distribution, $P(\nabla Z(\s_0),vech(\nabla^2Z(\s_0))\given\Y)$. 
{We address this in Section~\ref{sec::model} in the context of curvature wombling.}

\section{Wombling with Curvature Processes}\label{sec::bcw}
Bayesian wombling deals with inference for line integrals

\begin{equation}\label{eq::curve-bayes-int}
	{\Gamma}(C)=\int_C g\left({\cal L}Y\right)\,{\rm d}\ell ~~\text{ or}, ~~ \overline{\Gamma}(C)=\frac{1}{\ell(C)}\int_C g\left({\cal L}Y\right)\,{\rm d}\ell\;,
\end{equation}
where $C$ is a geometric structure of interest, such as lines or planar curves, residing within the spatial domain of reference, 
$\ell$ is an appropriate measure, often taken to 
{be the arc-length measure}, $g$ is a linear function (or functional) of the differential operator ${\cal L}Y$. 
$\Gamma$ and $\overline{\Gamma}$ are referred to as the total and average {\em wombling measures} respectively. The structure $C$ is defined to be a {\em wombling boundary} if it yields a large total (or average) wombling measure. 
Depending on 
the spatial domain, geometric structures of interest constructed within them may vary. For example, if we are dealing with surfaces in $\mathbb{R}^3$, choices of $C$ are curves and lines within the surface, with the local co-ordinate being $\mathbb{R}^2$. In higher dimensions they would be planes (curves) or hyperplanes (hypercurves). Specifically, Bayesian curvilinear wombling involves estimating integrals in (\ref{eq::curve-bayes-int}) over curves, which tracks rapid change over the spatial domain by determining boundaries (curves) with large gradients normal to the curve \citep[see for e.g.,][]{banerjee2006bayesian}.  
{The integrand in (\ref{eq::curve-bayes-int}) inherently involves a direction, in particular change measured is always in a direction normal to $C$. Hence, $g({\cal L}Y)$ can equivalently be expressed as a linear function (functional) of $\displaystyle{{\cal L}_{\n}Y(\s)}$, where $\n=\n(\s)$ denotes the unit normal vector to $C$ at $\s$. The next few paragraphs provide more detail.}

{With wombling measures for directional gradients discussed 
the Supplement, Section S1.2, 
we construct wombling measures for curvature.} 
Given $C$, depending on the smoothness of the surface, the rate at which gradients change along the curve may present sufficient heterogeneity while traversing the curve. If $C$ forms a wombling boundary with respect to the gradient, then wombling boundaries for curvature are subsets of $C$ that feature segments with large positive (negative) directional curvature along a normal direction to the curve. Leveraging only gradients, we develop wombling measures for curvature that further characterize such boundaries located for gradients. 
The wombling measure for curvature in $Y(\s)$ along $C$ ascertains whether $C$ also forms a wombling boundary with respect to curvature. 
{We associate a directional curvature to each $\s \in C$,} $\displaystyle{g({\cal L}Y(\s))=D^{(2)}_{\n,\n}Y(\s)=\c_{\n,\n}\trans vech(\nabla^2Y(\s))}$ 
{(a linear function of} $\displaystyle{{\cal L}_{\n}Y(\s)}$
{)} 
{ along the direction of a unit normal $\n=\n(\s)$ to $C$ at $\s$. 
Using (\ref{eq::curve-bayes-int}) we define {\em wombling measures} for total and average curvature as,}
	\begin{equation}
		\Gamma^{(2)}(C)=\int_C D^{(2)}_{\n, \n}Y(\s)d\ell=\int_C \n(\s)\trans\nabla^2Y(\s)\n(\s)d\ell\;,\qquad\overline{\Gamma}^{(2)}(C)=\Gamma^{(2)}(C)/\ell(C)\;,\label{eq::tot-avg-curve}
	\end{equation}
{respectively, where $\ell(C)$ denotes the arc-length of $C$. Parameterized curves, $C=\{\s(t)=(s_1(t),s_2(t)): t \in {\cal T}\subset \mathbb{R}\}$, offer further insights.
} 
As $t$ varies over its domain, $\s(t)$ outlines the curve $C$. Implicitly assuming that $C$ is regular, i.e., $||\s'(t)||\ne0$, allows the tangent and normal to exist at all points on the curve. The unit tangent and normal at each point of the curve are $\s'(t)/||\s'(t)||$ and $\n=\n(\s(t))=(s'_2(t),-s'_1(t))\trans/||\s'(t)||$, respectively, while $\displaystyle \c_{\n,\n}=\c_{\n(\s(t)),\n(\s(t))}=\left(\n(\s(t))\otimes\n(\s(t))\right)\trans {\cal D}_d$ from Section \ref{sec::sgcp}. 

The arc-length 
{of $C$ is $\ell(C)=\int_{\cal T}||\s'(t)||\,dt$ or ${\rm d}\ell=||\s'(t)||\,dt$. If ${\cal T}=[t_0,t_1]$, then $\ell(C)=\int_{t_0}^{t_1}||\s'(t)||\,dt$ and $\Gamma^{(2)}(C)=\int_{t_0}^{t_1} \n(\s(t))\trans\nabla^2Y(\s(t))\n(\s(t))||\s'(t)||\,dt$.}
{If $C$ is an open curve, then $\ell(C)^{-1}\int_C\n(\s)\trans\nabla^2Y(\s)\n(\s)d\s=\ell(C)^{-1}\int_C\n(\s(t))\trans\nabla^2Y(\s(t))\n(\s(t))||\s'(t)||\,dt$ is the average directional curvature.
    For example, $C=\{\s(t)=(r\cos t, r\sin t), t\in [0,\pi/4]\}$ is the arc of a parameterized circle of radius $r$. It follows that $||\s'(t)||=r$, $\n(\s(t))=(\cos t, \sin t)\trans$ and ${\ell(C)}^{-1}\int_0^{\pi/4}\n(\s(t))\trans\nabla^2Y(\s(t))\n(\s(t))r\,dt=\frac{4}{\pi}\int_0^{\pi/4}\n(\s(t))\trans\nabla^2Y(\s(t))\n(\s(t))\,dt$. 
	The average curvature in the tangential direction of $C$ is $\frac{1}{\ell(C)}\int_C\u(\s(t))\trans\nabla^2Y(\s(t))\u(\s(t))||\s'(t)||\,dt=\displaystyle{{\ell(C)}^{-1}\int_{t_0}^{t_1}\frac{\s'(t)}{||\s'(t)||}\trans\nabla^2Y(\s(t))\frac{\s'(t)}{||\s'(t)||}||\s'(t)||\,dt}=
    \u(\s(t_1))\trans\nabla Y(\s(t_1))-\u(\s(t_0))\trans\nabla Y(\s(t_0))$.
    Hence, the average directional curvature remains path independent and is the difference of directional gradient at the end points of $C$.} 
    
	{For a closed curve $C$, $\oint_C\n(\s)\trans\nabla^2Y(\s)\n(\s)d\s=\oint_C\n(\s(t))\trans\nabla^2Y(\s(t))\n(\s(t))||\s'(t)||\,dt$.
	If the surface admits up to three derivatives, i.e. $\nabla^3Y(\s)$ exists, the average curvature of the region, ${\cal D}$, enclosed by $C$, is free of $t$. If $\F(\s)=\nabla^2Y(\s)=(F_{ij}(\s))_{i,j=1,2}$, 
    with $F_{12}(\s) = F_{21}(\s)$ and $F_{ij}=F_{ij}(\s)=\deltwom{s_i}{s_j}Y(\s)$ then, $\oint_C\n(\s)\trans\nabla^2Y(\s)\n(\s)d\s=\oint_C\n(\s)\trans \F(\s)\n(\s)d\s=\oint_C\n(\s(t))\trans \F(\s)\n(\s(t))||\s'(t)||\,dt=\oint_C||\s'(t)||^{-1}\left(F_{11}s_2'(t)^2-2F_{12}s_1'(t)s_2'(t)+F_{22}s_1'(t)^2\right)\,dt=
    \iint\limits_{\cal D}\left\{\frac{\partial F_{11}n_2}{\partial s_1}+\left(\frac{\partial F_{12}n_2}{\partial s_2}+\frac{\partial F_{12}n_1}{\partial s_1}\right)+\frac{\partial F_{22}n_1}{\partial s_2}\right\}ds_1ds_2$.
	The last equality is obtained using Green's theorem \citep[see for e.g.,][]{rudin1976principles}. This can be interpreted as ``flux" in the gradient within ${\cal D}$. Since, $F_{ij}(\s)=\nabla^2_{ij}Y(\s)$, the integrand in the last equality require the existence of $\nabla^3_{ijk}Y(\s)$, $i,j,k=1,2$. Denoting, $\widetilde{\nabla}^3Y(\s)=(\nabla^3_{ijk}Y(\s))_{i,j,k=1,2}\trans$, vector of unique third derivatives, and $\n_0(\s)=(n_2(\s),n_2(\s),n_1(\s),n_1(\s))\trans$ then,
	\begin{equation}\label{eq::closed-curve-n}
	        \frac{1}{\ell(C)}\oint_C\c_{\n,\n}\trans vech(\nabla^2 Y(\s))\,d\s
            =\frac{1}{\ell(C)}\iint\limits_{\cal D}\n_0(\s)\trans\widetilde{\nabla}^3Y(\s)\,d\s.
	\end{equation}
    This extends the development in Section 3.2 of  \cite{banerjee2006bayesian} to study the behavior of spatial curvature over closed curves on surfaces in $\mathbb{R}^3$. Sampling along $C$ is generally harder than sampling inside ${\cal D}$. Hence, the computational implications of (\ref{eq::closed-curve-n}) are more appealing. When studying the same behavior along a tangential direction to $C$ with $\s(t_0)=\s(t_1)=\s_0$, $\displaystyle \oint_C\u(\s)\trans\nabla^2Y(\s)\u(\s)d\s=\oint_{\s(t_0)}^{\s(t_1)} F_{11}(\s)n_1ds_1+F_{12}(\s)n_1ds_2+ F_{21}(\s)n_2ds_1+F_{22}(\s)n_2ds_2=\u(\s(t_1))\trans\nabla Y(\s(t_1))-\u(\s(t_0))\trans\nabla Y(\s(t_0))=0$, again a consequence of path independence.
    This validates the choice of a normal direction to $C$ when measuring change in the gradient. Using the rectilinear approximation to curvature wombling, as discussed later, provides a more computationally tractable and simpler approach, where double integrals manifest when computing variances of the wombling measures.}

{Curvature wombling} requires predictive inference performed using gradient measures on the interval ${\cal T}$, to include $\Gamma^{(2)}({C})$ (or $\overline{\Gamma}^{(2)}({C})$) in (\ref{eq::tot-avg-curve}). Leveraging inference for differential processes in Section~\ref{sec::sgcp}, we obtain joint inference on the wombling measures. Suppose $C=\{\s(t):t \in [0,T]\}$ is generated over ${\cal T}=[0,T]$. For any $t^*\in [0,T]$, let $C_{t^*}$ denote the curve restricted to $[0,t^*]$ and $\ell({C_{t^*}})$ its arc-length. Line integrals for curvilinear gradient and curvature wombling measures are $\Gamma^{(1)}({C_{t^*}})=\int_0^{t^*}D^{(1)}_{\n}Y(\s(t))||\s'(t)||\,dt,~ \overline{\Gamma}^{(1)}({C_{t^*}})=\frac{1}{\ell({C_{t^*}})}\Gamma^{(1)}({C_{t^*}})$,
$\Gamma^{(2)}({C_{t^*}})=\int_0^{t^*}D^{(2)}_{\n,\n}Y(\s(t))||\s'(t)||\,dt$ and $\overline{\Gamma}^{(2)}({C_{t^*}})=\frac{1}{\ell({C_{t^*}})}\Gamma^{(2)}({C_{t^*}})$.
Since %
$D^{(1)}_{\n}Y(\s(t))$ and $D^{(2)}_{\n,\n}Y(\s(t))$ 
are %
Gaussian processes on ${\cal T}$
, $\Gamma^{(1)}({C_{t^*}})$ and $\Gamma^{(2)}({C_{t^*}})$ 
{are valid {\em dependent} Gaussian processes on ${\cal T}$}. Therefore, $\bGamma({C_{t^*}}) = (\Gamma^{(1)}({C_{t^*}}), \Gamma^{(2)}({C_{t^*}}))\trans \sim {\cal N}_2\big(\bmu_{\bGamma}(t^*),\K_{\bGamma}(t^*,t^*)\big)$, where $\bmu_{\bGamma}(t^*)= \left(\int_0^{t^*}D^{(1)}_{\n}\mu(\s(t))||\s'(t)||\,dt\;, \int_0^{t^*}D^{(2)}_{\n,\n}\mu(\s(t))||\s'(t)||\,dt\right)\trans=(m_1(t^*),m_2(t^*))\trans$ and $\K_{\bGamma}(t^*,t^*)=\{k_{ij}(t^*,t^*)\}_{i,j=1,2}$ whose elements are evaluated as
\begin{equation}\label{eq::womb-var}
    k_{ij}(t^*,t^*)=(-1)^j\int_0^{t^*}\int_0^{t^*}\a_i\trans(t_1)\nabla^{i+j}K(\Delta(t_1,t_2))\a_j(t_2)||\s'(t_1)||||\s'(t_2)||\,dt_1\,dt_2\;,
\end{equation}
{where $\a_1(t)=\n(\s(t))$ and $\a_2(t)=\c_{\n(\s(t)),\n(\s(t))}$.}
Simplifications arise in $d=2$. For example, $\c_{\n,\n}(t)=(s'_2(t)^2,-2s'_2(t)s'_1(t),s'_1(t)^2)\trans$, while $\nabla^kK$, for $k=2,3,4$, are matrices of orders $2\times2$, $2\times 3$ and $3\times 3$, respectively, of partial and mixed second, third and fourth derivatives of $K$ and $\Delta(t_1,t_2)=\s(t_2)-\s(t_1)$. For any two points $t_1^*,t_2^*\in {\cal T}$, the dependence 
is specified through $\begin{pmatrix}\bGamma({C_{t_1^*}})\\\bGamma({C_{t_2^*}})\end{pmatrix}\sim {\cal N}_{4}\left(\begin{pmatrix}\m_1\\\m_2\end{pmatrix},\begin{pmatrix}\k_{11}&\k_{12}\\\
	\k_{21}&\k_{22}\end{pmatrix}\right)$,
where $\m_i=(m_i(t_1^*),m_i(t_2^*))\trans$, $\k_{ij}=\begin{pmatrix}k_{ij}(t_1^*,t_1^*)&k_{ij}(t_1^*,t_2^*)\\k_{ij}(t_2^*,t_1^*)&k_{ij}(t_2^*,t_2^*)\end{pmatrix}$, $i,j=1,2$. 
Generally, for $n_P$ points partitioning ${\cal T}$ the above can be analogously extended. Clearly, $\bGamma({C_{t^*}})$ is a mean squared continuous process. However, stationarity of 
$Y(\s)$ does not imply stationarity of $\bGamma({C_{t^*}})$. 
For any $\s_j \in {\cal S}$ with $\Cov(Y(\s_j),\bGamma({C_{t^*}}))=\bgamma_j(t^*)$ and $\Delta_j(t)=\s(t)-\s_j$ we have,

\begin{equation}
\bgamma_j(t^*)=\left(\int_0^{t^*}D^{(1)}_{\n}K(\Delta_j(t))||\s'(t)||\,dt,\int_0^{t^*}D^{(2)}_{\n,\n}K(\Delta_j(t))||\s'(t)||\,dt\right)\trans\;.\label{eq::cross-womb}
\end{equation}
A valid {\em joint distribution} can be specified over ${\cal T}$ by,

\begin{equation}\label{eq::joint-dist-womb}
\begin{pmatrix}\Y \\  \bGamma({C_{t^*}}) \end{pmatrix} \sim {\cal N}_{L+2}\left(\begin{pmatrix}\bmu\\\mu_{\bGamma}(t^*)\end{pmatrix},\begin{pmatrix}\Sigma_{\Y}&\bgamma_{\bGamma}(t^*)\\\bgamma\trans_{\bGamma}(t^*) & \K_{\bGamma}(t^*,t^*) \end{pmatrix}\right)\;,
\end{equation}
where $\bgamma\trans_{\bGamma}(t^*)=\left[\bgamma_{1}(t^*) \; \bgamma_{2}(t^*)\;\cdots\;\bgamma_{L}(t^*)\right]$ is the $2\times L$ cross-covariance matrix. 

{In practical applications curvilinear wombling is performed by approximating the curve $C$ 
using linear segments.} 
These measures at the segment level are then aggregated to produce a wombling measure for the curve. The curve is segmented using a partition. Consequently, the accuracy of estimated wombling measures for the curve depend on the choice of partition. Figures S2 and S3 
in the online Supplement illustrate this concept. 
{Explicitly, let $C$ 
be a regular rectifiable curve 
and $[a,b]\subset {\cal T}$ be a compact interval. Let $g$ be a uniformly continuous function. 
For any partition, $P$ of $[a,b]$, $a=t'_0<t'_1<\ldots<t'_{n_P}=b$, with its norm defined as $|P|=\max\limits_{i=1,\ldots,n_P}(t'_i-t'_{i-1})$. A polygonal (piecewise-linear) approximation to the curve is, $\widetilde{C}_P=\bigcup\limits_{i=1}^{n_P}C_{t_i}$, where $C_{t_i}=\{\s(t'_{i-1})+t\u_i,t\in[0,t_i]\}$, $t_i=||\s(t'_i)-\s(t'_{i-1})||$ and $\u_i=||\s(t'_i)-\s(t'_{i-1})||^{-1}(\s(t'_i)-\s(t'_{i-1}))\trans$. 
Note that $\s(t)=\s(t'_{i-1})+t\u_i$ for $t\in[0,t_i]$ and, hence, $||\s'(t)||=||\u_i||=1$. Wombling measure for $\widetilde{C}_P$ is, $\Gamma(\widetilde{C}_P)=\sum\limits_{i=1}^{n_P}\int_{C_{t_i}}g\left({\cal L} Y(\s(t))\right)||\s'(t)||\,dt$. 
As $|P|\to 0$ we have, $\Gamma(\widetilde{C}_P)\overset{a.s.}{\longrightarrow}\Gamma(C)=\int_a^b g\left({\cal L}Y(\s(t))\right)||\s'(t)||\,dt$.} This provides us with an estimate, $\Gamma(\widetilde{C}_P)$ for curvilinear wombling measures associated with any general curve $C$. 
{Further details are provided in the Supplement, at the end of Section S5
.} 

{The choices of $g$ for our wombling measures result in, $\u\trans\nabla Y$ and $\c_{\u,\u}\trans vech(\nabla^2Y)$, which are linear and therefore uniformly continuous over any compact interval.} 
Since predictive inference is performed iteratively on individual line segments, 
it is sufficient to show the inferential procedure for an arbitrary curve segment 
{$C_{t_i}$. The normal to $C_{t_i}$ is free of $t$ and denoted as, $\u_i^{\perp}$, which is the normal to $\u_i$.} The associated wombling measures with $C_{t_i}$ are $\displaystyle\bGamma(t_i)=\left(\int_0^{t_i}D^{(1)}_{\u_i^{\perp}}Y(\s(t))\,dt,\int_0^{t_i}D^{(2)}_{\u_i^{\perp},\u_i^{\perp}}Y(\s(t))\,dt\right)\trans$.
		For a point $\s_j$ define $\Delta_{i-1,j}=\s_{i-1}-\s_j$, $j=1,2,\ldots,L$. 
		{Their joint distribution is specified by (\ref{eq::joint-dist-womb}), where $\bgamma_{j}(t_i)$ is obtained from (\ref{eq::cross-womb}) by replacing $\Delta_j(t)$ with $\Delta_{i-1,j}+t\u_i$ and} 
{$\K_{\bGamma}(t_i,t_i)$ is obtained from (\ref{eq::womb-var}) replacing $\Delta(t_1,t_2)=(t_2-t_1)\u_i$ in the integrand.} 
The analytic tractability of the line integrals in $\bgamma_{j}(t_i)$ is not a concern. Given choices of $\mu(\cdot)$ and $K(\cdot)$, they are all one or two dimensional integrals which are efficiently computed using simple quadrature. 
For example, 
let $Y(\s)$ be the isotropic Gaussian process with mean $\mu(\s)=\mu$ 
and $K(||\Delta||;\sigma^2,\phi)=\sigma^2\exp(-\phi||\Delta||^2)$, where $\Delta=(\delta_1,\delta_2)\trans$. $\nabla^kK(\Delta)$, $k=2,3,4$ is obtained from (\ref{eq::nabla-k}) and related results.
$\bgamma_{j}(t_i)=\bgamma_{j}(t_i;\sigma^2,\phi)=\left\{\Phi\left(\sqrt{2\phi}\left(t_{i}+\u_i\trans\Delta_{i-1,j}\right)\right)-\Phi\left(\sqrt{2\phi}\u_i\trans\Delta_{i-1,j}\right)\right\}(c_1,c_2)\trans$ where, $c_1=c_1(\sigma^2,\phi,\u_i^\perp,\Delta_{i-1,j})=-2\sigma^2\sqrt{\pi\phi}{\u_i^\perp}\trans\Delta_{i-1,j}e^{-\phi\left({\u_i^\perp}\trans\Delta_{i-1,j}\right)^2}$, $c_2=c_2(\sigma^2,\phi,\u_i^\perp,\Delta_{i-1,j})=c_1(1-2\phi{\u_i^\perp}\trans\Delta_{i-1,j}\Delta_{i-1,j}\trans\u_i^\perp)$, and $\Phi(\cdot)$ denotes the standard Gaussian cumulative distribution function. These are simple computations with 
quadrature required only for computing $K_{\bGamma}(t_i,t_i)$.

\section{Bayesian Hierarchical Model}\label{sec::model}
We operate under a Bayesian hierarchical model, which is specified as

\begin{equation}\label{eq::hier-sp-mod}
		Y(\s)=\mu(\s,\bbeta)+Z(\s)+\epsilon(\s)\;,
\end{equation}
where $Z(\s)\sim GP(0,K(\cdot;\sigma^2,\phi))$ is a Gaussian process, and $\epsilon(\s)\sim N(0,\tau^2)$ is a white noise process, termed as the nugget \citep[see][and references therein]{banerjee2014hierarchical}. The process parameters are $\btheta=\{\bbeta,\sigma^2,\phi,\tau^2\}$. More generally, we can consider a latent specification for response arising from exponential families, $\alpha(\bleta(\s))=\x\trans(\s)\bbeta+Z(\s)+\epsilon(\s)$, $Z(\s)\sim GP(0,K(\cdot;\sigma^2,\phi))$ and $Y(\s)\sim \pi\left(\bleta(\s),\cdot\right)$, where $\alpha$ is a monotonic link function, $\pi$ is a member of the exponential family and $\bleta$ is the natural parameter. 
Predictive inference on differential processes and curvature wombling proceeds on the latent surface through $P({\cal L}Z\given\Y)$. The joint posterior for differential processes is obtained through, $P(\nabla Z\trans,vech(\nabla^2Z)\trans\given\Y)=\int P(\nabla Z\trans,vech(\nabla^2Z)\trans\given\Z,\btheta)P(\Z\given\Y,\btheta)P(\btheta\given\Y)\,d\btheta \,d\Z$, while wombling measures $\bGamma_Z(t^*)$ for a curve $C_{t^*}$ within the estimated posterior surface for $\Z$, are sampled from the posterior, $P(\bGamma_Z(t^*)\given\Y)=\int P(\bGamma_Z(t^*)\given\Z,\btheta)P(\Z\given\Y,\btheta)P(\btheta\given\Y)\,d\btheta\, d\Z$. 
Customary prior specifications 
for $\btheta$ yield 
\begin{equation}\label{eq::pure-sp}
\begin{split}
P(\btheta, \Z\given\Y) &\propto U(\phi\given a_{\phi},b_{\phi})\times IG(\sigma^2\given a_\sigma,b_\sigma)\times IG(\tau^2\given a_\tau,b_\tau) \times  \mathcal{N}_{L}(\Z\given\mathbf{0},\sigma^2\R_Z)\\
& \qquad \times  \mathcal{N}_{p}(\bbeta\given\mu_{\beta},\Sigma_{\beta})\times \prod\limits_{l=1}^{L}\mathcal{N}_1\big(Y(\s_l)\given\x(\s_l)\trans\bbeta+Z(\s_l),\tau^2\big)\;,
\end{split}
\end{equation}
where $IG$ denotes the inverse-gamma distribution with a shape-rate parameterization, $U$ is a uniform distribution and $\R_Z$ 
is the correlation matrix corresponding to $K(\cdot;\sigma^2,\phi)$. The resulting full conditionals are $\bbeta\given\tau^2,\Z,\Y \sim  \mathcal{N}_{p}(M_{\beta}m_{\beta},M_{\beta})$, $\sigma^2\given\phi,\Z \sim IG(a_\sigma+\frac{L}{2},b_\sigma+\frac{1}{2}\Z\trans \R^{-1}_Z(\cdot;\phi)\Z)$, $\tau^2\given\bbeta,\Z,\Y \sim IG\left(a_\tau+\frac{L}{2},b_\tau+\frac{1}{2}||\Y-\X\bbeta-\Z||_2^2\right)$,  $\Z\given\Y,\btheta \sim \mathcal{N}_{L}(M_{Z}m_{Z},\tau^2M_{Z})$, where $\X$ is the $L\times p$ matrix with $\x(\s_i)\trans$ as rows, $M_{\beta}^{-1} = \Sigma_{\beta}^{-1}+\tau^{-2}\X\trans\X$, $m_{\beta}=\Sigma_{\beta}^{-1}\mu_{\beta}+\tau^{-2}\X\trans(\Y-\Z)$, $M_{Z}^{-1}=\tau^{-2}\big(\tau^{-2}I_L+\sigma^{-2}\R_Z^{-1}(\cdot;\phi)\big)$, and $m_{Z}=\Y-\X\bbeta$. $\phi$ is updated using Metropolis 
steps with a normal proposal and an adaptive variance. 

Under this setup posterior samples for the differential processes and wombling measures result from (\ref{eq::cond-pred-1}) and (\ref{eq::cond-pred-2}). For each posterior sample of $\{\Z,\btheta\}$, we draw $\bGamma_Z(t^*)\given \Z, \btheta \sim {\cal N}_2\big(\mu_{\bGamma_ Z}(t^*)-\bgamma\trans_{\bGamma_Z}(t^*)\Sigma_{\Z}^{-1}\Z,K_{\bGamma_Z}(t^*,t^*)-\bgamma\trans_{\bGamma_Z}(t^*)\Sigma_{\Z}^{-1}\bgamma_{\bGamma_Z}(t^*)\big)$,
where $\mu_{\bGamma_Z}(t^*)$, $\bgamma_{\bGamma_Z}(t^*)$, and $K_{\bGamma_Z}(t^*,t^*)$ are computed from (\ref{eq::womb-var}) and (\ref{eq::cross-womb}). Algorithms~1 and 2 
in the Supplement, Section~S4, 
present further details for posterior sampling. 
Next, we turn to numerical experiments and data analyses. Codes required for reproducing and emulating the analyses presented in the manuscript are produced for the \texttt{R} statistical programming environment and available for download in the public domain 
at {\if1\blind \url{https://github.com/arh926/spWombling}\fi}  {\if0\blind \textcolor{purple}{[Redacted for blinded version]}\fi}.

\section{Simulation Experiments}\label{sec::syn-exp}
\subsection{Data generation}
The proposed differential processes are not observed in reality, 
but %
are induced by an observed spatially indexed parent process. 
To 
evaluate statistical learning of the curvature process
we perform simulation experiments within a setup where true values of the differential process and wombling measures are available. 
We consider locations $\s=(s_1,s_2)\trans\in \mathbb{R}^2$ over the unit square
$[0,1]\times[0,1]\subset\mathbb{R}^2$.  We generate synthetic data from two distributions: (a) Pattern 1: $y_1(\s) \sim N(10[\sin(3\pi s_1)+\cos(3\pi s_2)],\tau^2)$; (b) Pattern 2: $y_2(\s) \sim N(10[\sin(3\pi s_1)\cdot\cos(3\pi s_2)],\tau^2)$,
\begin{figure}[t]
	\centering
	\includegraphics[width=0.8\linewidth , height=0.27\linewidth]{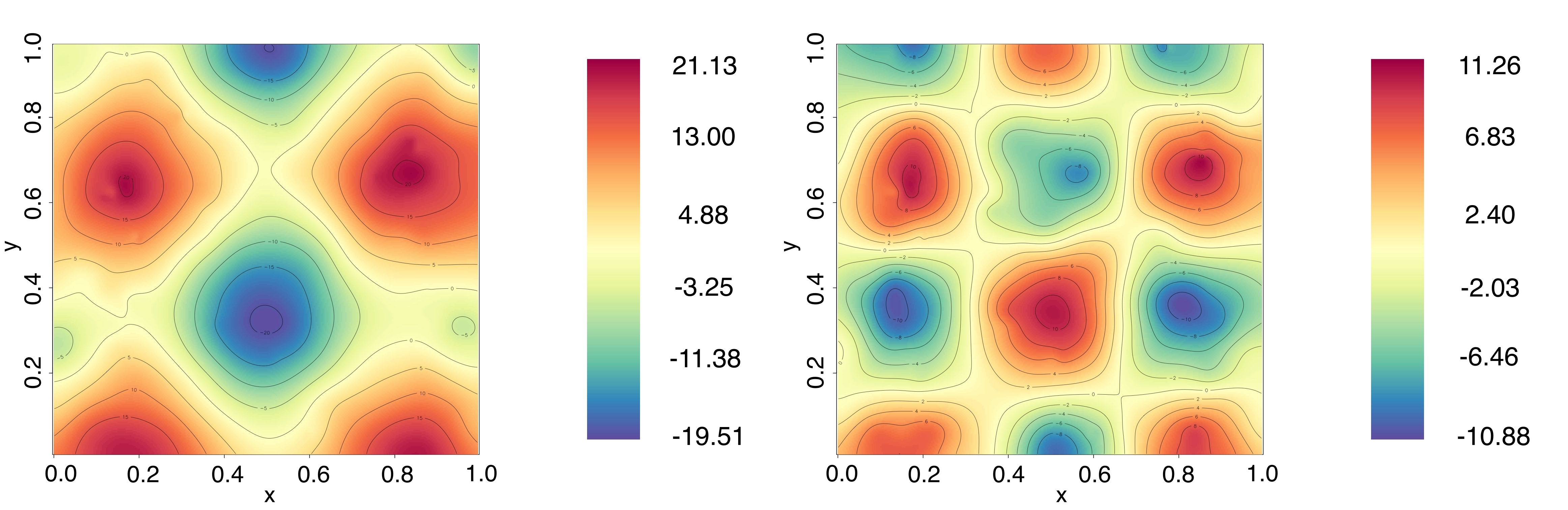}
	\caption{Spatial plots for synthetic patterns, from Pattern 1 (left) and Pattern 2 (right). Scales are shown in the legend alongside.}\label{fig::sim-pattern}
\end{figure}
where the value of $\tau^2=1$. Figure \ref{fig::sim-pattern} presents spatial plots of the generated synthetic response from these patterns. 
The rationale behind selecting these distributions is: (i) synthetic data is more practical and not from the model in (\ref{eq::hier-sp-mod}), and (ii) true gradient and curvature can be computed at every location $\s$. 

The synthetic patterns chosen feature two different scenarios that may arise. In the first pattern 
expressions for differentials along the principal directions $\e_1=(1,0)\trans$ and $\e_2=(0,1)\trans$ are functions of either $s_1$ or $s_2$, $\nabla\mu_1(\s)=30\pi(\cos(3\pi s_1),-\sin(3\pi s_2))\trans$, $\nabla^2\mu_1(\s)=-90\pi^2{\rm diag}\{\sin(3\pi s_1), \cos(3\pi s_2)\}$.
The curvature along $s_1$ does not influence curvature along $s_2$, $\left(\nabla^2\mu_1(\s)\right)_{12}=0$ for all $\s$. While 
$\nabla\mu_2(\s)=30\pi(\cos(3\pi s_1)\cos(3\pi s_2),-\sin(3\pi s_1)\sin(3\pi s_2))\trans$, $\nabla^2\mu_2(\s)= -90\pi^2M(\s)$, where $M(\s)$ is a $2\times 2$ matrix with, $m_{11}=\sin(3\pi s_1)\cos(3\pi s_2)$, $m_{12} = m_{21}= \cos(3\pi s_1)\sin(3\pi s_2)$ and $m_{22}=\sin(3\pi s_1)\cos(3\pi s_2)$ with differentials being functions of both $s_1$ and $s_2$ and $\left(\nabla^2\mu(\s)\right)_{12}\ne 0$ for some $\s$.
While setting up the experiments we vary 
$L\in \{100,500,1000\}$ with 10 replicated instances under each setting. 

\subsection{Bayesian model fitting}
We fit the model in (\ref{eq::pure-sp}) with only an intercept allowing the spatial process to learn the functional patterns in the synthetic response. We use the following hyper-parameter values in (\ref{eq::pure-sp}): $a_{\phi}=3/\max||\Delta||$, $b_{\phi}=30$, $a_\sigma=2$, $b_\sigma=1$, 
$a_\tau=2$, $b_\tau=0.1$ 
$\mu_\beta=0$ and $\Sigma_\beta=10^6I_p$. These choices comprise reasonable weakly informative priors. While a $\rm Uniform(2,3)$ prior on $\nu$ can be specified (and was implemented as part of this experiment) to ensure the existence of the curvature process, here our choice of scales in the data generating patterns ensured that $\nu = 5/2$ provided the best model fit when compared with values of $\nu\in \{1/2, 3/2, 5/2\}$. Hence, we present the results with $\nu=5/2$. 

The parameter estimates for $\btheta$ are computed using posterior medians and their highest posterior density (HPD) intervals \citep[][]{chen1999monte,plummer2015package}. For each replicate, we assess our ability to estimate the local geometry of the resulting posterior surface. For this we overlay a grid spanning the unit square. 
We perform posterior predictive inference for the differential processes at each grid location following Section~\ref{sec::sgcp}. 
Posterior predictive medians (accompanied by $95$\% HPD intervals) 
summarize inference for the differential processes over the grid locations (Section~\ref{sec::supp} offers supplementary analysis). 

\subsection{Bayesian wombling with curvature processes}\label{sec::bcw-curvep}
\begin{figure}[t]
	\begin{subfigure}{.3\textwidth}
		\centering
		\includegraphics[width=1\linewidth , height=1.1\linewidth]{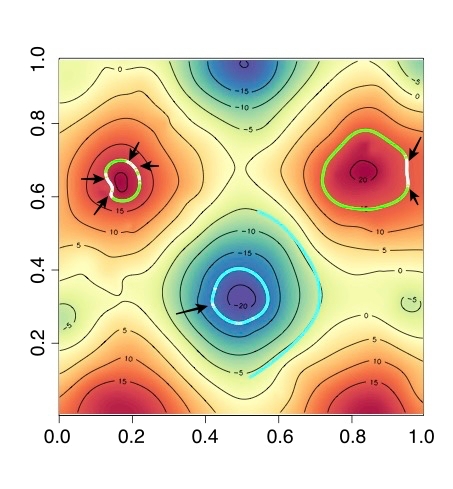}
	\end{subfigure}%
	\begin{subfigure}{.3\textwidth}
		\centering
		\includegraphics[width=1\linewidth , height=1.1\linewidth]{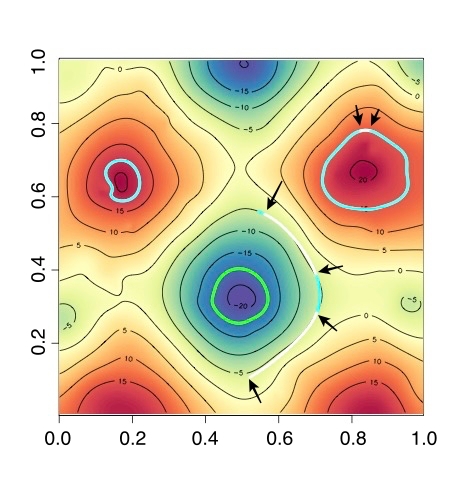}
	\end{subfigure}
	\begin{subfigure}{0.3\textwidth}
		\centering
		\includegraphics[width=1.4\linewidth , height=1.1\linewidth]{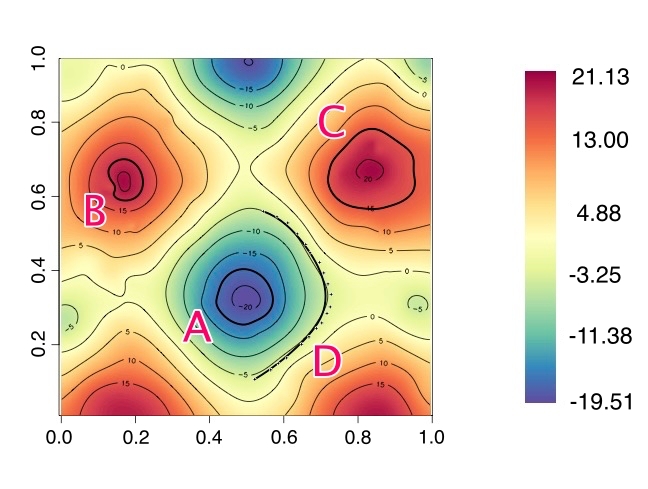}
	\end{subfigure}
	\caption{(left) shows color coded directional gradients for segments, (center) shows color coded directional curvature for segments in the direction normal to the curve, (right) shows curves selected for performing curvature wombling. {\tt green} indicates a positive significance, {\tt cyan} indicates negative significance and {\tt white} indicates no significance.}\label{fig::womble}
\end{figure}

For wombling with curvature processes, or \emph{curvature wombling}, we focus on locating curves that track rapid change within the simulated random surfaces. For example, consider the surface produced by the first pattern. 
If a curve is provided to us, we can evaluate the posterior distribution of the average or total curvature wombling measures to assess their statistical significance. On the other hand, without a given curve, we consider three different approaches for constructing them from a boundary analysis or wombling perspective: (a) level curves: $C_{y_0}=\{\s:Y(\s)=y_0\}$: Bayesian wombling literature finds that curves parallel to contours often form wombling boundaries \citep[see, e.g.,][]{banerjee2006bayesian} and level curves on a surface are parallel to local contours by definition; (b) smooth curves: produces a smooth curve using B\'ezier splines \citep[see, e.g.,][]{gallier2000curves} from a set of {\em annotated} points that are of interest within the surface; and (c) rectilinear curves: produces a rectilinear curve joining adjacent {\em annotated} points of interest within the surface using straight lines, performs curvature wombling using a {Riemann sum approximation} (see (S1
) in the Supplement).
Curves of types (b) and (c) allow the investigator to specify a region of interest that house possible wombling boundaries. For the surface realization produced by Pattern 1, we consider four different types of curves on the response surface, (A) a closed curve enclosing a trough corresponding to a level curve, $C_{y_0=-18}$, (B) a closed curve enclosing a peak corresponding to a level curve, $C_{y_0=+18}$, (C) a closed curve that outlines a contour corresponding to a level curve, $C_{y_0=+15}$ and (D) an open curve along a contour constructed using a B\'ezier spline. These curves are marked in Figure \ref{fig::womble}c. 

Curvature wombling is performed using methods outlined in Section \ref{sec::bcw}.
{Referring to the discussion on rectilinear approximation, for each curve, given a partition, we compute $t_i$ and $\u_i$. 
Combining the segments produces a vector $\t$ and a matrix of directions, $\U$ that represents the curve. 
Algorithm 2 
in the Supplement, Section S4 
devises efficient computation using $\t$ and $\U$.}
The total (and average) wombling measures $\overline{\bGamma}(C)$ are sampled from their posteriors using (\ref{eq::joint-dist-womb}). 
For curves A, B, C and D, we use partitions with sufficiently small norms ($|P|$) to achieve accuracy ($3.99\times10^{-3}$, $3.97\times 10^{-3}$, $4.42\times 10^{-3}$ and $2.66\times 10^{-2}$ respectively). 
One and two dimensional line integrals (refer to (\ref{eq::womb-var}) and (\ref{eq::cross-womb})) are computed via quadrature using grids of size 10 on $[0,t_i]$, and size 100 on $[0,t_i]\times [0,t_i]$ respectively, for $i=1,2,\ldots,n_P$. The median of sampled $\overline{\bGamma}(\widetilde{C}_P)$ is our estimated wombling measure for the curve. Significance at the curve-segment level is assessed based on the inclusion of 0 within the HPD intervals. 
{Our design allows us to compute true values of average wombling measures for each rectilinear segment in the curve. They are computed using, $\mu_{\small \bGamma}^{true}(\widetilde{C}_P)=\left(\sum_{i=1}^{n_P}t_i\right)^{-1}\left(\sum_{i=1}^{n_P}\int_0^{t_i}{\u_i^{\perp}}\trans\nabla\mu_1(\s(t))\,dt,\sum_{i=1}^{n_P}\int_0^{t_i}{\u_i^{\perp}}\trans\nabla^2\mu_1(\s(t))\u_i^{\perp}\,dt\right)\trans$. We compute HPD intervals  
for the wombling measures 
at the segment level.  
Coverage probabilities (CPs) are then constructed by aggregating coverage of true values by HPD intervals over segments.} 
\begin{table}[t]
	\renewcommand{\arraystretch}{0.8}
	\caption{Results from curvature wombling performed on curves A, B, C and D as shown in Figure \ref{fig::womble}. The estimated average directional gradient and curvature are accompanied by their respective HPD intervals in brackets. HPD intervals {\em containing 0} are marked in bold.}\label{tab::womb-results-2}
	\centering
	\resizebox*{\linewidth}{!}{
		\begin{tabular}{l|@{\extracolsep{40pt}}cccc@{}}
			\hline
			\hline
			\multirow{4}{*}{Curves $(C)$} & \multicolumn{2}{c}{\multirow{2}{*}{Average Gradient ($\Gamma^{(1)}(C)$)}} & \multicolumn{2}{c}{\multirow{2}{*}{Average Curvature ($\Gamma^{(2)}(C)$)}}\\
			&&&&\\
			\cline{2-3}\cline{4-5}
			&\multirow{2}{*}{True}& \multirow{2}{*}{Estimated}&\multirow{2}{*}{True} &\multirow{2}{*}{Estimated}\\
			&&&&\\
			\hline
			\multirow{2}{*}{Curve A} & \multirow{2}{*}{-61.54} & -64.97 &  \multirow{2}{*}{731.94}  & 768.03 \\
			&&(-92.37, -38.57) & &  (599.30, 913.70)\\
			\multirow{2}{*}{Curve B} & \multirow{2}{*}{40.85} & 49.19 &  \multirow{2}{*}{-808.04}  & -850.84 \\
			&&(20.45, 73.12) & &  (-1066.98,  -630.09)\\
			\multirow{2}{*}{Curve C} & \multirow{2}{*}{84.03} & 85.65 &  \multirow{2}{*}{-558.58}  & -504.98 \\
			&&(59.81, 109.97) & &  (-767.55, -241.61)\\
			\multirow{2}{*}{Curve D} & \multirow{2}{*}{-110.84} & -113.27 &  \multirow{2}{*}{11.32}  & \bf -94.64 \\
			&&(-153.23, -77.01) & &  ({\bf -386.94, 233.78})\\
			\hline
			\hline
		\end{tabular}
	}
\end{table}

Curve A encloses a trough and a local minima for the surface, while B and C enclose peaks and local maximums (referring to corresponding locations in Figures S8c and S9c). 
Along all  segments of A we expect negative gradients owing to the decreasing nature of the response in that region, while for B and C we expect positive gradients. Each of them would be expected to yield significant wombling measures for gradients. Referring to the Laplacian surface (see Supplement, Figures S8e and S9e) 
A, B, and C are located in regions manifesting rapid change in the gradient surface, implying they should yield large positive (curve A) or negative (curves B and C) curvature, forming curvature wombling boundaries. These are all aligned with our findings presented in Table \ref{tab::womb-results-2}, which presents measures of quality assessment for wombling. The magnitude and sign of wombling measures also allow us to differentiate between the type of curvature for the different wombling boundaries. For instance, B is located in a region of higher convexity compared to C, while the nature of convexity for regions enclosed by them are different compared to A. Plots in Figure \ref{fig::womble} (left and center) show line segment level inference for average wombling measures. Arrows indicate segments which were not significant with respect to gradient or curvature, while regions of significance are color coded. D is located in a ``relatively flat" region of the surface (see Figures S8e and S9e) 
and is expected to have gradients but no curvature, which aligns with results shown in Table \ref{tab::womb-results-2}. 
We conclude by noting that the true values, $\mu_{\small \bGamma}^{true}(C)$ of the wombling measures for the curves considered, are all covered by the estimated HPD intervals for respective curves. Additionally, at the line segment level we achieved a CP of 1.0 across all curves.

\subsection{Supplementary analysis}\label{sec::supp}
We present additional results in the online supplement. Tables S1 
and 
S2 present parameter estimates, measures of goodness of fit for the fitted process, and assessment of derivative process characteristics for each pattern considered. We compute root mean square errors (RMSE) across observed locations averaged over 10 replicates for each sample size setting for the fitted process $\widehat{Y}(\s) = \widehat{\beta_0}+\widehat{Z}(\s)$, and $\widehat{\nabla Y(\s)}$, $\widehat{vech(\nabla^2 Y(\s)})$. We report standard deviations across replicates. With increasing number of observed locations we are able to effectively learn the underlying process and induced differential processes
. Figures 
S4, S5, S6 and S7 present spatial plots of posterior medians of gradient and curvature processes, for $L=100$ locations. These plots demonstrate the effectiveness of our methods in learning about the differential processes from the underlying patterns. Similarly plots shown in Figures 
S8, S9, S10 and S11 demonstrate the same for derived quantities and operators of ${\cal L}Y(\s)$---
{principal curvature (eigenvalues), Gaussian curvature (determinant) \citep[see, e.g.,][]{spivak1970comprehensive,do2016differential}, divergence and Laplacian, which pertain to geometric analysis of curvature for the random surface resulting from the underlying patterns.} Statistical significance is assessed at every grid point by checking the inclusion of 0 in their HPD intervals. Significantly positive (negative) points are color coded. We compute average CPs at every grid location to measure the accuracy of our assessment. These CPs are then averaged over replicates. We observed high CPs across the grid for parent and differential processes. Figures 
S12 and S13 compare observed against estimated differential processes coupled with their HPD regions. 

\section{Applications}\label{sec::app}

Frameworks developed  for differential assessment and boundary analysis in spatially indexed response are applied to multiple data sets with the aim of locating curvature wombling boundaries that track rapid change in response. The chosen data arise from varied areas of scientific interest, we briefly describe the origin and significance of each with respect to our methods before performing our analysis. 
Response is modeled using the hierarchical model in  (\ref{eq::hier-sp-mod}). Prior specifications used in (\ref{eq::pure-sp}) are, $\phi\sim {\rm Unif}\left(3/\max_{\s\in {\cal S}}{||\Delta||},300\right)$, $\sigma^2\sim IG(2,1)$, $\tau^2\sim IG(2,1)$ (mean 1, infinite variance), $\bbeta\sim N(0,10^6I_p)$, $p$ being the number of covariates and $\nu=5/2$ for the Mat\'ern kernel ensuring existence of the differential processes.

\begin{figure}[t]
	\centering
	\includegraphics[width=0.8\linewidth , height=0.3\linewidth]{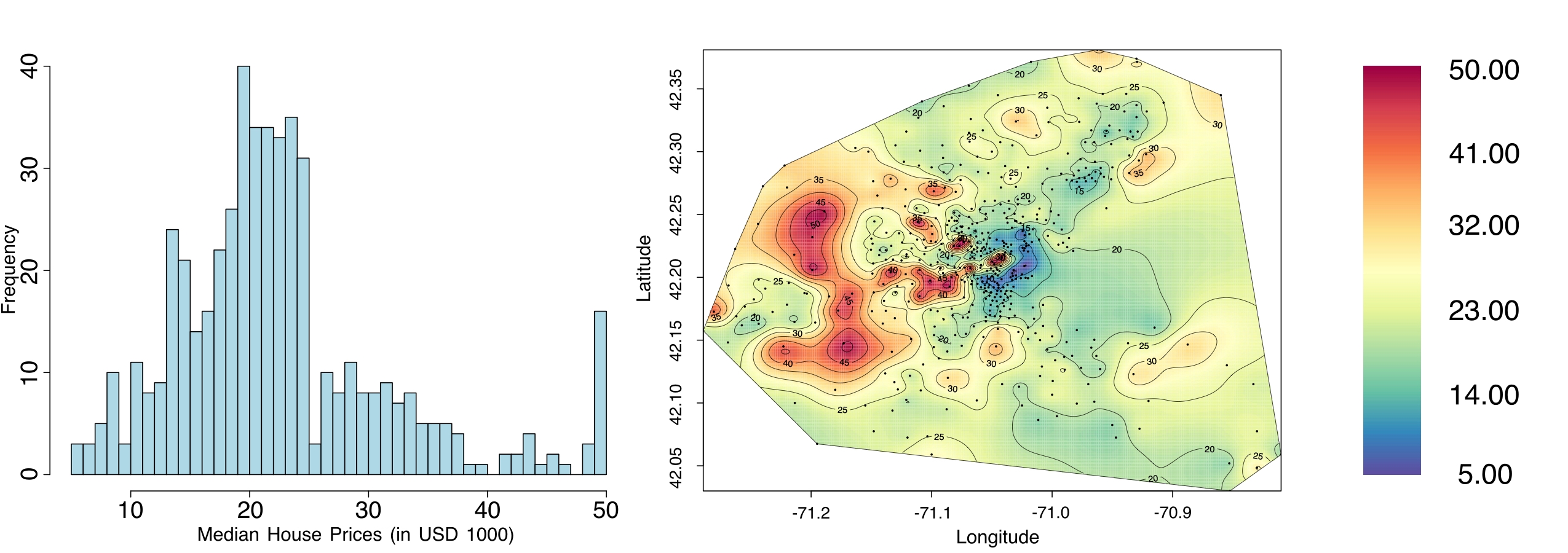}
	\caption{Plots showing (left) probability density of median house prices (in USD 1000) (right) spatial plot of median owner occupied house prices in Boston.}
	\label{fig::boston-data}
\end{figure}

\begin{figure}[b]
	\centering
	\includegraphics[width=1\linewidth , height=0.3\linewidth]{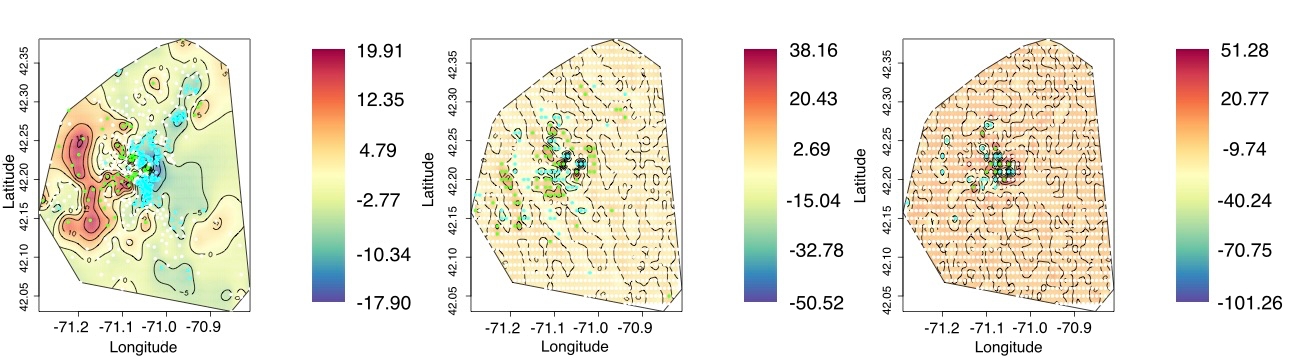}
	\caption{Plots (left to right) showing fitted process, divergence and Laplacian for the median house price surface.}
	\label{fig::boston-surf}
\end{figure}

\noindent {\em Boston Housing:} 
{The Boston housing data \citep[see, e.g.,][]{harrison1978hedonic} was collected by the United States Census Service featuring median house prices for tracts and towns in Boston, Massachusetts area. The purpose was to study heterogeneity in the market caused by the need for residents to have clean air. To study such heterogeneity, modern equitable housing policies are incorporating statistical modeling to quantify such behavior. Often they are a result of unobserved effects of rapidly shifting socioeconomic conditions \citep[see, e.g.,][]{hu2019monitoring}.} Within a spatial map this manifests as neighboring regions of disparity. 
{Figure \ref{fig::boston-data} shows two such regions: high priced including Downtown Boston, Cambridge, Newton, Wellesley, Brookline etc. and low priced including South and East End. For effective policy implementation, identifying such regions becomes crucial. Spatial variation in the median house prices is evidenced in Figure \ref{fig::boston-surf}. Curvature wombling effected on the house price surface would locate regions that feature such change}. 

The data contains median house price values for 506 census-tracts along with demographic data. Latitude-longitude centers of the census-tracts are used for spatial referencing. To allow $Z(\s)$ to capture all the spatial variation, we include only an intercept in the 
{model. 
Table \ref{tab::post-boston}} shows posterior estimates and HPD intervals for process parameters. We observe that $\frac{\sigma^2}{\sigma^2+\tau^2}\approx 78.75\%$---larger portion of total variance being explained by varying location. 

\begin{table}[t]
\centering
\caption{Posterior Estimates from the hierarchical linear model in (\ref{eq::hier-sp-mod}) to Boston housing}\label{tab::post-boston}
\resizebox*{\linewidth}{!}{
	\begin{tabular}{l|@{\extracolsep{100pt}}cc@{}}
		\hline
		\hline
		\multirow{2}{*}{Parameters ($\btheta$)}& \multirow{2}{*}{Posterior Estimates ($\widehat{\btheta}$)} & \multirow{2}{*}{HPD}\\ 
		&&\\
		\hline
		$\phi$ & 0.96 & (0.83, 1.11) \\ 
		$\sigma^2$ & 55.18 & (43.91, 68.06)\\ 
		$\tau^2$ & 14.89 & (11.77, 18.71) \\ 
		$\beta_0$ & 25.58 & (24.29, 27.34) \\ 
		\hline
		\hline
	\end{tabular}
}
\end{table}
Modeled spatial variation in the response is shown in 
Figure \ref{fig::boston-data} (left). Significance for the estimate, $\widehat{Z}(\s)$, is assessed using the inclusion of 0 in its posterior HPD. 
Using posterior samples we estimate the derivative processes for $Z(\s)$. A grid, (${\cal G}=\{\s_g: \s_g\in {\tt convex-hull}({\cal S})\}$, containing 1229 equally spaced locations) is overlaid over the region with the same purpose. 
To effect posterior surface analysis on the estimated surface we use posterior predictive distributions of ${\rm div} (Z)$ 
and $\Delta(Z)$ 
revealing zones that manifest rapid change in response and gradients respectively. They are shown in Figures \ref{fig::boston-surf} (center and right).
\begin{figure}[t]
	\centering
	\includegraphics[width=0.7\linewidth , height=0.25\linewidth]{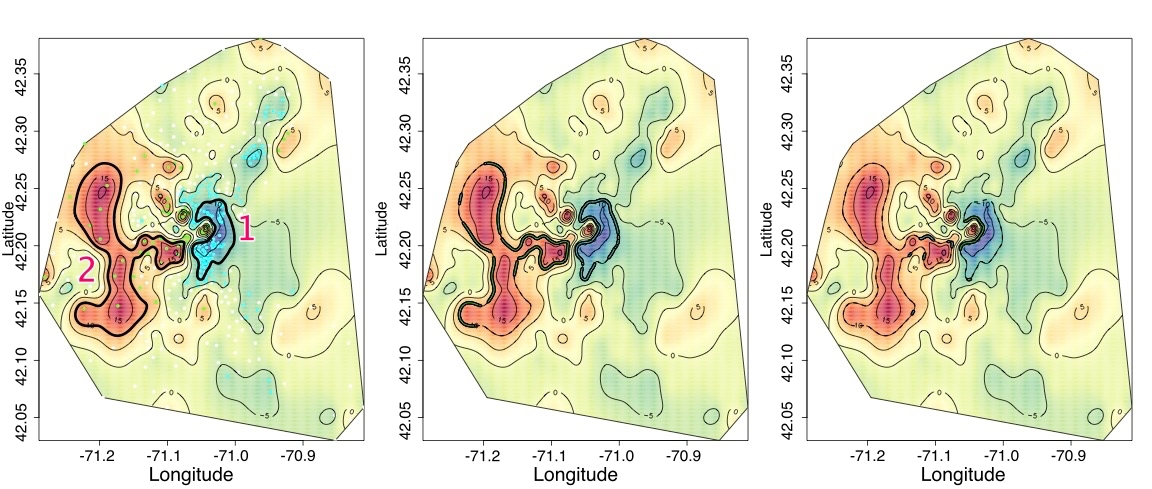}
	\includegraphics[width=0.7\linewidth , height=0.25\linewidth]{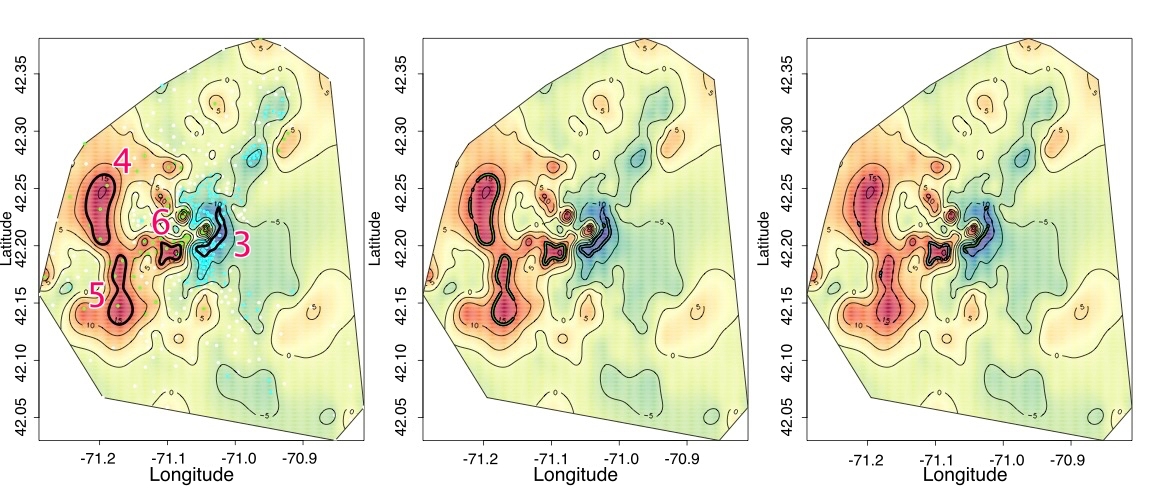}
	\caption{Curvature wombling on the Boston Housing Data.}\label{fig::boston-cwomb}
\end{figure}
Next, we focus on performing curvature wombling on the estimated surface. Strategic posterior surface analysis is used to locate level-sets of interest within the surface that could possibly contain wombling boundaries. We start with contours shown in Figure \ref{fig::boston-cwomb} (left column). Boundary 1 (2) bounds a region where the fitted process has positive (negative) significant estimates. Evidently, the chosen curves should house significant gradients along most segments, but significant curvature should only be detected for segments located at the center (lat-long: $(42.18,42.23)\times(-71.05,-70.05)$) of the surface in Figures \ref{fig::boston-surf} (center and right). Estimated average wombling measures for these curves are shown in Table \ref{tab::wmeasure-boston}. Figures \ref{fig::boston-cwomb} (center and right) correspond to segment level posterior inference for the curves, line segments with significant directional differentials are indicated in bold. 
Summarizing, we observe that the gradient, curvature and posterior surface analysis allow us to highlight towns (with census-tracts) within Boston that exhibit heterogeneity in prices. 
Curvature wombling performed on the surface allows us to delineate zones that house such heterogeneity. 
{For instance, towns located within boundaries 3 (South and East End) and 6 (Newton and Brookline) show significant 
change in price gradients, compared to towns within boundaries 4 (Lincoln and Weston) and 5 (Wellesley and Dover). These findings can be verified referring back to price dynamics for real estate in Boston during 1978 \citep[see e.g.,][]{schnare1976segmentation}. The same regions are scrutinized for studying segmentation---towns within curves 1 and 3 are accessible to lower income groups willing to sacrifice air quality.}
\begin{table}[h]
\centering
\caption{Curvature wombling measures for boundaries in Boston housing accompanied by corresponding HPD intervals in brackets below. Estimates corresponding to HPD intervals containing 0 are marked in bold.}\label{tab::wmeasure-boston}
\resizebox{\linewidth}{!}{
\begin{tabular}{l|@{\extracolsep{100pt}}cc@{}}
  \hline
  \hline
 \multirow{2}{*}{Curve ($C$)} & \multirow{2}{*}{Average Gradient ($\overline{\Gamma}^{(1)}(C)$)} & \multirow{2}{*}{Average Curvature ($\overline{\Gamma}^{(2)}(C)$)}  \\ 
 &&\\
  \hline
\multirow{2}{*}{Boundary 1} & -8.91 & 10.14 \\ 
& (-11.31, -6.65) & (2.84, 18.34)\\
\multirow{2}{*}{Boundary 2} & 6.18 & \bf -0.09 \\ 
&(4.75, 7.49)& \bf (-3.45, 3.35)\\
\hline
\multirow{2}{*}{Boundary 3} & -6.47 & 12.69 \\ 
&(-9.74, -3.27)& (2.65, 22.48)\\
\multirow{2}{*}{Boundary 4} & 6.92 & \bf 1.26 \\ 
&(4.63, 9.19)& \bf (-5.04, 7.14)\\
\multirow{2}{*}{Boundary 5} & 5.47 & \bf 1.36 \\ 
&(2.95, 7.86)& \bf (-4.33, 7.42)\\
\multirow{2}{*}{Boundary 6} & 11.82 & -16.27 \\ 
&(7.28, 16.14)& (-26.68,-6.57)\\
   \hline
   \hline
\end{tabular}
}
\end{table}

\begin{figure}[t]
	\centering
	\includegraphics[width=0.8\linewidth , height=0.33\linewidth]{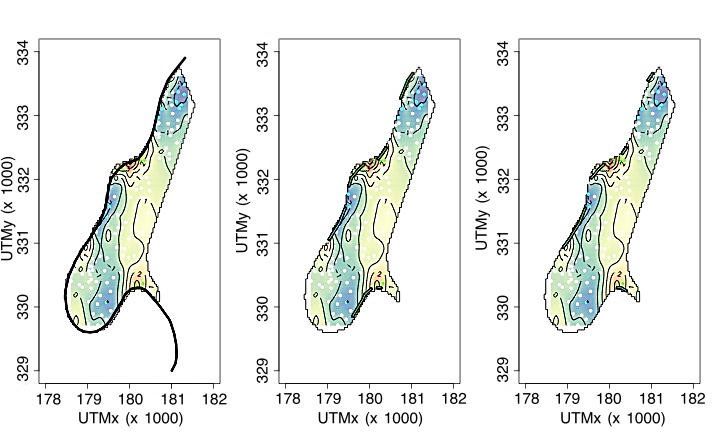}
	\caption{Plots showing results for curvature wombling on the Meuse river for Cadmium (Cd) concentration. 
	{Plots showing (left) the resulting fitted process (center) the contiguous segments that display significant gradients (right) the contiguous segments with significant curvature.}}
	\label{fig::meuse-wombling-cd}
\end{figure}

\noindent {\em Meuse River Data:} The Meuse river data features in \cite{pebesma2012package}. It provides locations of topsoil heavy metal concentrations, along with soil and landscape variables at the observed locations, collected in a flood plain of the river Meuse, near the village of Stein, Netherlands. The heavy metal concentrations recorded include Cadmium (Cd), Copper (Cu), Lead (Pb) and Zinc (Zn). A distinguishing feature is the naturally occurring boundary---the Meuse. From a boundary analysis standpoint we are interested in examining differentials in heavy metal concentrations along the flood plain of the river to understand the heterogeneous effect of the river on the topsoil. \begin{figure}[t]
	\centering
	\includegraphics[width=\linewidth , height=0.3\linewidth]{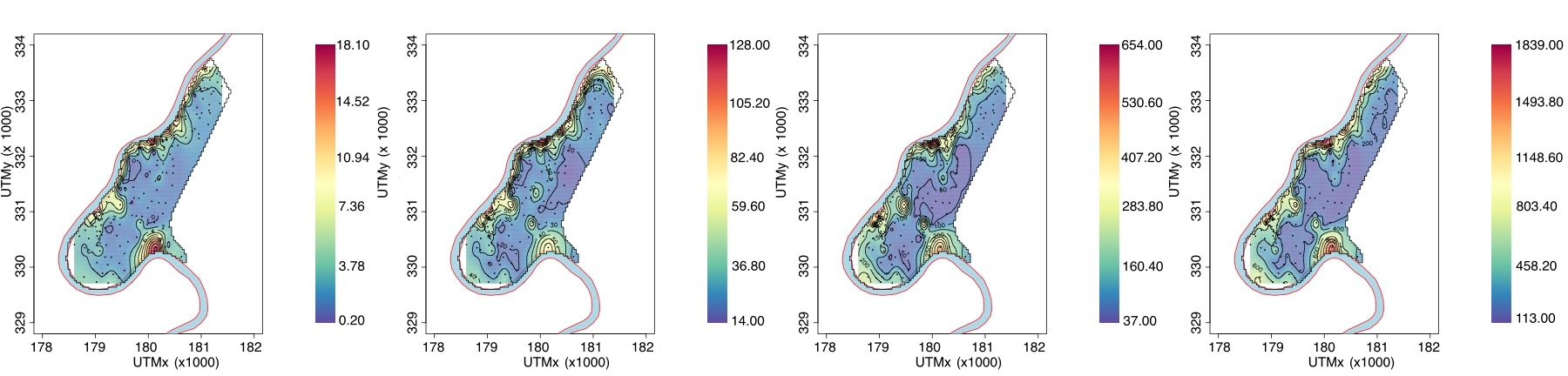}
	\caption{Plots showing heavy metal concentrations in the topsoil of a flood plain at 155 locations for (from left to right) Cadmium (Cd), Copper (Cu), Lead (Pb) and Zinc (Zn) (in mg/kg of soil).}
	\label{fig::meuse-data}
\end{figure}
The soils of the floodplain are commonly used for agriculture. 
Crops grown on the floodplain of the river banks of the Meuse may be consumed by man and/or livestock. 
\begin{table}[t]
	\centering
	\caption{Posterior estimates of process parameters and covariates for the Meuse river data accompanied by their corresponding HPD intervals in brackets below. Effects with HPDs containing 0 are marked in bold.}\label{tab::post-meuse}
	\resizebox*{\linewidth}{!}{
		\begin{tabular}{l|@{\extracolsep{20pt}}cccc@{}}
			\hline\hline
			\multirow{2}{*}{Parameters ($\btheta$)} & \multirow{2}{*}{Cadmium (Cd)} & \multirow{2}{*}{Copper (Cu)} & \multirow{2}{*}{Lead (Pb)} & \multirow{2}{*}{Zinc (Zn)} \\ 
			&&&&\\
			\hline
			\multirow{2}{*}{$\phi$} & 0.0379 &  0.1138 & 0.0399 &  0.0472 \\ 
			& (0.0207, 0.0618) & (0.0871, 0.1471) & (0.0131, 0.1900) &  (0.0230, 0.0744)\\
			\multirow{2}{*}{$\sigma^2$} & 2.9566  & 3.2044  & 0.9303  & 38.3538  \\ 
			&(1.2803, 5.2227) & (2.3955, 4.0892) & (0.2763, 1.7641) & (16.7815, 65.1450)\\
			\multirow{2}{*}{$\tau^2$} & 1.7771 & 0.0067 &  0.8555 & 23.2226  \\ 
			& (0.9107, 2.6328) & (0.0012, 0.0244) & (0.0010, 1.2280) & (9.3743, 35.6867)\\  \hline
			\multirow{2}{*}{\tt Intercept} & 9.4973  & 4.8503 & 6.1120  & 37.1315 \\ 
			& (5.9750, 13.3704) & (3.1392, 6.8308) & (3.4615, 8.1910) & (25.0903, 53.0870)\\
			\multirow{2}{*}{\tt elev} & -0.7672  & -0.4065 &  -0.5413 & -2.8781\\ 
			& (-1.2531, -0.3574) & (-0.7418, -0.1656) & (-0.7853, -0.1442) &  (-4.7805, -1.2834)\\
			\multirow{2}{*}{\tt om} & 0.4011 & 0.4293 & 0.3434 & 0.8606 \\ 
			& (0.2616, 0.5233) & (0.3276, 0.4728) & (0.2490, 0.4253) & (0.3681, 1.3166)\\
			\multirow{2}{*}{\tt dist} & \bf -0.0033  & \bf -0.0025 & \bf -0.0011 & \bf -0.0081\\ 
			& (-0.0061, 0.0000) & (-0.0043, -0.0014) & (-0.0029, 0.0006) & (-0.0197, 0.0038)\\
			\multirow{2}{*}{\tt ffreq (=2)} & -1.4176 & -2.4727 & -0.8483 & -4.3182  \\ 
			& (-2.3202, -0.3432) &  (-3.1794, -1.6716) & (-1.6109, -0.2598) & (-7.9184, -0.6220)\\
			\multirow{2}{*}{\tt ffreq (=3)} & \bf -0.7322 & -1.4298 & \bf -0.1865 & \bf -3.3159 \\ 
			& (-2.0520, 0.6248) & (-2.4443, -0.5157) & (-1.2972, 0.6861) & (-7.9307, 1.9128)\\
			\multirow{2}{*}{\tt soil (=2)} & \bf-0.3337 & \bf0.2236  &\bf 0.5988  &\bf -2.2213 \\ 
			& (-1.4661, 0.7491) & (-0.7248, 0.9799) & (-0.0345, 1.2956) & (-6.1446, 2.0831)\\
			\multirow{2}{*}{\tt soil (=3)} & \bf -0.3884 & \bf 0.6344 & \bf 0.3707 &\bf -2.9922 \\ 
			& (-2.0891, 1.2628) & (-0.2309, 1.8474) &  (-0.7108, 1.4029) & (-9.0918, 3.6289) \\
			\multirow{2}{*}{\tt lime (=1)} & \bf 0.5752 & 1.3223 & 0.7759 &\bf -0.4759  \\ 
			& (-0.3509, 1.4341) &  (0.7152, 1.9427) & (0.1173, 1.4645) &(-3.9057, 2.6510)\\
			\hline
			\hline
		\end{tabular}
	}
\end{table}
The spatial variation in heavy metal concentration can be seen in Figure \ref{fig::meuse-data}. The path of the Meuse river is shown in each of the spatial plots. Evidently, the heavy metal concentrations decreases with increasing distance from the river. We model the concentrations as independent Gaussian processes. 
Covariates used are relative elevation above local river bed ({\tt elev}, measured in meters), organic matter ({\tt om} measured in kg/(100kg) of soil), distance to Meuse ({\tt dist}),  frequency of flooding, 
soil type ({\tt soil}), 
and lime content in soil ($p = 9$). 
Table \ref{tab::post-meuse} shows the posterior estimates of process parameters and model coefficients, $\bbeta$ for each of the heavy metals in question. We observe that $\sigma^2/(\sigma^2+\tau^2)\approx$ 62.45\%, 99.79\%, 52.09\%, 62.29\% for Cd, Cu, Pb and Zn respectively, indicating larger portions of total variation being explained by spatial heterogeneity, except for Pb. Variation in Cd and Zn concentration is significantly affected by elevation, organic matter and flooding frequency, while variations in Cu and Pb concentration is significantly affected by elevation, organic matter and flooding frequency and lime content.
The estimated residual surface is shown in Figure \ref{fig::meuse-wombling-cd} (left) for Cd concentrations. 
We observe significant positive gradients with varying curvature depending on segments of the river bed for all heavy metals.
We perform curvature wombling on the Meuse using the residual surface, $\Z$. The results of curvature wombling for cadmium are shown in Figure \ref{fig::meuse-wombling-cd}. Results and plots for other metals can be found in the Supplement, Section S7, 
Figure 
S14. The accompanying wombling measures are shown in Table \ref{tab::cwomb-meuse}. 
We observe sufficient heterogeneity in the signs of the wombling measures, yielding contiguous positive (negative) segments. For example, in Cd concentration, boundaries located for average gradients in the northern and southern region are positive, as opposed to boundaries located in the north western region. Therefore, while displaying the wombling measures, in Table \ref{tab::cwomb-meuse}, we separate them by their sign.

We conclude 
that effects of river Meuse on regions of the flood plain exhibit significant heterogeneity when considered across heavy metals. 
Compared to other metals, Pb concentrations are limited to northern regions of the flood plain. 
Concentrations of Cd and Zn concentrations along the river are similar. 
Compared to the northern region, in the northwestern region Zn concentrations decrease significantly as we move inland. 
{Studies corroborating such evidence can be found in \cite{leenaers1988variability} and \cite{albering1999human}.}
\section{Discussion and Future Work}
We developed a fully model-based Bayesian inferential framework for differential process assessment and curvature-based boundary analysis for spatial processes.  
Introducing the directional curvature process and its associated inferential framework supplements the directional gradients with inference for their rates of change, while its induction
into the folds of Bayesian curvilinear wombling allows for further characterization of difference boundaries. 
Adopting a Bayesian hierarchical model allows for Gaussian calibration when characterizing points, regions and boundaries 
within a surface. 
This framework is widely applicable; our applications arise from selected disciplines indicating the utilities of mapping curvature process boundaries to understand spatial data generating patterns. Substantive case studies will be reported separately. 
\begin{table}[t]
	\centering
	\caption{Curvature wombling measures for the Meuse, separated by zones of positive and negative signs, they are accompanied by their corresponding HPD intervals in brackets below.}\label{tab::cwomb-meuse}
	\resizebox*{\linewidth}{!}{
		\begin{tabular}{l|cccc}
			\hline\hline
			\multirow{2}{*}{Wombling Measures} & \multirow{2}{*}{Cd} & \multirow{2}{*}{Cu} & \multirow{2}{*}{Pb} & \multirow{2}{*}{Zn} \\ 
			&&&&\\
			\hline
			\multirow{2}{*}{$\overline{\Gamma}^{(1)}(>0)$} &   0.0510 &  0.1273 &  0.0375 &    0.1984 \\ 
			& (0.0298, 0.07401) & (0.0913, 0.1729) & (0.0019, 0.1561) & (0.0876, 0.3162)\\
			\multirow{2}{*}{$\overline{\Gamma}^{(1)}(<0)$} &  -0.0400 & -0.2561 & --  &  -0.1890 \\
			& (-0.0635, -0.0170) & (-0.3187, -0.1997 ) & -- & (-0.2967, -0.0669)\\
			\multirow{2}{*}{$\overline{\Gamma}^{(2)}(>0)$} &  0.0074 &    -- & -- & 0.0422  \\ 
			& (0.0019, 0.0158) &  -- & --  & (0.0111, 0.0879)\\
			\multirow{2}{*}{$\overline{\Gamma}^{(2)}(<0)$} & -0.0078 & -0.1247 & -0.0039 &-0.0473 \\ 
			& (-0.0223, -0.0024) & (-0.1979, -0.0860) &  (-0.1076, -0.0006)  & (-0.1114, -0.0095)\\
			\hline\hline
		\end{tabular}
	}
\end{table}

Several avenues hold scope for future developments. 
{
A more generalized theoretical framework can be developed for studying joint behavior of the principal curvature (direction of maximum (or minimum) curvature) 
and the aspect (direction of maximum gradient) \citep[see, e.g.,][]{wang2018process} leveraging dependent circular uniform distributions \citep[see, e.g.,][]{kent2008modelling}. 
We offer some brief remarks
. To obtain the direction of maximum curvature for a spatial surface, 
we solve $\displaystyle{\max_{\u\in \mathbb{R}^2} \left|\u\trans \nabla^2Y(\s) \u\right|}$, such that $||\u|| = 1$, at an arbitrary point $\s$. Using Lagrange multipliers and denoting $\kappa(\u) = |\u\trans \nabla^2Y(\s)\u|$, define ${\cal O}(\u) = \kappa(\u) - \lambda(||\u||^2 - 1)$ hence, $\partial{\cal O}(\u)/\partial u_i = \kappa(\u)^{-1}(\nabla^2_{ii}Y(\s)u_i + \nabla^2_{ij}Y(\s)u_j) - \lambda u_i = 0$, $i, j = 1,2$. With $u_2/u_1 = \tan\theta_{pc}$, eliminating $\lambda$ we get $\displaystyle{\tan\theta_{pc} = \frac{\nabla^2_{22}Y(\s)\tan\theta_{pc} +\nabla^2_{12}Y(\s)}{\nabla^2_{11}Y(\s) + \nabla^2_{12}Y(\s)\tan\theta_{pc}}}$.
    Defining $h_1 = h_1(\s)= (\nabla^2_{11}Y(\s)-\nabla^2_{22}Y(\s))/\nabla^2_{12}Y(\s)$ given $\nabla^2_{12}Y(\s) \ne 0$ and solving 
    $\theta_{pc} = \tan^{-1} \frac{1}{2}\left[-h_1\pm \sqrt{h_1^2+4}\right]$.
    If $\nabla^2_{12}Y(\s)=0$ then, $\nabla^2Y(\s)$ is diagonal and $\theta_{pc}$ corresponds to the direction of $\max\{\nabla^2_{11}Y(\s),\nabla^2_{22}Y(\s)\}$. We propose that $\Theta = (\theta_{asp},\theta_{pc})\trans$ follows a dependent circular uniform distribution over $[0,2\pi]\times [0,2\pi]$. Further developments with circular regression methods can proceed to examine the effect of covariates on $\Theta$.} 
    Multivariate extensions 
    would involve formulating these differential processes on arbitrary manifolds. 
    This requires simulating a Gaussian process on manifolds and inspecting the covariant derivative. Bayesian curvilinear wombling could then be implemented on curves of interest to the investigator. 
    This would not only involve an inferential framework for normal curvature, but also geodesic curvature for such curves. Spatiotemporal curvature processes can build upon \cite{quick2015bayesian} 
    to study evolutionary behavior of the curvature processes with respect to variations in the response across time. Finally, we remark that while there have been substantial recent developments in scalable spatial processes for massive data sets---a comprehensive review is beyond the scope of the current article \citep[see, e.g.,][]{heaton2019case}---not all scalable processes admit the correct degree of smoothness for curvature processes to exist. Constructing scalable processes for curvilinear wombling, and subsequent inference, remains a problem of interest in the wombling community. 

\section*{Supplementary Materials}
The following supplement includes additional theoretical derivations, computing details, additional simulation experiments and wombling for Northeastern US temperatures.


\spacingset{1.0}
\title{\bf  Supplementary Materials for\\ {\em ``Bayesian Modeling with Spatial Curvature Processes"}}
		\maketitle
  
\spacingset{1.9}


\section{Review of Directional Gradients and Wombling}\label{sec::review}
    \subsection{Directional Gradients}\label{sec::grad}
    For the scalar $h$ and unit vector $\u$ we define $ Y^{(1)}_{\u,h}=\left(Y(\s+h\u)-Y(\s)\right)/h$ to be the first order finite difference processes at location $\s$ in the directions of $\u$. Being a linear function of stationary processes this is well-defined. Passing to limits, we define $D^{(1)}_{\u}Y(\s)=\lim_{h\to 0}Y^{(1)}_{\u,h}(\s)$. Provided the limit exist, $D^{(1)}_{\u}Y(\s)$ is defined as the directional gradient process. If $Y(\s)$ is a mean square differentiable process in $\Rd$ for every $\s_0 \in \Rd$ then $D^{(1)}_{\u}Y(\s)=\u\trans\nabla Y(\s)$. Then, if $\u=\sum_{i=1}^{d}u_i\e_i$, we can compute $D^{(1)}_{\u}Y(\s)=\sum_{i=1}^{d}u_iD^{(1)}_{\e_i}Y(\s)$. The directional gradient process is linear in $\u$, hence $D^{(1)}_{-\u}Y(\s)=-D^{(1)}_{\u}Y(\s)$ and for any vector $\w=||\w||\u$, $D^{(1)}_{\w}Y(\s)=||\w||D^{(1)}_{\u}Y(\s)$.  The directional gradient at $\s_0$ in the direction $\w$ is the slope at $\s_0$ of the curve traced out by slicing $Y(\s)$ in the direction $\w$ \citep[see e.g.,][Section 2, for more details]{banerjee2006bayesian}.
    \subsection{Wombling Measures}\label{sec::grad-womb}
        Wombling measures constructed from (7) 
        for total and average gradient are associated with curves to characterize the magnitude of change. To each point $\s\in C$, a directional gradient is associated, $g\left({\cal L}Y(\s)\right)=D^{(1)}_{\n}Y(\s)=\n(\s)\trans\nabla Y(\s)$ (also a linear function of $\displaystyle{{\cal L}_{\n}Y(\s)}$), along the direction of a unit normal $\u=\n(\s)$ to the curve. 
        For a curve tracking rapid change in the surface, choice of the normal direction to a curve is motivated by sharp directional gradients 
        orthogonal 
        to the curve; $\ell$ is chosen to be the arc-length measure. The rationale behind this choice is to measure change in response with respect to distance traversed on the curve. With reference to (7) 
        the total and average gradients are, ${{\Gamma}^{(1)}( C)=\int_{t_0}^{t_1}\nabla Y(\s(t))\trans\n(\s(t))||\s'(t)||dt}$ and $\overline{\Gamma}^{(1)}({C})={\Gamma}^{(1)}( C)/\ell({C})$ respectively  \citep[see e.g.,][Section 3, for more details]{banerjee2006bayesian}.

        \section{Interpretation of Spatial Curvature}\label{sec::interpret} Pursuits in geospatial analysis generally encounter surfaces which have canonical coordinate systems (e.g. latitude-longitude, easting-northing etc.). This facilitates a parameterization for the surface that leverages the coordinate system, commonly known as the Monge parameterization \citep[also called a Monge patch, named after Gaspard Monge, see e.g.,][]{o2006elementary,pressley2010elementary}---a surface, $S$, embedded in $\Rthree$, is parameterized by giving its height $Y$ over some plane as a function of the orthonormal co-ordinates $s_1$ and $s_2$ in the plane,	$S=\{{\cal S}\subset \Rtwo\mapsto \Rthree: \s=(s_1,s_2) \mapsto Y(s_1,s_2)=Y(\s)\}$. A point is then, $(\s, Y(\s))=(s_1,s_2,Y(s_1,s_2))$. The two tangent vectors at $\s$ are, $\E_1(\s)=(1,0,\nabla_1Y(\s))\trans$ and $\E_2(\s)=(0,1,\nabla_2Y(\s))\trans$, where $\nabla_iY(\s)=\displaystyle{\del{\s_i}Y(\s)},\;i=1,2$. Let $\nabla Y(\s)=(\nabla_1 Y(\s), \nabla_2 Y(\s))\trans$ denote the gradient vector, consider a unit direction vector, $\u=(u_1,u_2)\trans \in S\subset \Rtwo$, then $u_1\E_1(\s)+u_2\E_2(\s)=(\u\trans,\u\trans\nabla Y(\s))\trans\in T_{S}(\s)$ corresponds to the {\em directional derivative} of $Y$ along the direction $\u$, where $T_{S}(\s)$ is the local tangent plane at $\s$, that is generated by $\{\E_1(\s),\E_2(\s)\}$. The outward pointing normal to the surface $S$, denoted by $\N(\s)=\E_1(\s)\times\E_2(\s)=(-\nabla_1Y(\s),-\nabla_2Y(\s),1)\trans$, where $\times$ denotes the usual cross-product of vectors. Evidently, $\N(\s)$ is orthogonal to the local tangent plane at that point, $T_{S}(\s)$. Quantifying the local geometry of a surface, we are interested in how $\N(\s)$ changes (``tips") as we move in the direction $\u$ from the point $\s$ on the surface---derivatives for $\N(\s)$ at the point $\s$, which lie in $T_{S}(\s)$. This is quantified by the {\em normal curvature} of $S$ along a direction $\u$. Before defining normal curvature for a surface, we digress briefly to investigate effects of surface curvature on curves---for a curve parameterized by $t$, $C=\{\s(t)=(s_1(t),s_2(t)):t\in [a,b]\}$, passing through $\s$, if the curvature of $C$ is, $\kappa$, the tangent to $C$ at $\s$, $\t(\s)$, and the principal unit normal, i.e. the normal to $C$ on $S$, $\n(\s)=\t'(\s)/\kappa$, then we have, $\n(\s)\cdot \N(\s)=\cos(\theta)$, $\t'(\s)=\kappa\n(\s)$, which implies $\kappa\cos(\theta)=\t'(\s)\cdot\N(\s)$. We observe that, 
        \begin{figure}[t]
    	\centering
	\def\svgwidth{0.7\linewidth}
	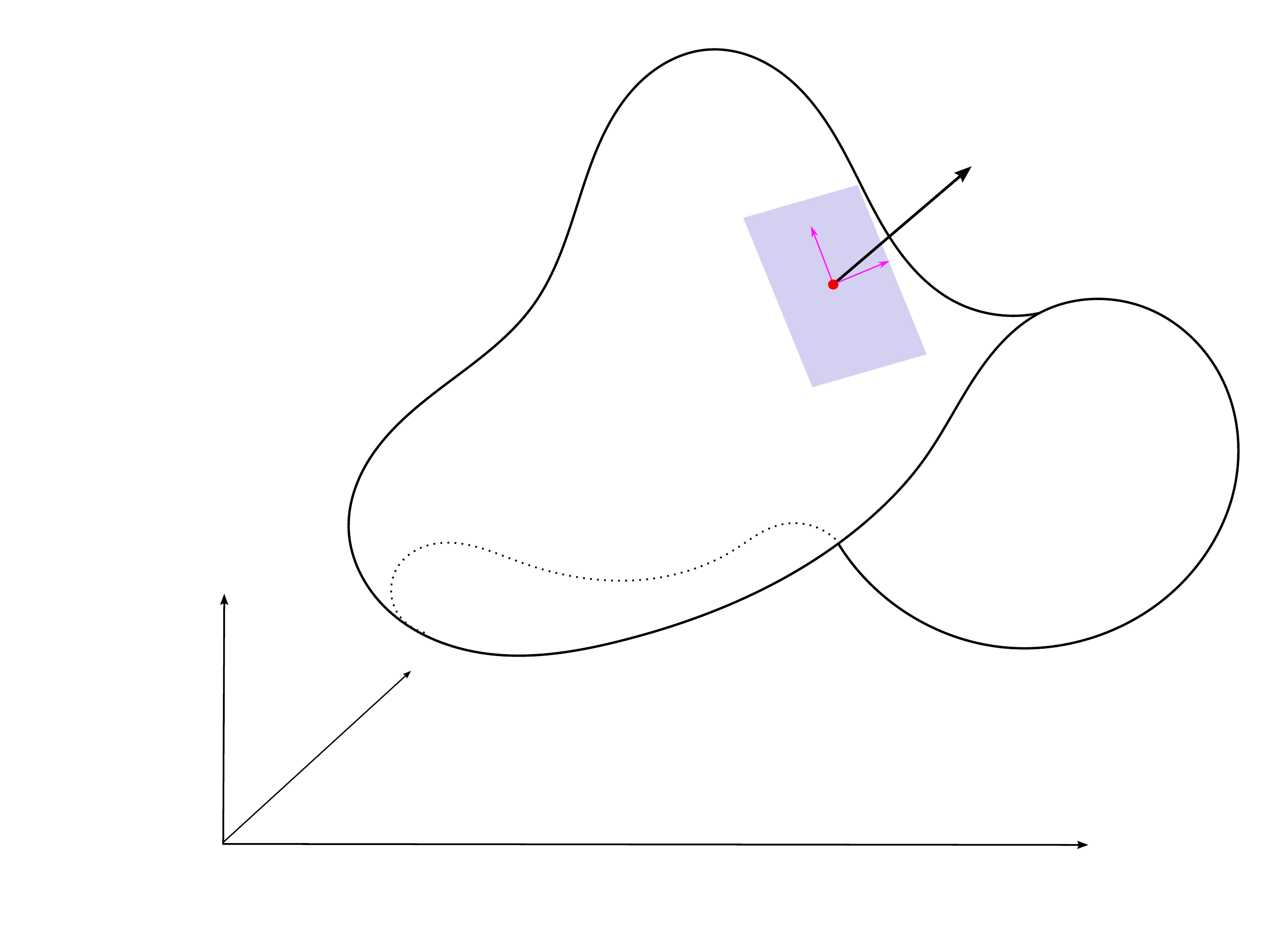

    	\caption{A Monge patch, $S=(s_1,s_2,Y(s_1,s_2))\trans=(\s\trans,Y(\s))\trans$, showing a point $\s$, a curve $C$ passing through $\s$, normal to the surface, $\N(\s)$, normal to the curve $\n(s)$, $\theta$ is the angle between them, and the local tangent plane to the surface, $T_{S}(\s)$. The thin perpendicular pink arrows are tangent vectors, $\E_1(\s)$ and $\E_2(\s)$. The thin outward pointing black arrows around $\N(\s)$ demonstrate change in $\N$ as we move along the direction (dotted line) on the surface.}	\label{fig::diff-surf}
        \end{figure}
        \begin{equation*}
        \t'(\s(t))=\deltwo{t}\left(\s(t),Y(\s(t))\right)\trans=\del{t}\E_i(\s(t))s_i'(t)=\E_{ij}(\s(t))s_i'(t)s_j'(t)+\E_i(\s(t))s_i''(t)\;,
        \end{equation*}
        since $\E_i(\s)\cdot \N(\s)=0$, $\kappa\cos(\theta)=\t'(\s)\cdot\N(\s)=(\E_{ij}(\s)\cdot\N(\s))\partial s_i\partial s_j$, $i,j=1,2$, where the expression in the parenthesis is a property of the surface, independent of curve $C$, and is defined as the {\em second fundamental form}, 
        \begin{equation*}
            \Pi(\s)=\begin{pmatrix} \E_{11}(\s)\cdot\N(\s) & \E_{12}(\s)\cdot\N(\s)\\\E_{21}(\s)\cdot\N(\s) & \E_{22}(\s)\cdot\N(\s)\end{pmatrix}=\begin{pmatrix}\nabla^2_{11}Y(\s) & \nabla^2_{12}Y(\s)\\\nabla^2_{21}Y(\s) & \nabla^2_{22}Y(\s)\end{pmatrix}=\nabla^2Y(\s)\;,
        \end{equation*}
        where $\E_{ij}$ and $\nabla^2_{ij}Y$ are the partial differentiation of $\E$  and $Y$ with respect to $s_i,s_j$ respectively, $\cdot$ is the usual dot product for vectors and $\nabla^2_{12}Y=\nabla^2_{21}Y$. The second to last equality is obtained under the Monge parameterization. The second fundamental form is invariant with respect to transformations of the local co-ordinate which preserves the sense of $\N$, i.e. the transformation does not change an outward (inward) pointing normal to an inward (outward) pointing normal for $S$. Such surfaces are termed as {\em orientable surfaces}. The M\"obius transformation is an example of non-orientable surfaces. The individual terms of $\Pi$, quantify the local geometry of a surface (or curvature) along orthonormal coordinates.

        Curvature of the $C$ can be attributed to (a) the curvature of the curve itself, and (b) the curvature of the surface on which $C$ lies. $\kappa$ is the curvature of $C$, termed as {\em geodesic curvature}. The curvature of the surface is termed as {\em normal curvature}, computed along a direction $\u=\u(\s)$ is denoted by $\kappa_n(\u)$. The normal curvature, which is an intrinsic property of the surface independent of $C$, is of primary interest to us, $\kappa_n(\u)=\u\trans\Pi(\s)\u/\u\trans\u=\u\trans\Pi\u$, under $\u\trans\u=1$.  $\kappa_n(\u)$ is also the {\em directional curvature} of the $Y$ along $\u$. For our exploits, $\u=\n$, the normal direction to $C$. The sign of $\kappa_n(\u)$, or equivalently eigen-values of $\Pi$ inform about the nature of curvature at $\s$---for example, if $\kappa_1$ and $\kappa_2$ denote eigenvalues of $\Pi$, with $K=\det\Pi=\kappa_1\kappa_2$, if $K>0$, it implies that the surface is bending away from $T_{S}(\s)$; depending on whether $\kappa_1,\kappa_2<0$ (or $>0$), $\s$ can be locally classified as a concave (convex) ellipsoid \citep[for more details see][]{stevens1981visual}.
        
        For a purely differential geometric treatment of this discussion see---\cite{gauss1902general,spivak1970comprehensive,do2016differential,kreyszig2019introduction}. Figure \ref{fig::diff-surf} illustrates this discussion.
        
        \section{Examples for selected Covariance Functions}\label{sec::examples} The detailed calculations for closed form expression of selected covariance functions are presented. We start with the power exponential family of isotropic covariance functions, $\tK(||\Delta||)=\alpha\exp(-\phi||\Delta||^\nu)$, $0<\nu\leq 2$. It is clear that $\nabla^4\tK(||\Delta||)$ and exists only for $\nu=2$, we have the following form, $\tK(||\Delta||)=\sigma^2\exp(-\phi||\Delta||^2)$. For this choice we have,
{\allowdisplaybreaks
	\begin{align*}
		&\left(\nabla\tK(\Delta)\right)_i=-2\sigma^2\phi\exp(-\phi||\Delta||^2)\delta_i,\\
		&\left(\nabla^2\tK(\Delta)\right)_{ii}=-2\sigma^2\phi\exp(-\phi||\Delta||^2)(1-2\phi\delta_i^2),\\
		&\left(\nabla^2\tK(\Delta)\right)_{ij}=4\sigma^2\phi^2\exp(-\phi||\Delta||^2)\delta_i\delta_j,\\
		&\left(\nabla^3\tK(\Delta)\right)_{iii}=4\sigma^2\phi^2\exp(-\phi||\Delta||^2)(3-2\phi\delta_i^2)\delta_i,\\
		&\left(\nabla^3\tK(\Delta)\right)_{iij}=4\sigma^2\phi^2\exp(-\phi||\Delta||^2)(1-2\phi\delta_i^2)\delta_j,\\
		&\left(\nabla^3\tK(\Delta)\right)_{ijk}=-8\sigma^2\phi^3\exp(-\phi||\Delta||^2)\delta_i\delta_j\delta_k,\\
		&\left(\nabla^4\tK(\Delta)\right)_{iiii}=4\sigma^2\phi^2\exp(-\phi||\Delta||^2)(3-12\phi\delta_i^2+4\phi^2\delta_i^4),\\
		&\left(\nabla^4\tK(\Delta)\right)_{iiij}=-8\sigma^2\phi^3\exp(-\phi||\Delta||^2)(3-2\phi\delta_i^2)\delta_i\delta_j,\\
		&\left(\nabla^4\tK(\Delta)\right)_{iijj}=4\sigma^2\phi^2\exp(-\phi||\Delta||^2)(1-2\phi\delta_i^2)(1-2\phi\delta_j^2),\\
		&\left(\nabla^4\tK(\Delta)\right)_{ijkl}=16\sigma^2\phi^4\exp(-\phi||\Delta||^2)\delta_i\delta_j\delta_k\delta_l,
	\end{align*}
}
where $i,j,k,l=1,2,\ldots,d$. The squared exponential or the Gaussian covariance kernel is the only member of its class that admits such derivatives, although they have been critiqued to produce realizations that are too smooth to be of practical use in modeling (see \cite{stein1999interpolation}).

Turning to the Mat\'ern class we see that with, $\tK||\Delta||=\alpha(\phi||\Delta||)^\nu K_{\nu}(\phi||\Delta||)$, where $\nu$ is a parameter controlling the smoothness of realizations, that is mean square differentiability and $K_{\nu}$ is the modified Bessel function of order $\nu$. At $\nu=3/2$ and $\nu=5/2$, $\tK(||\Delta||)$ takes the forms,
\begin{align*}
	\tK(||\Delta||)=\begin{cases}
		\sigma^2(1+\sqrt{3}\phi||\Delta||)e^{-\sqrt{3}\phi||\Delta||},&\nu=3/2\\
		\sigma^2\left(1+\sqrt{5}\phi||\Delta||+\mfrac{5}{3}\phi^2||\Delta||^2\right)e^{-\sqrt{5}\phi||\Delta||},& \nu=5/2
	\end{cases},
\end{align*}
where $\sigma^2$ is the overall process variance. Mat\'ern with $\nu=3/2$ is once mean square differentiable, where as Mat\'ern with $\nu=5/2$ is twice mean square differentiable at 0. As $\nu \to \infty$, Mat\'ern covariances tend to the Gaussian covariance. Unlike the Gaussian covariance, they do not yield overly smoothed process realizations. For $\nu=3/2$ we have,
\begin{align*}
	&\left(\nabla\tK(\Delta)\right)_i=-3\sigma^2\phi^2e^{-\sqrt{3}\phi||\Delta||}\delta_i,\\
	&\left(\nabla^2\tK(\Delta)\right)_{ii}=-3\sigma^2\phi^2e^{-\sqrt{3}\phi||\Delta||}\left(1-\sqrt{3}\phi\mfrac{\delta_i^2}{||\Delta||}\right),\\
	&\left(\nabla^2\tK(\Delta)\right)_{ij}=3\sqrt{3}\sigma^2\phi^3e^{-\sqrt{3}\phi||\Delta||}\mfrac{\delta_i\delta_j}{||\Delta||}.
\end{align*}
where $i,j=1,2,\ldots,d$. Since the process is just once mean square differentiable higher order derivatives do not exist. However, for $\nu=5/2$ we have,
{\allowdisplaybreaks
	\begin{align*}
		&\left(\nabla\tK(\Delta)\right)_i=-\mfrac{5}{3}\sigma^2\phi^2e^{-\sqrt{5}\phi||\Delta||}\left(1+\sqrt{5}\phi||\Delta||\right)\delta_i,\\
		&\left(\nabla^2\tK(\Delta)\right)_{ii}=-\mfrac{5}{3}\sigma^2\phi^2e^{-\sqrt{5}\phi||\Delta||}\left(1+\sqrt{5}\phi||\Delta||-5\phi^2\delta_i^2\right),\\
		&\left(\nabla^2\tK(\Delta)\right)_{ij}=\mfrac{25}{3}\sigma^2\phi^4e^{-\sqrt{5}\phi||\Delta||}\delta_i\delta_j,\\
		&\left(\nabla^3\tK(\Delta)\right)_{iii}=\mfrac{25}{3}\sigma^2\phi^4e^{-\sqrt{5}\phi||\Delta||}\left(3-\sqrt{5}\phi\mfrac{\delta_i^2}{||\Delta||}\right)\delta_i,\\
		&\left(\nabla^3\tK(\Delta)\right)_{iij}=\mfrac{25}{3}\sigma^2\phi^4e^{-\sqrt{5}\phi||\Delta||}\left(1-\sqrt{5}\phi\mfrac{\delta_i^2}{||\Delta||}\right)\delta_j,\\
		&\left(\nabla^3\tK(\Delta)\right)_{ijk}=-\mfrac{25\sqrt{5}}{3}\sigma^2\phi^5e^{-\sqrt{5}\phi||\Delta||}\mfrac{\delta_i\delta_j\delta_k}{||\Delta||},\\
		&\left(\nabla^4\tK(\Delta)\right)_{iiii}=\mfrac{25}{3}\sigma^2\phi^4e^{-\sqrt{5}\phi||\Delta||}\left[3-6\sqrt{5}\phi\mfrac{\delta_i^2}{||\Delta||}+\sqrt{5}\phi\left(\sqrt{5}\phi+\mfrac{1}{||\Delta||}\right)\mfrac{\delta_i^4}{||\Delta||^2}\right],\\
		&\left(\nabla^4\tK(\Delta)\right)_{iiij}=\mfrac{25\sqrt{5}}{3}\sigma^2\phi^5e^{-\sqrt{5}\phi||\Delta||}\left[\mfrac{\delta_i}{||\Delta||^3}-\mfrac{1}{||\Delta||}\left(3-\sqrt{5}\phi\mfrac{\delta_i^2}{||\Delta||}\right)\right]\delta_i^2\delta_j,\\
		&\left(\nabla^4\tK(\Delta)\right)_{iijj}=\mfrac{25}{3}\sigma^2\phi^4e^{-\sqrt{5}\phi||\Delta||}\left[\left(1-\sqrt{5}\phi\mfrac{\delta_i^2}{||\Delta||}\right)\left(1-\sqrt{5}\phi\mfrac{\delta_j^2}{||\Delta||}\right)+\sqrt{5}\phi\mfrac{\delta_i^2\delta_j^2}{||\Delta||^3}\right],\\
		&\left(\nabla^4\tK(\Delta)\right)_{ijkl}=-\mfrac{25\sqrt{5}}{3}\sigma^2\phi^5e^{-\sqrt{5}\phi||\Delta||}\left(\sqrt{5}\phi+\mfrac{1}{||\Delta||}\right)\mfrac{\delta_i\delta_j\delta_k\delta_l}{||\Delta||^2},
	\end{align*}
}
where $i,j,k,l=1,2,\ldots,d$. 

The above expressions for entries of the cross-covariance matrices correspond to the joint process, ${\cal L}Y(\s)=(Y(\s),\nabla Y(\s)\trans, vech(\nabla^2 Y(\s))\trans)\trans$ with respect to our kernel choices. The following cross-covariance matrices are evaluated at $||\Delta||\to 0$.
\begin{enumerate}
	\item {\em Squared Exponential:} In $\Rd$, we have,
	\begin{align*}
		V_{{\cal L}Y}({\bf 0})=\sigma^2\begin{pmatrix}
			1 & {\bf 0}\trans & -2\phi vech(I_d)\trans\\
			{\bf 0} & 2\phi I_d & {\bf O}\\
			-2\phi vech(I_d) & {\bf O} & 4\phi^2\mathrm{diag}\{3,1\ldots,1,3,1,\ldots,1,\ldots,3\}
		\end{pmatrix},
	\end{align*}
	\item {\em Mate\'rn} $(\nu=3/2)$: For this kernel the existence of  only the gradient process is guaranteed, therefore the covariance is for the process, ${\cal L}Y(\s)=(Y(\s),\nabla Y(\s)\trans)\trans$. In $\Rd$ we have, $V_{{\cal L}Y}({\bf 0})=\sigma^2\begin{pmatrix}
			1 & {\bf 0}\trans\\
			{\bf 0} & 3\phi^2I_d
		\end{pmatrix}$.
	\item {\em Mate\'rn} $(\nu=5/2)$: In $\Rd$ we have,
	\begin{align*}
		V_{{\cal L}Y}({\bf 0})=\sigma^2\begin{pmatrix}
			1 & {\bf 0}\trans & -\frac{5\phi^2}{3} vech(I_d)\trans\\
			{\bf 0} & \frac{5\phi^2}{3} I_d & {\bf O}\\
			-\frac{5\phi^2}{3} vech(I_d) & {\bf O} & \frac{25\phi^4}{3}\mathrm{diag}\{3,1\ldots,1,3,1,\ldots,1,\ldots,3\}
		\end{pmatrix}.
	\end{align*}
\end{enumerate}

\section{Algorithms}\label{sec::algos}
In what follows, we provide the required algorithms for sampling gradients and wombling measures. Although listed separately to highlight the requirement of only posterior samples, the required steps could be included within the MCMC subroutine devised for spatial learning (or fitting the model) of $Y(\s)$.

\paragraph{Sampling Gradients and Curvature:} The choice for $K$ varies between Gaussian, Mat\'ern with $\nu=3/2$ and $\nu=5/2$. There is scope for parallel computation across grid locations. Additionally if the inverse of estimated covariance matrices are stored for the MCMC runs from the model fit, sufficient gains in compilation can be achieved while sampling gradients. If $(\nu=3/2)$ is chosen the $\nabla^2$ terms are not computed.
\begin{algorithm}[h]
	\caption{Algorithm for Sampling Gradients and Curvature}\label{algo::gradient}
	\KwIn{${\cal S}$, A Grid ${\cal G}$ spanning ${\cal S}$, posterior MCMC samples $\bbeta$, $\btheta_K=\{\sigma^2,\phi\}$, $\Z$}
	\KwResult{Posterior samples for gradients $\nabla Y(\s_g)$ and curvature $\nabla^2 Y(\s_g)$ for $\s_g\in {\cal G}$}
	\For{$i=1,2,\ldots,L$}{
		\For{$j=1,2,\ldots,n_G$}{
			$\Delta[i,j] = \s_g[j]-\s[i]$ \Comment{Compute distances of grid locations to observed process}
		}
	}
	\For{$i=1,2,\ldots,n_{\rm MCMC}$}{
		$K[i] = K(\cdot;\btheta_K[i])$\\
		$K.inv[i] = (K(\cdot;\btheta_K[i]))^{-1}$\\
		\For{$j =1,2,\ldots,n_G$}{
			$\nabla K[i,j] = (\nabla K(\Delta[,j];\btheta_K[i])\trans,vech(\nabla^2 K(\Delta[,j];\btheta_K[i]))\trans)\trans$\\
			$V[i,j] = V_{{\cal L}Y}({\bf 0})$\\
			$\mu[i,j] = \nabla\mu(\s;\bbeta[i])-\nabla K[i,j]\trans K.inv[i]\Z[i]$  \Comment*[r]{$\mu(\s;\bbeta[i])=\X\bbeta[i]$}
			$\Sigma[i,j] = V[i,j]-\nabla K[i,j]\trans K.inv[i]\nabla K[i,j]$\\
			${\cal L}Y[i,j] = {\cal N}(\mu[i,j],\Sigma[i,j])$
			\Comment{Posterior sample of Gradients and Curvature}
		}
	}
	\Return{${\cal L}Y$}
\end{algorithm}

\begin{algorithm}[h]
		\caption{Algorithm for Sampling Wombling Measures}\label{algo::wombling}
		\KwIn{${\cal S}$, A curve $C$, posterior MCMC samples $\bbeta$, $\btheta_K=\{\sigma^2,\phi\}$, $\Z$}
		\KwResult{Posterior samples for wombling measures $\bGamma(\widetilde{C}_P)$.}
		\For{$j=1,2,\ldots,(n_P-1)$}{
			$t[j] = ||C[j]-C[j+1]||$\\
			$\u[j] = (C[j]-C[j+1])/t[j]$ \Comment{Compute $\t$ and $\U$}
		}
		\For{$i=1,2,\ldots,L$}{
			\For{$j=1,2,\ldots,n_P$}{
				$\Delta[i,j] = C[j]-\s[i]$ \Comment{Compute distances of points in curve to observed process and norms: $||\Delta||[i,j]= ||\Delta[i,j]||$}
			}
		}
		\For{$i=1,2,\ldots,n_{\rm MCMC}$}{
			$K[i] = K(\cdot;\btheta_K[i])$\\
			$K.inv[i] = (K(\cdot;\btheta_K[i]))^{-1}$\\
			\For{$j =1,2,\ldots,n_P$}{
				$\nabla K[i,j] =\begin{pmatrix}{\tt q}_1(0,t[j],D^{(1)}K(\Delta[,j]+t\u[j];\btheta_K[i]))\\{\tt q}_1(0,t[j],D^{(2)}K(\Delta[,j]+t\u[j];\btheta_K[i]))\end{pmatrix}\trans$,\\
				$V[i,j] = \begin{pmatrix}{\tt q}_2(0,0,t[j],t[j],k_{11}(\btheta_K[i]))& {\tt q}_2(0,0,t[j],t[j],k_{12}(\btheta_K[i]))\\{\tt q}_2(0,0,t[j],t[j],k_{21}(\btheta_K[i]))&{\tt q}_2(0,0,t[j],t[j],k_{22}(\btheta_K[i]))\end{pmatrix}$\\
				$\mu[i,j] = \mu_{\bGamma}(t[j])-\nabla K[i,j]\trans K.inv[i]\Z[i]$ 
				$\Sigma[i,j] = V[i,j]-\nabla K[i,j]\trans K.inv[i]\nabla K[i,j]$\\
				$\bGamma[i,j]={\cal N}(\mu[i,j],\Sigma[i,j])$\Comment{Posterior sample of Wombling Measures}
			}
		}
		\Return{$\bGamma$}
\end{algorithm}

\paragraph{Sampling Wombling Measures} The choices for $K$ are again between Gaussian, Mat\'ern with $\nu=3/2$ and $\nu=5/2$. Choices for curves to be evaluated for wombling boundaries range from those outlined in Section 5.2. In case $\nu=3/2$ wombling measures for curvature are not computed. Choices for approximations include computing Riemann sums replacing quadrature for line integrals. There is scope for parallel computation with the curve being broken into segments evaluated in parallel for wombling boundaries.

The functions ${\tt q}_1$ and ${\tt q}_2$ denote one and two-dimensional quadrature respectively. In case a Riemann sum (see (\ref{eq::ineq-1}), Section \ref{sec::proofs}) approximation is chosen, the points partitioning $C$ are treated as grid points and the algorithm for sampling gradients and curvature is used for predictive inference on the differential process. The Riemann sums are computed using $\t$ and $\U$ and returned.

\section{Proofs and Discussion}\label{sec::proofs}
For the curvature process formulated in Section 2, we aim to show that the covariance matrix associated with the process $D^{(2)}_{\u,\v}Y(\s)=\c_{\u,\v}\trans vech\nabla^2Y(\s)$ is valid (pg. 5 last paragraph). We obtain the expression for the covariance matrix by leveraging the directional finite difference process $Y^{(2)}_{\u,\v,h}(\s)$. For points $\s,\s'$, we denote $\Delta=\s-\s'$ and $\u$, $\v$ are unit vectors specifying direction and $\bdelta(x,y)=\Delta+x\u+y\v$ as a map from $\Rtwo\to \Rd$, after suppressing dependence on  $\Delta$ and $\u$, $\v$, let $g(x,y)=K(\Delta(x,y))=K(\Delta+x\u+y\v)$ denote a map from $\Rtwo\to \Rone$ we compute the covariance ,
{\allowdisplaybreaks
	\begin{align*}
		C^{(2)}_{\u,\v}(\s,\s')&=\lim_{h\to 0}\lim_{k\to 0}E\left[Y^{(2)}_{\u,\v,h}(\s)Y^{(2)}_{\u,\v,k}(\s')\right],\\
		&=\lim_{h\to 0}\lim_{k\to 0}\frac{1}{h^2k^2}[g(h-k,h-k)-g(h-k,h)-g(h,h-k)+g(h,h)\\
		&\hspace{3cm} -g(h-k,-k)+g(h-k,0)+g(h,-k)-g(h,0)\\
		&\hspace{3cm} -g(-k,h-k)+g(-k,h)+g(0,h-k)-g(0,h)\\
		&\hspace{3cm} +g(-k,-k)-g(-k,0)-g(0,-k)+g(0,0)],\\
		&=\lim_{h\to0}\frac{g''(h,h)-g''(h,0)-g''(0,h)+g''(0,0)}{h^2}=g^{(iv)}(0,0).
	\end{align*}
}
On repeated application of the chain rule and noting that $\bdelta_x(x,y)=\u\trans$, $\bdelta_y(x,y)=\v\trans$ with all other higher order derivatives being 0 we have,
{\allowdisplaybreaks
	\begin{align*}
		g_x(x,y)&=\bdelta_x(x,y)\nabla K(\Delta(x,y))=\u\trans\nabla K(\Delta(x,y)),\\
		g_y(x,y)&=\bdelta_y(x,y)\nabla K(\Delta(x,y))=\v\trans\nabla K(\Delta(x,y)),\\
		g_{xx}(x,y)&=\bdelta_x(x,y)\nabla^2K(\Delta(x,y))\bdelta_x(x,y)\trans=\u\trans\nabla^2K(\Delta(x,y))\u,\\
		g_{xy}(x,y)&=\bdelta_x(x,y)\nabla^2K(\Delta(x,y))\bdelta_y(x,y)\trans=\u\trans\nabla^2K(\Delta(x,y))\v,\\
		g_{yy}(x,y)&=\bdelta_y(x,y)\nabla^2K(\Delta(x,y))\bdelta_y(x,y)\trans=\v\trans\nabla^2K(\Delta(x,y))\v,\\
		g_{xxx}(x)&=\sum_{i=1}^{d}\delta_{i,xxx}(x,y)\left(\frac{\partial K}{\partial\delta_i}\right)+3\sum\limits_{i,j=1}^{d}\delta_{i,xx}(x,y)\frac{\partial^2K}{\partial\delta_i\partial\delta_j}\delta_{j,x}(x,y)\\&\hspace{2cm}+\sum\limits_{i,j,k=1}^{d}\delta_{i,x}(x,y)\delta_{j,x}(x,y)\delta_{k,x}(x,y)\frac{\partial^3K}{\partial\delta_i\partial\delta_j\partial\delta_k},\\
		&=\sum\limits_{i,j,k=1}^{d}\delta_{i,x}(x,y)\delta_{j,x}(x,y)\delta_{k,x}(x,y)\frac{\partial^3K}{\partial\delta_i\partial\delta_j\partial\delta_k}=\c_{\u,\u}\trans\nabla^3K(\Delta(x,y))\u.
	\end{align*}
}
Similarly $g_{xxy}(x)=\c_{\u,\u}\trans\nabla^3K(\Delta(x,y))\v$, $g_{yyx}(x)=\c_{\v,\v}\trans\nabla^3K(\Delta(x,y))\u$ and\\ $g_{yyy}(x)=\c_{\v,\v}\trans\nabla^3K(\Delta(x,y))\v$. Next,
{\allowdisplaybreaks
	\begin{align*}
		g_{xxxx}(x,y)&=\sum_{i=1}^{d}\delta_{i,xxxx}(x,y)\left(\frac{\partial K}{\partial\delta_i}\right)+\sum\limits_{i,j=1}^{d}\delta_{i,xxx}(x,y)\frac{\partial^2K}{\partial\delta_i\partial\delta_j}\delta_{j,x}(x,y)\\&+3\bigg[\sum\limits_{i,j=1}^{d}\delta_{i,xxx}(x,y)\frac{\partial^2K}{\partial\delta_i\partial\delta_j}\delta_{j,x}(x,y)+\sum\limits_{i,j=1}^{d}\delta_{i,xx}(x,y)\frac{\partial^2K}{\partial\delta_i\partial\delta_j}\delta_{j,xx}(x,y)\\&\hspace{1cm}+\sum_{i,j,k=1}^{d}\delta_{i,xx}(x,y)\delta_{j,x}(x,y)\delta_{k,x}(x,y)\frac{\partial^3K}{\partial\delta_i\partial\delta_j\partial\delta_k}\bigg]\\&\hspace{0.5cm}
		+3\sum\limits_{i,j,k=1}^{d}\delta_{i,xx}(x,y)\delta_{j,x}(x,y)\delta_{k,x}(x,y)\frac{\partial^3K}{\partial\delta_i\partial\delta_j\partial\delta_k}\\&+\sum_{i,j,k,l=1}^{d}\delta_{i,x}(x,y)\delta_{j,x}(x,y)\delta_{k,x}(x,y)\delta_l'(x)\frac{\partial^4K}{\partial\delta_i\partial\delta_j\partial\delta_k\partial\delta_l},\\
		&=\sum_{i,j,k,l=1}^{d}\delta_{i,x}(x,y)\delta_{j,x}(x,y)\delta_{k,x}(x,y)\delta_l'(x)\frac{\partial^4K}{\partial\delta_i\partial\delta_j\partial\delta_k\partial\delta_l}=\c_{\u,\u}\trans\nabla^4K(\bdelta(x))\c_{\u,\u}.
	\end{align*}
}
Similarly, $g_{xxxy}(x,y)=\c_{\u,\u}\trans\nabla^4K(\bdelta(x))\c_{\u,\v}$, $g_{yyyx}(x,y)=\c_{\v,\v}\trans\nabla^4K(\bdelta(x))\c_{\v,\u}$, $g_{xxyy}(x,y)=\c_{\u,\v}\trans\nabla^4K(\bdelta(x))\c_{\u,\v}$ and $g_{yyyy}(x,y)=\c_{\v,\v}\trans\nabla^4K(\bdelta(x))\c_{\v,\v}$. 
Evaluated at $x,y=0$, i.e. $\bdelta(0,0)=\Delta$, $g_{xxyy}$ provides us with the required expression, $C^{(2)}_{\u,\v}(\s,\s')=\c_{\u,\v}\trans\nabla^4K(\Delta)\c_{\u,\v}$ and $var(D^{(2)}_{\u,\v}Y(\s))=\lim_{h\to0}E(Y^{(2)}_{\u,\v,h}(\s),Y^{(2)}_{\u,\v,k}(\s))=\c_{\u,\v}\trans\nabla^4K({\bf 0})\c_{\u,\v}$ which exists if $K^{(iv)}(\Delta)$ exists for all $\Delta$, including $\Delta={\bf 0}$.

\paragraph{\em Note:} {\em In the above proof we make some abuse of notation for brevity of mathematical expressions involved. To clarify, $\delta_x=\del x\delta(x,y)$, $\delta_{xx}=\deltwo x\delta(x,y)$ and so on, $g_x=\del x g(x,y)$, $g_{xx}=\deltwo x g(x,y)$ etc., $\sum_{i,j=1}^d=\sum_{i=1}^{d}\sum_{j=1}^{d}$ etc.}

To derive (2) 
and the covariance for the directional curvature process, we assume that $Y(\s)$ is isotropic, i.e. $K(\Delta)=\tK(||\Delta||)$ therefore,
\begin{align*}
		C^{(2)}_{\medmath\u,\medmath\u}(\s,\s')&=\lim_{h\to 0}\lim_{k\to 0}E\left[Y^{(2)}_{\u,\v,h}(\s)Y^{(2)}_{\u,\v,k}(\s')\right],\\&=	\lim_{h\to 0}\lim_{k\to 0}\frac{1}{h^2k^2}\bigg[	E\left(Y(\s+h(\u+\v))Y^{(2)}_{\u,\v,k}(\s')\right)-E(Y(\s+h\u)Y^{(2)}_{\u,\v,k}(\s'))\\&\hspace{3cm}-E(Y(\s+k\v)Y^{(2)}_{\u,\v,k}(\s'))+E(Y(\s)Y^{(2)}_{\u,\v,k}(\s'))\bigg]
\end{align*}
where,
\begin{align*}
	E\left(Y(\s+h(\u+\v))Y^{(2)}_{\u,\v,k}(\s')\right)&=\tK(||\Delta+(h-k)(\u+\v)||)-\tK(||\Delta+(h-k)\u+h\v||)-\\&~~\tK(||\Delta+h\u+(h-k)\v||)+\tK(||\Delta+h(\u+\v)||)\\&~~-\tK(||\Delta+(h-k)\u-k\v||),\\
	E(Y(\s+h\u)Y^{(2)}_{\u,\v,k}(\s'))&=\tK(||\Delta+(h-k)\u-k\v||)-\tK(||\Delta+(h-k)\u||)\\&~~-\tK(||\Delta+h\u-k\v||)+\tK(||\Delta+h\u||)\\
	E(Y(\s+k\v)Y^{(2)}_{\u,\v,k}(\s'))&=\tK(||\Delta-k\u+(h-k)\v||)-\tK(||\Delta-k\u+h\v||)\\&~~-\tK(||\Delta+(h-k)\v||)+\tK(||\Delta+h\v||)\\
	E(Y(\s)Y^{(2)}_{\u,\v,k}(\s'))&=\tK(||\Delta-k(\u+\v)||)-\tK(||\Delta-k\u||)-\tK(||\Delta-k\v||)+\tK(||\Delta||)
\end{align*}
suppressing dependence on $\Delta$, $\u$ and $\v$ , we define $\rho(h,k)=||\Delta(h,k)||=||\Delta+h\u+k\v||$ and let $g(h,k)=K(\rho(h,k))$. Hence,

{\allowdisplaybreaks
	\begin{align*}
		C^{(2)}_{\u,\v}(\s,\s')&=\lim_{h\to 0}\lim_{k\to 0}\frac{1}{h^2k^2}[\rho(h-k,h-k)-\rho(h-k,h)-\rho(h,h-k)+\rho(h,h)\\
		&\hspace{3cm} -\rho(h-k,-k)+\rho(h-k,0)+\rho(h,-k)-\rho(h,0)\\
		&\hspace{3cm} -\rho(-k,h-k)+\rho(-k,h)+\rho(0,h-k)-\rho(0,h)\\
		&\hspace{3cm} +\rho(-k,-k)-\rho(-k,0)-\rho(0,-k)+\rho(0,0)],\\
		&=\lim_{h\to0}\frac{\rho''(h,h)-\rho''(h,0)-\rho''(0,h)+\rho''(0,0)}{h^2}=\rho^{(iv)}(0,0)\;.
	\end{align*}
}
Since, $\rho(0,0)=||\Delta||$, from the previous proof we can see that, $\rho^{(iv)}(0,0)=\nabla\tK(||\Delta||)\rho^{(iv)}(0,0)+\nabla^2\tK(||\Delta||)\left(4\rho'''(0,0)\rho'(0,0)+3[\rho''(0,0)]^2\right)+6\nabla^3\tK(||\Delta||)\rho''(0,0)[\rho'(0,0)]^2+\nabla^4\tK(||\Delta||)[\rho'(0,0)]^4$. After some algebra, $\rho'(0)=\frac{\u\trans\Delta}{||\Delta||}$, $\rho''(0)=\frac{1}{||\Delta||}\left(1-\frac{(\u\trans\Delta)^2}{||\Delta||^2}\right)$, $\rho'''(0)=-3\frac{\u\trans\Delta}{||\Delta||^3}\left(1-\frac{(\u\trans\Delta)^2}{||\Delta||^2}\right)$ and $\rho^{(iv)}(0)=\frac{3}{||\Delta||^3}\left(5\frac{(\u\trans\Delta)^2}{||\Delta||^2}-1\right)\left(1-\frac{(\u\trans\Delta)^2}{||\Delta||^2}\right)$. Substituting and grouping terms corresponding to $\left(\nabla^2\tK(||\Delta||)-\frac{\nabla\tK(||\Delta||)}{||\Delta||}\right)$, $\nabla^3\tK(||\Delta||)$ and $\nabla^4\tK(||\Delta||)$ we get,

\begin{equation*}
    \begin{split}
	g^{(iv)}(0)&=\frac{3}{||\Delta||^2}\left\{1-5\frac{(\u\trans\Delta)^2}{||\Delta||^2}\right\}\left[1-\frac{(\u\trans\Delta)^2}{||\Delta||^2}\right]\left(\nabla^2\tK(||\Delta||)-\frac{\nabla\tK(||\Delta||)}{||\Delta||}\right)\\&\hspace{.5cm}+\frac{6}{||\Delta||}\frac{(\u\trans\Delta)^2}{||\Delta||^2}\left[1-\frac{(\u\trans\Delta)^2}{||\Delta||^2}\right]\nabla^3\tK(||\Delta||)+\left(\frac{(\u\trans\Delta)^2}{||\Delta||^2}\right)^2\nabla^4\tK(||\Delta||),
 \end{split}
\end{equation*}
which is the required expression. 
We discuss validity of the curvature process for surfaces in $\Rthree$. It can be easily extended to surfaces in $\Rd$. Define,
{\allowdisplaybreaks
	\begin{align*}
		{\cal L}_{\e_1,\e_2,h}Y(\s)&=\begin{pmatrix}1&0&0&0&0&0\\-\frac{1}{h}&\frac{1}{h}&0&0&0&0\\-\frac{1}{h}&0&\frac{1}{h}&0&0&0\\\frac{1}{h^2}&-\frac{2}{h^2}&0&\frac{1}{h^2}&0&0\\\frac{1}{h^2}&-\frac{1}{h^2}&-\frac{1}{h^2}&0&\frac{1}{h^2}&0\\\frac{1}{h^2}&0&-\frac{2}{h^2}&0&0&\frac{1}{h^2}\end{pmatrix}\begin{pmatrix}Y(\s)\\Y(\s+h\e_1)\\Y(\s+h\e_2)\\Y(\s+2h\e_1)\\Y(\s+h(\e_1+\e_2))\\Y(\s+2h\e_2)\end{pmatrix}\\&=\A_h\L_h(\s)=\left(Y(\s),Y^{(1)}_{\e_1,h}(\s),Y^{(1)}_{\e_2,h}(\s),Y^{(2)}_{\e_1,\e_1,h}(\s),Y^{(2)}_{\e_1,\e_2,h}(\s),Y^{(2)}_{\e_2,\e_2,h}(\s)\right)\trans,
	\end{align*}
}
as finite difference corresponding to the differential operator ${\cal L}Y$ using the expressions of $Y^{(1)}_{\e_i,h}(\s)$ and $Y^{(2)}_{\e_i,\e_j,h}(\s)$, $i,j=1,2$. For every $h>0$ this defines a linear transformation, since the determinant of $\A_h$ is $h^{-8}$. We denote the differenced differential process on the right of $\A_h$, suppressing dependence on $\e_1,\e_2$ by $\L_h(\s)$. The associated covariance matrix is given by, $Cov({\cal L}_{\e_1,\e_2h}Y(\s),{\cal L}_{\e_1,\e_2,k}Y(\s'))=\A_h{\cal K}_{\e_1,\e_2,h,k}(\Delta)\A_k\trans$, where elements of ${\cal K}_{\e_1,\e_2,h,k}(\Delta)$ are obtained from $Cov(\L_h(\s),\L_k(\s'))$. Hence, as $h\downarrow 0$, $	{\cal L}_{\e_1,\e_2,h}Y(\s)\to {\cal L}Y(\s)$ and $\lim_{h\downarrow 0,k\downarrow 0} Cov({\cal L}_{\e_1,\e_2h}Y(\s),{\cal L}_{\e_1,\e_2,k}Y(\s')) = V_{{\cal L}Y}(\Delta)$, where the limits operate element-wise on the matrix $\A_h{\cal K}_{\e_1,\e_2,h,k}(\Delta)\A_k\trans$ and, the expression for each element is obtained from previous computations, by setting $\u=\e_1$ and $\v=\e_2$.

For the directional operator, this can be established by observing that the directional differential operator is obtained as follows,
\begin{align*}
		{\cal L}_{\u,h}Y(\s)=\begin{pmatrix}
		1&0&0&0&0&0\\
		0&u_1&u_2&0&0&0\\
		0&0&0&u_1^2&2u_1u_2&u_2^2
	\end{pmatrix}{\cal L}_{\e_1,\e_2,h}Y(\s)=\left(1\oplus\u\trans\oplus\c_{\u,\u}\trans\right){\cal L}_{\e_1,\e_2,h}Y(\s),
\end{align*}
where $\oplus$ denotes the direct sum for matrices, then as $h\to 0$, ${\cal L}_{\u,h}Y(\s)\to {\cal L}_{\u}Y(\s)$. The covariance matrix is obtained following similar arguments presented in the proof for the previous result.
In case the covariance is isotropic we have $K(\Delta)=\tK(||\Delta||)$, on repeated differentiation and noting that $\del\Delta||\Delta||=\frac{\Delta}{||\Delta||}$ we have,

\begin{align*}
	\nabla K(\Delta)&=\frac{\nabla\tK(||\Delta||)}{||\Delta||}\Delta,\\
	\nabla^2K(\Delta)&=\frac{\nabla\tK(||\Delta||)}{||\Delta||}I-\frac{\nabla\tK(||\Delta||)}{||\Delta||^3}\Delta\Delta\trans+\frac{\nabla^2\tK(||\Delta||)}{||\Delta||^2}\Delta\Delta\trans\\
	&=\frac{\nabla\tK(||\Delta||)}{||\Delta||}I+\left(\nabla^2\tK(||\Delta||)-\frac{\nabla\tK(||\Delta||)}{||\Delta||}\right)\frac{\Delta\Delta\trans}{||\Delta||^2}.
\end{align*}
Differentiating $\nabla^2K(\Delta)$ w.r.t. $\Delta$ we obtain,
{\allowdisplaybreaks
	\begin{align*}
		\nabla^3K(\Delta)&=\frac{\nabla^2\tK(||\Delta||)}{||\Delta||^2}vech(I)\trans\otimes\Delta-\frac{\nabla\tK(||\Delta||)}{||\Delta||^3}vech(I)\trans\otimes\Delta\\
		&~~+\left(\nabla^2\tK(||\Delta||)-\frac{\nabla\tK(||\Delta||)}{||\Delta||}\right)\frac{1}{||\Delta||^2}\frac{\partial vech(\Delta\Delta\trans)}{\partial\Delta}\\
		&~~-2\left(\nabla^2\tK(||\Delta||)-\frac{\nabla\tK(||\Delta||)}{||\Delta||}\right)\frac{1}{||\Delta||^4}\frac{\partial vech(\Delta\Delta\trans)}{\partial\Delta}\\
		&~~-\left(\nabla^2\tK(||\Delta||)-\frac{\nabla\tK(||\Delta||)}{||\Delta||}\right)\frac{1}{||\Delta||^4}\frac{\partial vech(\Delta\Delta\trans)}{\partial\Delta}\\
		&~~+\nabla^3\tK(||\Delta||)\cdot\frac{vech(\Delta\Delta\trans)\trans\otimes\Delta}{||\Delta||^3}.
	\end{align*}
}
On grouping terms for $\nabla^2\tK(||\Delta||)-\frac{\nabla\tK(||\Delta||)}{||\Delta||}$ and $\nabla^3\tK(||\Delta||)$ we obtain the required expression,
{\allowdisplaybreaks
	\begin{align*}
		\nabla^3K(\Delta)&=\left(\nabla^2\tK(||\Delta||)-\frac{\nabla\tK(||\Delta||)}{||\Delta||}\right)\Bigg\{\frac{vech(I)\trans\otimes\Delta}{||\Delta||^2}-3\frac{vech(\Delta\Delta\trans)\trans\otimes\Delta }{||\Delta||^4}\\&\hspace{7cm}+\frac{1}{||\Delta||^2}\frac{\partial vech(\Delta\Delta\trans)}{\partial\Delta}\Bigg\}\\&\hspace{3cm}+\nabla^3\tK(||\Delta||)\cdot\frac{vech(\Delta\Delta\trans)\trans\otimes\Delta}{||\Delta||^3}
	\end{align*}
}
To obtain $\nabla^4K(\Delta)$ we differentiate $\nabla^3K(\Delta)$ w.r.t. $\Delta$, we use notations $A_1=\frac{\partial\Delta\otimes vech(I)\trans}{\partial \Delta}$, $A_2=\frac{\partial\Delta\otimes vech(\Delta\Delta\trans)\trans}{\partial \Delta}$, $A_3=\frac{\partial}{\partial\Delta}\left(\frac{\partial vech(\Delta\Delta\trans)}{\partial\Delta}\right)$ for matricized tensors of order $d(d+1)/2\times d(d+1)/2$, where the order of matricization conforms to the listing order of the half-vectorization operator $vech$ and $A_4$ denotes the element-wise product of $\Delta$ with $\left(\frac{\partial vech(\Delta\Delta\trans)}{\partial\Delta}\right)$ in the same order as the matricized tensor. 

On differentiating the factor corresponding to the coefficient $\nabla^2\tK(||\Delta||)-\frac{\nabla\tK(||\Delta||)}{||\Delta||}$ we obtain,
\begin{gather*}
	\begin{aligned}
		\frac{1}{||\Delta||^2}A_1-3\frac{1}{||\Delta||^4}A_2+\frac{1}{||\Delta||^2}A_3-\frac{2}{||\Delta||^4}vech(\Delta\Delta\trans)vech(I)\trans\\+\frac{12}{||\Delta||^6}vech(\Delta\Delta\trans)vech(\Delta\Delta\trans)\trans-\frac{2}{||\Delta||^4}A_4.
	\end{aligned}
\end{gather*}
Differentiating $\nabla^2\tK(||\Delta||)-\frac{\nabla\tK(||\Delta||)}{||\Delta||}$,
\begin{align*}
	\frac{\nabla^3\tK(||\Delta||)}{||\Delta||}\Delta-\left(\nabla^2\tK(||\Delta||)-\frac{\nabla\tK(||\Delta||)}{||\Delta||}\right)\frac{\Delta\Delta\trans}{||\Delta||^2}.
\end{align*}
Differentiating the factor corresponding to the coefficient $\nabla^3\tK(||\Delta||)$ we obtain,
\begin{align*}
	-3\frac{vech(\Delta\Delta\trans)vech(\Delta\Delta\trans)\trans}{||\Delta||^5}+\frac{1}{||\Delta||^3}A_2,
\end{align*}
finally, differentiating $\nabla^3\tK(||\Delta||)=\nabla^4\tK(||\Delta||)\frac{\Delta}{||\Delta||}$. Grouping coefficients for $\nabla^2\tK(||\Delta||)-\frac{\nabla\tK(||\Delta||)}{||\Delta||}$, $\nabla^3\tK(||\Delta||)$ and $\nabla^4\tK(||\Delta||)$ we obtain the required expression, thereby completing the proof.

For proving the result focusing on spectral theory, note that since $f$ is symmetric about 0 by hypothesis,
\begin{align*}
	K(t)=\int_{\Rone} e^{i\lambda t}f(\lambda)d\lambda=\int_{\Rone} \cos(\lambda t) f(\lambda)d\lambda+i\int_{\Rone} \sin(\lambda t) f(\lambda)d\lambda=\int \cos(\lambda t) f(\lambda)d\lambda.
\end{align*}
Differentiating w.r.t. $t$ on both sides we have,
\begin{align*}
	\nabla K(t)=-\int \sin(\lambda t) \lambda f(\lambda)d\lambda,
\end{align*}
Since $|\sin(\lambda t)\lambda f(\lambda)|\leq|\lambda|f(\lambda)$ and $\int|\lambda|f(\lambda)d\lambda<\infty$ under hypothesis, differentiation under the integral sign is valid. We repeat the process to obtain,
\begin{align*}
	\nabla^2 K(t)=-\int \cos(\lambda t)\lambda^2 f(\lambda)d\lambda&,& \nabla^3K(t)=\int \sin(\lambda t)\lambda^3 f(\lambda)d\lambda&,& \nabla^4K(t)=\int \cos(\lambda t)\lambda^4 f(\lambda)d\lambda,
\end{align*}
Next we make the following observations for limits of these derivatives,

{\allowdisplaybreaks
    \begin{align*}
	&\lim_{t\to 0}\left(\nabla^2K(t)-\frac{\nabla K(t)}{t}\right)=\lim_{t\to 0}\int\left(\cos(\lambda t)-\frac{\sin(\lambda t)}{\lambda t}\right)\lambda^2f(\lambda)d\lambda=0,\\
	&\lim_{t\to 0} \nabla^2K(t)=-\lim_{t\to 0}\int \cos(\lambda t)\lambda^2 f(\lambda)d\lambda=-\int \lambda^2 f(\lambda)d\lambda<\infty,\\
	&\lim_{t\to 0}\nabla^3K(t)=\lim_{t\to 0}\int \sin(\lambda t)\lambda^3 f(\lambda)d\lambda=0,\\
	&\lim_{t\to 0}\nabla^4K(t)=\lim_{t\to 0}\int \cos(\lambda t)\lambda^4 f(\lambda)d\lambda=\int \lambda^4 f(\lambda)d\lambda<\infty,
\end{align*}
}
We evaluate the results obtained above under these observations. Making note of,
\begin{align*}
	\left\{\frac{vech(I)\trans\otimes\Delta}{||\Delta||^2}-3\frac{vech(\Delta\Delta\trans)\trans\otimes\Delta }{||\Delta||^4}+\frac{1}{||\Delta||^2}\frac{\partial vech(\Delta\Delta\trans)}{\partial\Delta}\right\}&,& \frac{vech(\Delta\Delta\trans)\trans\otimes\Delta}{||\Delta||^3}
\end{align*}
stay bounded as $\Delta\to 0$ implying $\nabla^3K(\Delta)\to 0$ as $\Delta\to 0$. For $\nabla^4K(\Delta)$, we observe that the factors corresponding to $\nabla^2\tK(||\Delta||)-\frac{\nabla\tK(||\Delta||)}{||\Delta||}$, $\nabla^3\tK(||\Delta||)$ and $\nabla^4\tK(||\Delta||)$ 
remain bounded as $\Delta\to 0$, additionally $\frac{vech(\Delta\Delta\trans)vech(\Delta\Delta\trans)\trans}{||\Delta||^4}\to I_{d(d+1)/2}$ as $\Delta\to 0$. For each diagonal element of $(\nabla^4K)_{ii}=a_i=\int \lambda^4 f_i(\lambda)d\lambda$ which completes the proof.

For results in page 8, we prove this for $\s \in \Rtwo$, the proof can be extended to $\Rd$ analogously. Under the hypothesis,

\begin{align*}
	Y_1(\s),  \sim GP({\bf 0},K(\cdot,\btheta^{1}_K)) \text { and } Y_2(\s)  \sim GP({\bf 0},K(\cdot,\btheta^{2}_K))
\end{align*}
independently. Without loss of generality, consider the finite difference differential process,

{\allowdisplaybreaks
	\begin{align*}
		{\cal L}_{\e_1,\e_2,h}Y_1(\s)=\A_h\L^1_{h}(\s)=\begin{pmatrix}1&0&0&0&0&0\\-\frac{1}{h}&\frac{1}{h}&0&0&0&0\\-\frac{1}{h}&0&\frac{1}{h}&0&0&0\\\frac{1}{h^2}&-\frac{2}{h^2}&0&\frac{1}{h^2}&0&0\\\frac{1}{h^2}&-\frac{1}{h^2}&-\frac{1}{h^2}&0&\frac{1}{h^2}&0\\\frac{1}{h^2}&0&-\frac{2}{h^2}&0&0&\frac{1}{h^2}\end{pmatrix}\begin{pmatrix}Y_1(\s)\\Y_1(\s+h\e_1)\\Y_1(\s+h\e_2)\\Y_1(\s+2h\e_1)\\Y_1(\s+h(\e_1+\e_2))\\Y_1(\s+2h\e_2)\end{pmatrix},
	\end{align*}
}
noting that for every $h>0$, $\L^1_h(\s)$ follows a $6$-dimensional normal distribution and $|\A_h|=h^{-8}\ne0$, making the above linear transformation non-singular. We know from properties of multivariate normal distributions that ${\cal L}_{\e_1,\e_2,h}Y_1(\s)\sim {\cal N}_5(\A_h.{\bf 0},\A_h\K_h({\bf 0},\btheta^1_K)\A_h\trans)$, where $\K_h({\bf 0},\btheta^1_K)=Var(\L^1_h(\s))$, the cross-covariance matrix for the process ${\cal L}_{\e_1,\e_2,h}Y_1(\s)$ is $\A_h{\cal K}_{h,k}(\Delta)\A_k\trans$, where ${\cal K}_{h,k}(\Delta)=Cov(\L^1_h(\s),\L^1_{k}(\s'))$, with $\Delta=\s-\s'$. As $h\to 0$, $\A_h\K_h({\bf 0},\btheta^1_K)\A_h\trans\to V_{{\cal L}Y_1}({\bf 0})$ and ${\cal L}_{\e_1,\e_2,h}Y_1(\s) \overset{d}{\to} {\cal L}Y_1(\s)\sim {\cal N}_5({\bf 0}, V_{{\cal L}Y_1}(\bf 0))$, where $\overset{d}{\to}$ indicates convergence in distribution. As $h,k\downarrow 0$ $\A_h{\cal K}_{h,k}(\Delta)\A_k\trans\to V_{{\cal L}Y_1}(\Delta)$, implying that ${\cal L}Y_1(\s) \sim GP({\bf 0},V_{{\cal L} Y_1}(\cdot, \btheta^1_K))$. The same arguments can be followed for showing ${\cal L}Y_2(\s) \sim GP({\bf 0},V_{{\cal L} Y_2}(\cdot, \btheta^2_K))$.
\begin{enumerate}
	\item for (a) consider the associated finite differential operators, ${\cal L}_{\e_1,\e_2,h}Y_1(\s)=\A_h\L^1_h(\s)$ and ${\cal L}_{\e_1,\e_2,h}Y_2(\s)=\A_h\L^2_h(\s)$. We note that for every $h>0$, $Cov(\L^1_h(\s),\L^2_h(\s))={\bf O}$, which is the zero matrix (of order $6\times 6$). Now consider the process,
	\begin{align*}
		&{\cal L}^{1,2}_h=({\cal L}_{\e_1,\e_2,h}Y_1(\s)\trans, {\cal L}_{\e_1,\e_2,h}Y_2(\s)\trans)\trans=\A_h\oplus\A_h(\L^1_h(\s)\trans,\L^2_h(\s)\trans)\trans\\
		&{\cal L}^{1,2}_h\sim {\cal N}_{12}({\bf 0},(\A_h\oplus\A_h) (\K_h(\cdot,\btheta^1_K)\oplus \K_h(\cdot,\btheta^2_K))(\A_h\oplus\A_h)\trans)
	\end{align*}
	as $h\downarrow 0$, ${\cal L}^{1,2}_h \overset{d}{\to} ({\cal L}Y_1\trans,{\cal L}Y_2\trans )\trans\sim {\cal N}_{12}({\bf 0},V_{{\cal L}Y_1}({\bf 0})\oplus V_{{\cal L}Y_2}({\bf 0}))$ which implies $Cov({\cal L}Y_1, {\cal L}Y_2)={\bf O}$, and since they jointly follow a multivariate Gaussian this implies that they are independent. Observing that the cross covariance for all $h,k\downarrow0$, $Cov(\L^1_h(\s),\L^2_k(\s'))={\bf O}$, we can establish that they are independent Gaussian processes.
	\item (b) and (c) follow from standard properties of the Gaussian processes.
\end{enumerate}

For the discussion on page 14 preceding (14) we suppress dependence on $\s$ and $Y$, we denote $g(t)=g\left({\cal L} Y(\s(t))\right)$. By definition of the integral, given $\epsilon>0$, there exists a $\delta_0>0$ such that if $|P|<\delta_0$, then
\begin{align}\label{eq::ineq-1}
		\left|\int_a^bg(t)||\s'(t)||dt-\sum\limits_{i=1}^{n_P}(t'_i-t'_{i-1})g(t'_i)||\s'(t'_i)||\right|<\frac{\epsilon}{2},
\end{align}
where $\sum\limits_{i=1}^{n_P}(t'_i-t'_{i-1})g(t'_i)||\s'(t'_i)||$ is a {\em Riemann sum approximation} of the integral. On the other hand, since $g(t)$ is uniformly continuous over $[a,b]$, given $\epsilon>0$, there exists $\delta_1>0$ such that if $x,y\in [a,b]$ with $|x-y|<\delta_1$,
\begin{align*}
	\big|g(x)||\s'(x)||-g(y)||\s'(y)||\big|<\frac{\epsilon}{2(b-a)}
\end{align*}
Set $\delta=\min\{\delta_0,\delta_1\}$, then $|P|<\delta$, using the mean value theorem we obtain,
\begin{align*}
	&\left|\sum_{i=1}^{n_P}\int_{C_{t_i}}g(t)||\s'(t)||dt -\sum_{i=1}^{n_P}(t'_i-t'_{i-1})g(t'_i)||\s'(t'_i)||\right|\\&\hspace{4cm}\leq\left|\sum_{i=1}^{n_P}(t'_i-t'_{i-1})\sup g(t)||\s'(t)||-\sum_{i=1}^{n_P}(t'_i-t'_{i-1})g(t'_i)||\s'(t'_i)||\right|\\&\hspace{4cm}\leq\left|\sum_{i=1}^{n_P}(t'_i-t'_{i-1})\sup\big|g(t)||\s'(t)||-g(t')||\s'(t')||\big|\right|\leq\frac{\epsilon}{2}\;,
\end{align*}
where the first inequality follows from the assumption of $C$ being regular. Together with the inequality in (\ref{eq::ineq-1}) we have,
\begin{align*}
	\left|\int_a^bg(t)||\s'(t)||dt-\sum_{i=1}^{n_P}\int_{C_{t_i}}g(t)||\s'(t)||dt\right|<\epsilon.
\end{align*}
Finally, almost sure convergence yields for every $\epsilon>0$,
\begin{align*}
	P\left[\left|\int_a^bg(t)||\s'(t)||dt-\sum_{i=1}^{n_P}\int_{C_{t_i}}g(t)||\s'(t)||dt\right|<\epsilon\right]=1.
\end{align*}
Considering a sequence of $\epsilon\downarrow 0$, and using the preceding arguments for each $\epsilon$ we can find $\delta\downarrow 0$ such that $|P|<\delta$ which concludes the proof.

\newpage

\section{Tables}\label{sec::tables_supp}

\begin{table}[h]
	\renewcommand{\arraystretch}{0.8}
	\caption{Table showing results for goodness of fit from the first synthetic experiment where the true response is generated from $y\sim N(10[\sin(3\pi s_1)+\cos(3\pi s_2)],1)$.}\label{tab::sim-results-1}
	\centering
	\resizebox*{\linewidth}{!}{
		\begin{tabular}{l|@{\extracolsep{25pt}}c|c|cccccc@{}}
			\hline
			\hline
			\multirow{4}{*}{Sample Size} & \multirow{4}{*}{Parameter} & \multirow{4}{*}{Estimate} & \multicolumn{6}{c}{\multirow{2}{*}{RMSE}}\\
			&&&&&&&&\\
			\cline{4-9}
			&&&\multirow{2}{*}{$Y$}&\multirow{2}{*}{$\nabla_1 Y$}&\multirow{2}{*}{$\nabla_2 Y$}&\multirow{2}{*}{$\nabla^2_{11} Y$}&\multirow{2}{*}{$\nabla^2_{12} Y$}&\multirow{2}{*}{$\nabla^2_{22} Y$}\\
			&&&&&&&&\\
			\hline
			\multirow{8}{*}{$L=100$} & \multirow{2}{*}{$\phi$} & 2.91 &&&&&&\\
			&&(2.13, 3.81)&&&&&&\\
			&\multirow{2}{*}{$\sigma^2$}& 218.35 &&&&&&\\
			&&(69.69, 588.86)&0.05&9.74&9.86&150.82&91.49&180.38\\
			&\multirow{2}{*}{$\tau^2$ (Truth = {\bf 1.00})}& \bf 0.96& (0.00) &(0.23)&(0.31)&(4.94)&(1.94)&(2.10)\\
			&&({\bf 0.63, 1.39})&&&&&&\\
			&\multirow{2}{*}{$\beta_0$}& 0.00 &&&&&&\\
			&&(-0.06, 0.06)&&&&&&\\
			\hline
			\multirow{8}{*}{$L=500$} & \multirow{2}{*}{$\phi$} & 2.30 &&&&&&\\
			&&(1.88, 2.73)&&&&&&\\
			&\multirow{2}{*}{$\sigma^2$}& 367.17 &&&&&&\\
			&&(166.27, 784.53)&0.04&6.66&6.84&127.99&68.11&126.71\\
			&\multirow{2}{*}{$\tau^2$ (Truth = {\bf 1.00})}& \bf 0.99& (0.00)&(0.10)&(0.12)&(3.40)&(2.56)&(2.69)\\
			&&({\bf 0.86, 1.13})&&&&&&\\
			&\multirow{2}{*}{$\beta_0$}& 0.00 &&&&&&\\
			&&(-0.06, 0.06)&&&&&&\\
			\hline
			\multirow{8}{*}{$L=1000$} & \multirow{2}{*}{$\phi$} & 2.00 &&&&&&\\
			&&(1.60, 2.36)&&&&&&\\
			&\multirow{2}{*}{$\sigma^2$}& 553.79 &&&&&&\\
			&&(231.95, 1385.62)&0.03&5.64&6.45&97.16&61.54&114.97\\
			&\multirow{2}{*}{$\tau^2$ (Truth = {\bf 1.00})}& \bf 1.04& (0.00)&(0.18)&(0.19)&(2.05)&(2.27)&(2.22)\\
			&&({\bf 0.95, 1.14})&&&&&&\\
			&\multirow{2}{*}{$\beta_0$}& 0.00 &&&&&&\\
			&&(-0.06, 0.06)&&&&&&\\
			\hline
			\hline
		\end{tabular}
	}
\end{table}

\begin{table}[!]
	\renewcommand{\arraystretch}{0.8}
	\caption{Table showing results for goodness of fit from the second synthetic experiment where the true response is generated from $y\sim N(10[\sin(3\pi s_1)\cdot\cos(3\pi s_2)],1)$.}\label{tab::sim-results-2}
	\centering
	\resizebox*{\linewidth}{!}{
		\begin{tabular}{l|@{\extracolsep{25pt}}c|c|cccccc@{}}
			\hline
			\hline
			\multirow{4}{*}{Sample Size} & \multirow{4}{*}{Parameter} & \multirow{4}{*}{Estimate} & \multicolumn{6}{c}{\multirow{2}{*}{RMSE}}\\
			&&&&&&&&\\
			\cline{4-9}
			&&&\multirow{2}{*}{$Y$}&\multirow{2}{*}{$\nabla_1 Y$}&\multirow{2}{*}{$\nabla_2 Y$}&\multirow{2}{*}{$\nabla^2_{11} Y$}&\multirow{2}{*}{$\nabla^2_{12} Y$}&\multirow{2}{*}{$\nabla^2_{22} Y$}\\
			&&&&&&&&\\
			\hline
			\multirow{8}{*}{$L=100$} & \multirow{2}{*}{$\phi$} & 5.06 &&&&&&\\
			&&(3.69, 6.72)&&&&&&\\
			&\multirow{2}{*}{$\sigma^2$}& 38.96 &&&&&&\\
			&&(15.52, 85.33)&0.06&16.12&11.49&258.91&159.54&183.54\\
			&\multirow{2}{*}{$\tau^2$ (Truth = {\bf 1.00})}& \bf 0.83 & (0.00) &(0.13)&(0.22)&(3.16)&(1.19)&(3.70)\\
			&&({\bf 0.53, 1.23})&&&&&&\\
			&\multirow{2}{*}{$\beta_0$}& -3.12 &&&&&&\\
			&&(-0.34, 5.04)&&&&&&\\
			\hline
			\multirow{8}{*}{$L=500$} & \multirow{2}{*}{$\phi$} & 3.63 &&&&&&\\
			&&(3.05, 4.38)&&&&&&\\
			&\multirow{2}{*}{$\sigma^2$}& 114.36 &&&&&&\\
			&&(44.59, 211.47)&0.04&7.04&7.39&179.23&98.39&169.19\\
			&\multirow{2}{*}{$\tau^2$ (Truth = {\bf 1.00})}& \bf 0.93 & (0.00)&(0.17)&(0.17)&(6.94)&(3.76)&(5.16)\\
			&&({\bf 0.80, 1.07})&&&&&&\\
			&\multirow{2}{*}{$\beta_0$}& -0.15 &&&&&&\\
			&&(-2.14, 2.49)&&&&&&\\
			\hline
			\multirow{8}{*}{$L=1000$} & \multirow{2}{*}{$\phi$} & 2.89 &&&&&&\\
			&&(2.55, 3.26)&&&&&&\\
			&\multirow{2}{*}{$\sigma^2$}& 173.19 &&&&&&\\
			&&(94.06, 297.76)&0.03&6.19&6.21&167.27&93.21&163.44\\
			&\multirow{2}{*}{$\tau^2$ (Truth = {\bf 1.00})}& \bf 1.01& (0.00)&(0.26)&(0.17)&(10.40)&(6.75)&(8.84)\\
			&&({\bf 0.91, 1.10})&&&&&&\\
			&\multirow{2}{*}{$\beta_0$}& -0.22 &&&&&&\\
			&&(-2.03, 2.15)&&&&&&\\
			\hline
			\hline
		\end{tabular}
	}
\end{table}

\newpage

\section{Plots}\label{sec::plots_supp}
\begin{figure}[H]
	\centering
	\def\svgwidth{0.7\linewidth}
	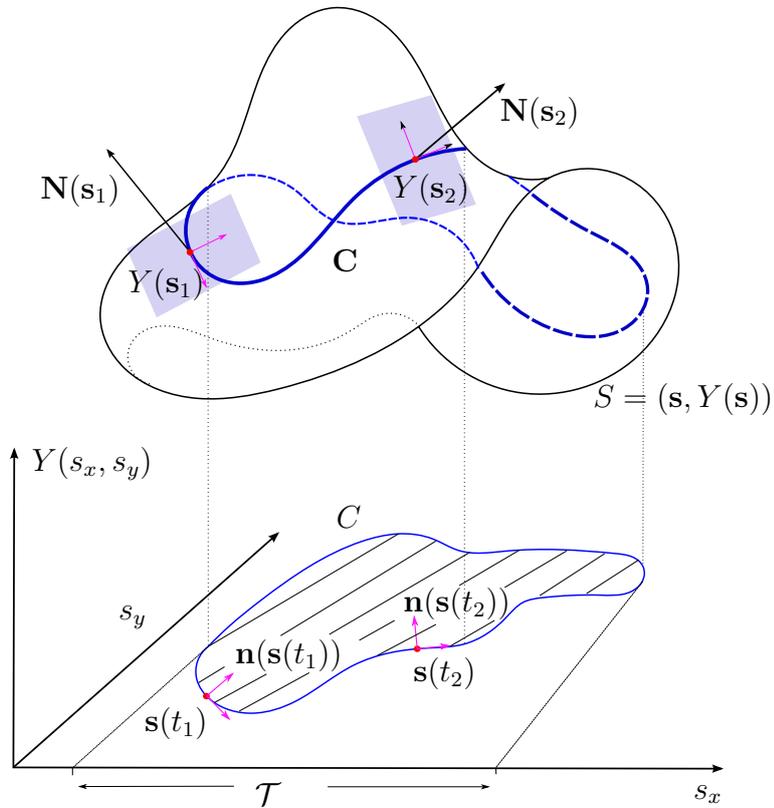

	\caption{Illustration showing geometric interpretation of curvilinear wombling. Local tangent planes are shaded around points ${\bf s}_1$, ${\bf s}_2$. Normals to the surface are marked as ${\bf N}({\bf s}_1)$ and ${\bf N}({\bf s}_2)$, locally projected principal unit normals to the projected curve ${\bf C}$ are marked as ${\bf n}({\bf s}(t_1))$ and ${\bf n}({\bf s}(t_2))$ respectively. The tangent vectors spanning the local tangent planes are shown with arrows.}\label{fig:geom-womb}
\end{figure}

\begin{figure}[H]
	\centering
	\def\svgwidth{0.7\linewidth}
	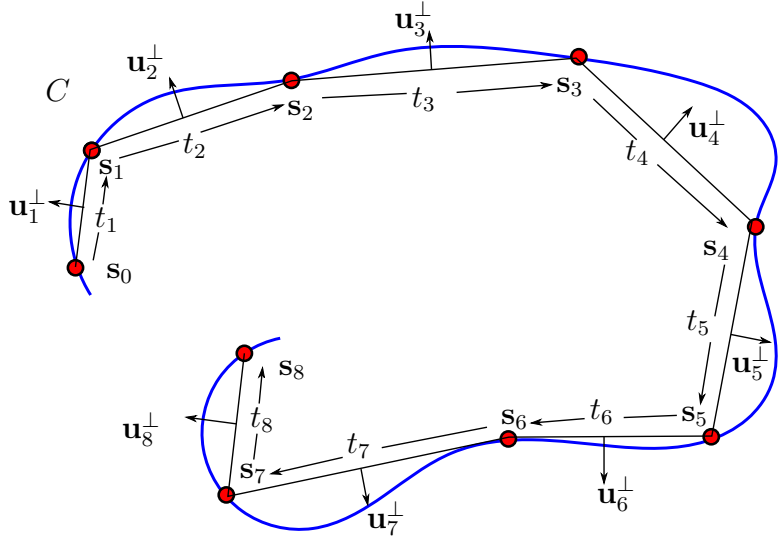

	\caption{Illustration of rectilinear wombling, showing a curve $C$, an initial starting point ${\bf s}_0$ on the curve, with following points, $\{{\bf s}_1,{\bf s}_2,\ldots,{\bf s}_8\}$ corresponding a partition ${\cal T}$, of the parameterized curve. Each linear segment consists of a norm-direction pair $(t,{\bf u})$, where $t$ specifies the length of the segment and ${\bf u}$ the direction of movement. The normal direction for each segment is indicated as ${\bf u}^{\perp}$.}	\label{fig::rect-womb}
\end{figure}

\begin{figure}[H]
	\centering
	\includegraphics[width=1\linewidth , height=0.5\linewidth]{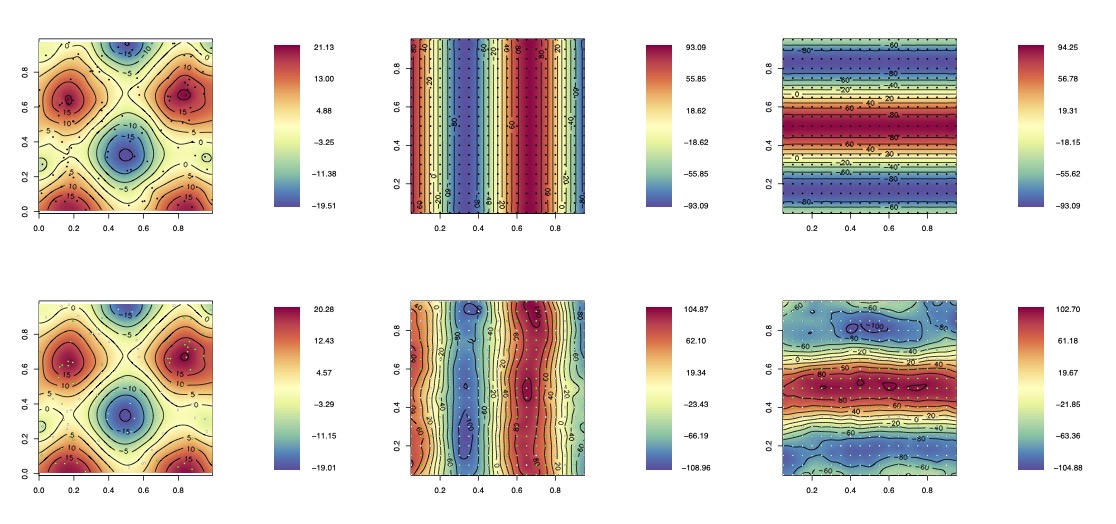}
	\caption{\small Plots showing the true surfaces for the (a) process, (b) gradients along $x$-axis, (c) gradients along $y$-axis, and estimated surfaces for the (d) process,  gradients (e), (f) w.r.t. synthetic response generated, $y\sim N(10[\sin(3\pi s_1)+\cos(3\pi s_2)],1)$}\label{fig::grad-1-t}
\end{figure}
\begin{figure}[H]
	\centering
	\includegraphics[width=1\linewidth , height=0.5\linewidth]{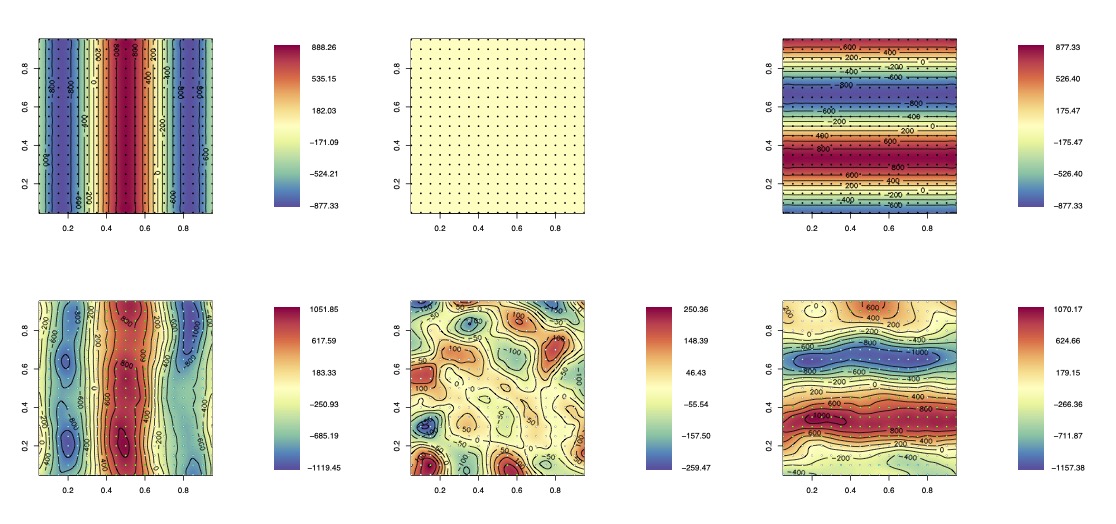}
	\caption{\small Plots showing the true surfaces for the (a) curvature along $x$-axis, (b) curvature along $x$-$y$-axis, (c) curvature along $y$-axis, and estimated surfaces for (d) curvature along $x$-axis, (e) mixed curvature along $x$-$y$-axis, (f) curvature along $y$-axis, for synthetic response generated from, $y\sim N(10[\sin(3\pi s_1)+\cos(3\pi s_2)],1)$.}\label{fig::curv-1-t}
\end{figure}
\begin{figure}[H]
	\centering
	\includegraphics[width=1\linewidth , height=0.5\linewidth]{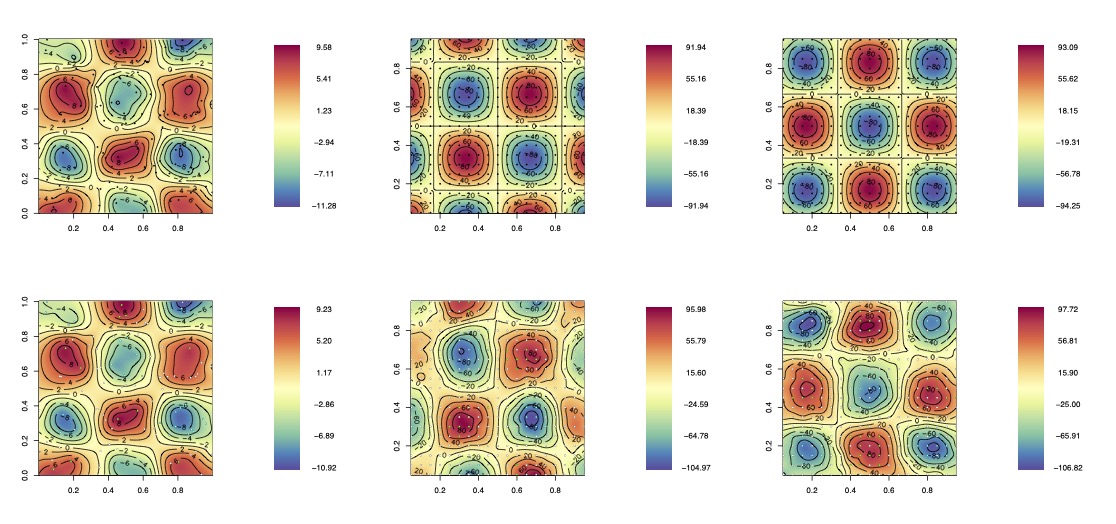}
	\caption{\small Plots showing the true surfaces for the (a) process, (b) gradients along $x$-axis, (c) gradients along $y$-axis, and estimated surfaces for the (d) process,  gradients (e), (f) for synthetic response generated from, $y\sim N(10[\sin(3\pi s_1)\cdot\cos(3\pi s_2)],1)$.}\label{fig::grad-2-t}
\end{figure}
\begin{figure}[H]
	\centering
	\includegraphics[width=1\linewidth , height=0.5\linewidth]{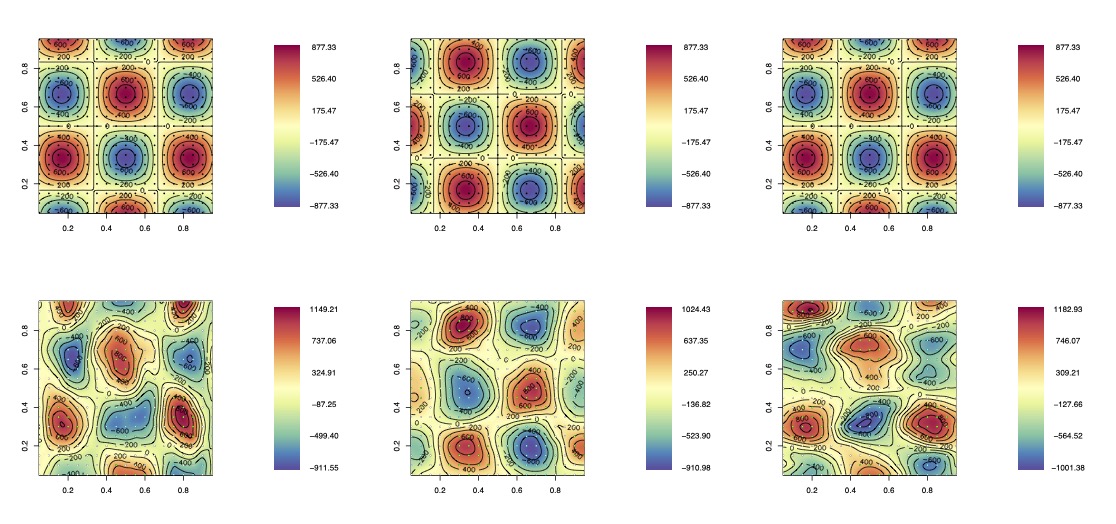}
	\caption{\small Plots showing the true surfaces for the (a) curvature along $x$-axis, (b) curvature along $x$-$y$-axis, (c) curvature along $y$-axis, and estimated surfaces for (d) curvature along $x$-axis, (e) mixed curvature along $x$-$y$-axis, (f) curvature along $y$-axis, for synthetic response generated from, $y\sim N(10[\sin(3\pi s_1)\cdot\cos(3\pi s_2)],1)$.}\label{fig::curv-2-t}
\end{figure}
\begin{figure}[H]
	\centering
	\includegraphics[width=1\linewidth , height=0.5\linewidth]{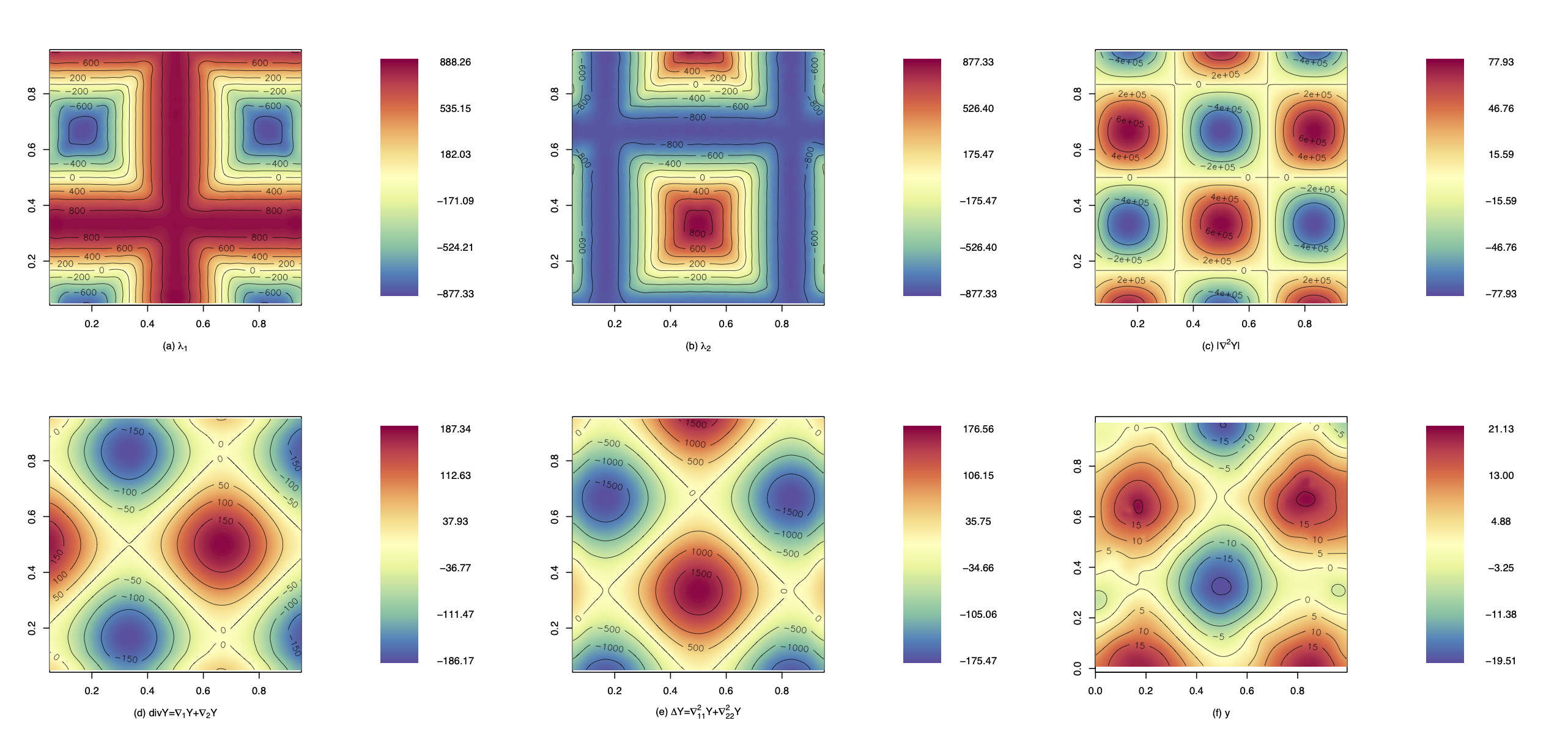}
	\caption{\small Plots showing the true surfaces for (a) eigen value, $\lambda_1$  (b) eigen value $\lambda_2$, (c) Gaussian curvature (scales in $\times 10^4$) (d) divergence (e) Laplacian (scales in $\times 10^1$) over grid points, and (f) fitted process. This is shown for synthetic response generated from, $y\sim N(10[\sin(3\pi s_1)+\cos(3\pi s_2)],1)$.}\label{fig::diff-forms-1-t}
\end{figure}
\vspace*{-.2in}
\begin{figure}[H]
	\centering
	\includegraphics[width=1\linewidth , height=0.5\linewidth]{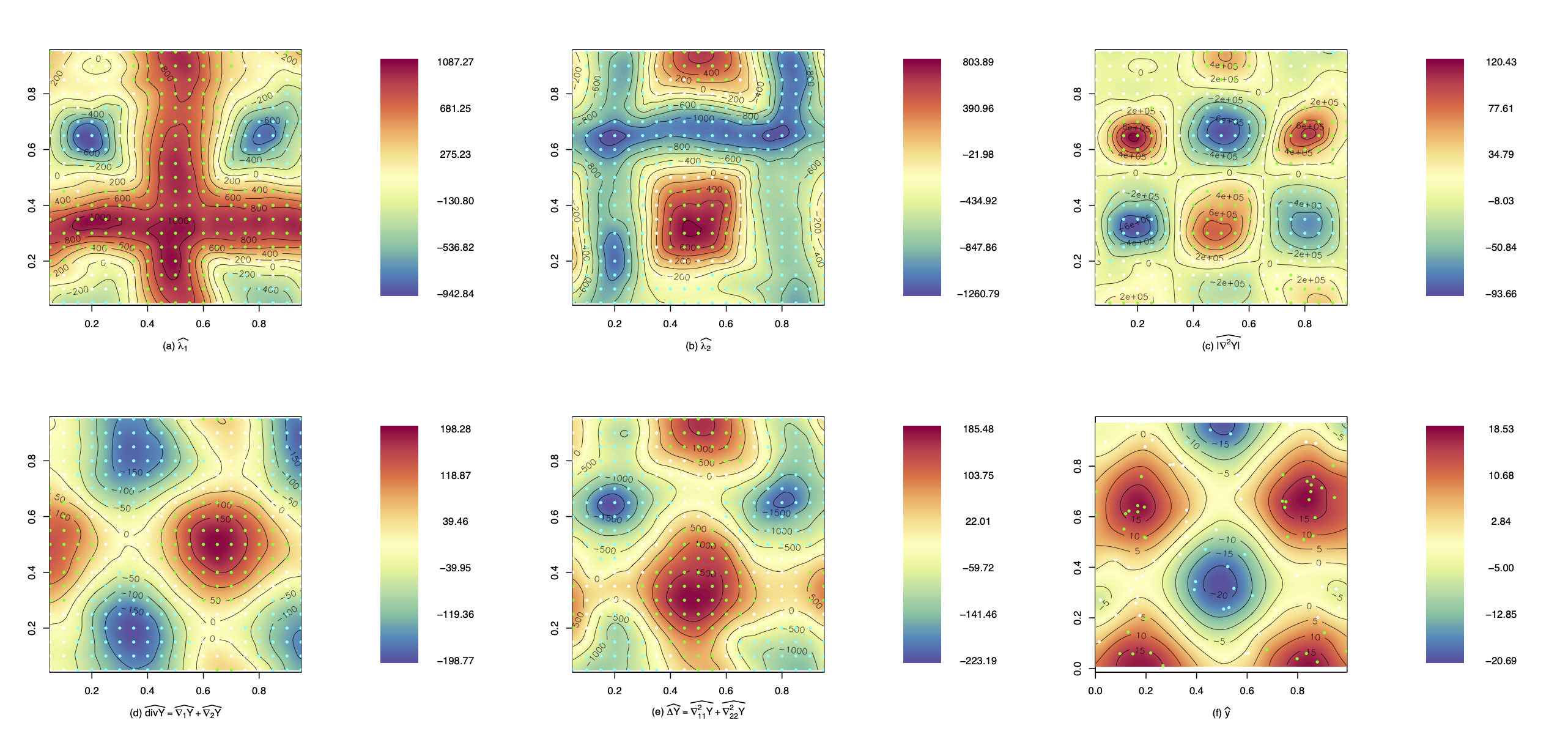}
	\caption{\small Plots showing the estimated surfaces for (a) eigen value, $\lambda_1$  (b) eigen value $\lambda_2$, (c) Gaussian curvature (scales in $\times 10^4$) (d) divergence (e) Laplacian (scales in $\times 10^1$) (f) fitted process over grid points. Each point is color coded; {\tt green} denoting the HPD intervals not containing 0, with positive end points, while {\tt cyan} denotes HPD intervals not containing 0, with negative end points.}\label{fig::diff-forms-1-s}
\end{figure}
\begin{figure}[H]
	\centering
	\includegraphics[width=1\linewidth , height=0.5\linewidth]{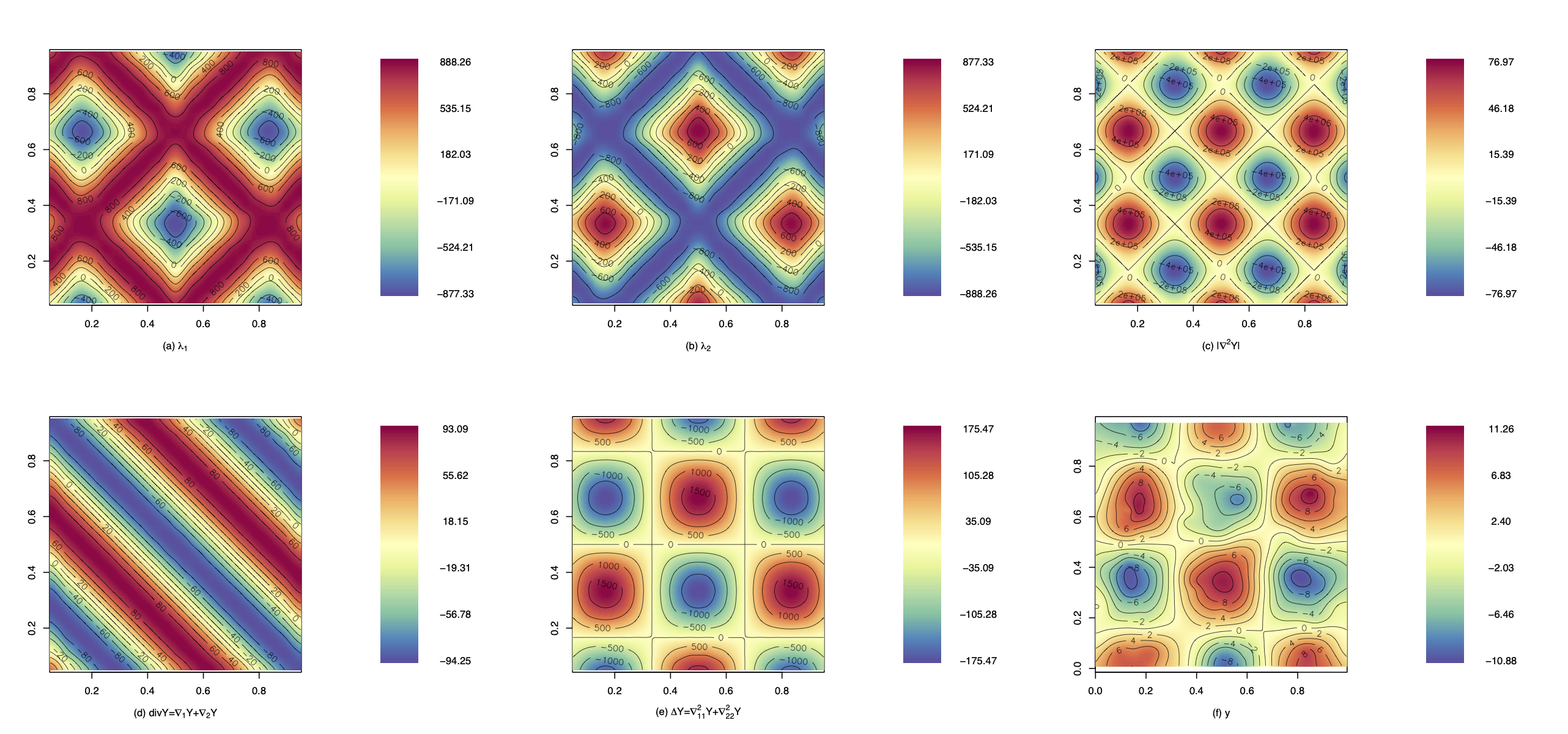}
	\caption{\small Plots showing the true surfaces for (a) eigen value, $\lambda_1$  (b) eigen value $\lambda_2$, (c) Gaussian curvature (scales in $\times 10^4$) (d) divergence (e) Laplacian (scales in $\times 10^1$) (f) fitted process over grid points. This is shown for synthetic response generated from, $y\sim N(10[\sin(3\pi s_1)\cdot\cos(3\pi s_2)],1)$}\label{fig::diff-forms-2-t}
\end{figure}
\vspace*{-.2in}
\begin{figure}[H]
	\centering
	\includegraphics[width=1\linewidth , height=0.5\linewidth]{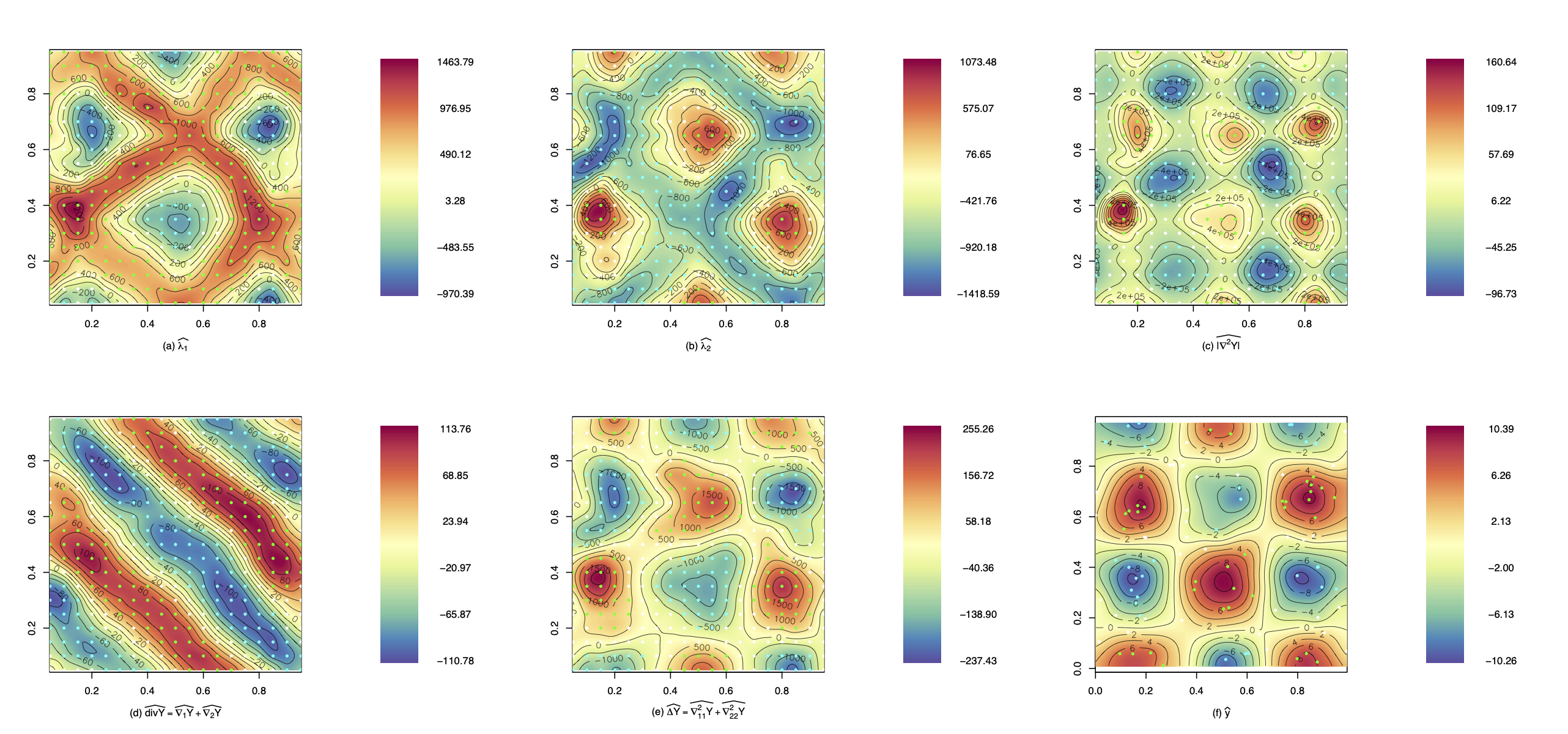}
	\caption{\small Plots showing the estimated surfaces for (a) eigen value, $\lambda_1$  (b) eigen value $\lambda_2$, (c) Gaussian curvature (scales in $\times 10^4$) (d) divergence (e) Laplacian (scales in $\times 10^1$) (f) fitted process over grid points. Each point is color coded; {\tt green} denoting the HPD intervals not containing 0, with positive end points, while {\tt cyan} denotes HPD intervals not containing 0, with negative end points.}\label{fig::diff-forms-2-s}
\end{figure}
\begin{figure}[H]
	\centering
	\includegraphics[width=1\linewidth , height=0.6\linewidth]{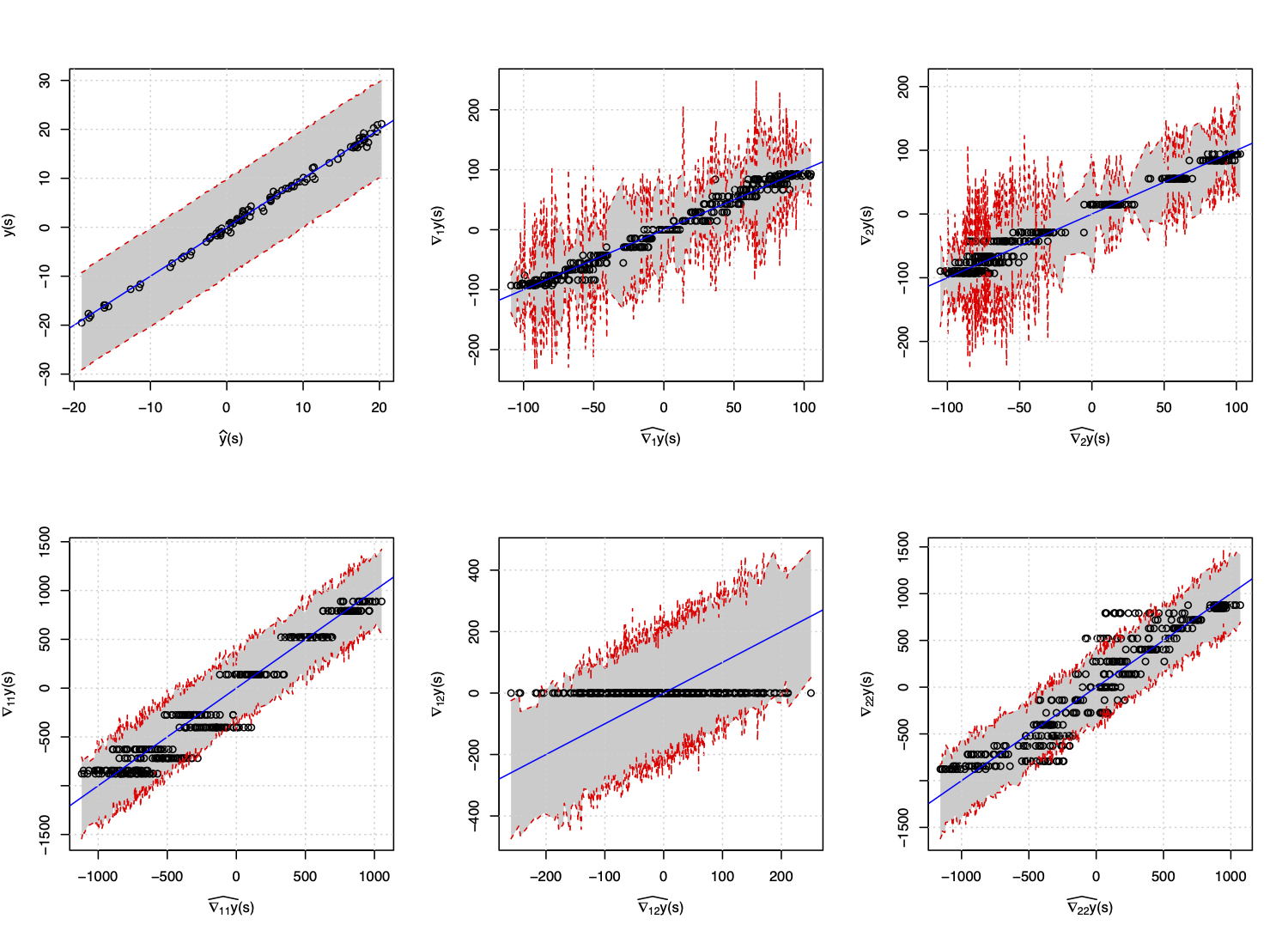}
	\caption{\small Plots showing observed versus fitted values for (a) the response variable $Y(s)$; (b) gradients with respect to $x$-axis; (c) gradients with respect to $y$-axis; (d) curvature with respect to $x$-axis; (e) mixed curvature over $x$-$y$; (f) curvature for $y\sim N(10[\sin(3\pi s_1)+\cos(3\pi s_2)],1)$ with respect to $y$-axis. The gray shades represent the 95\% HPD regions for each estimate.}\label{fig::ovf-1-t}
\end{figure}
\begin{figure}[H]
	\centering
	\includegraphics[width=1\linewidth , height=0.6\linewidth]{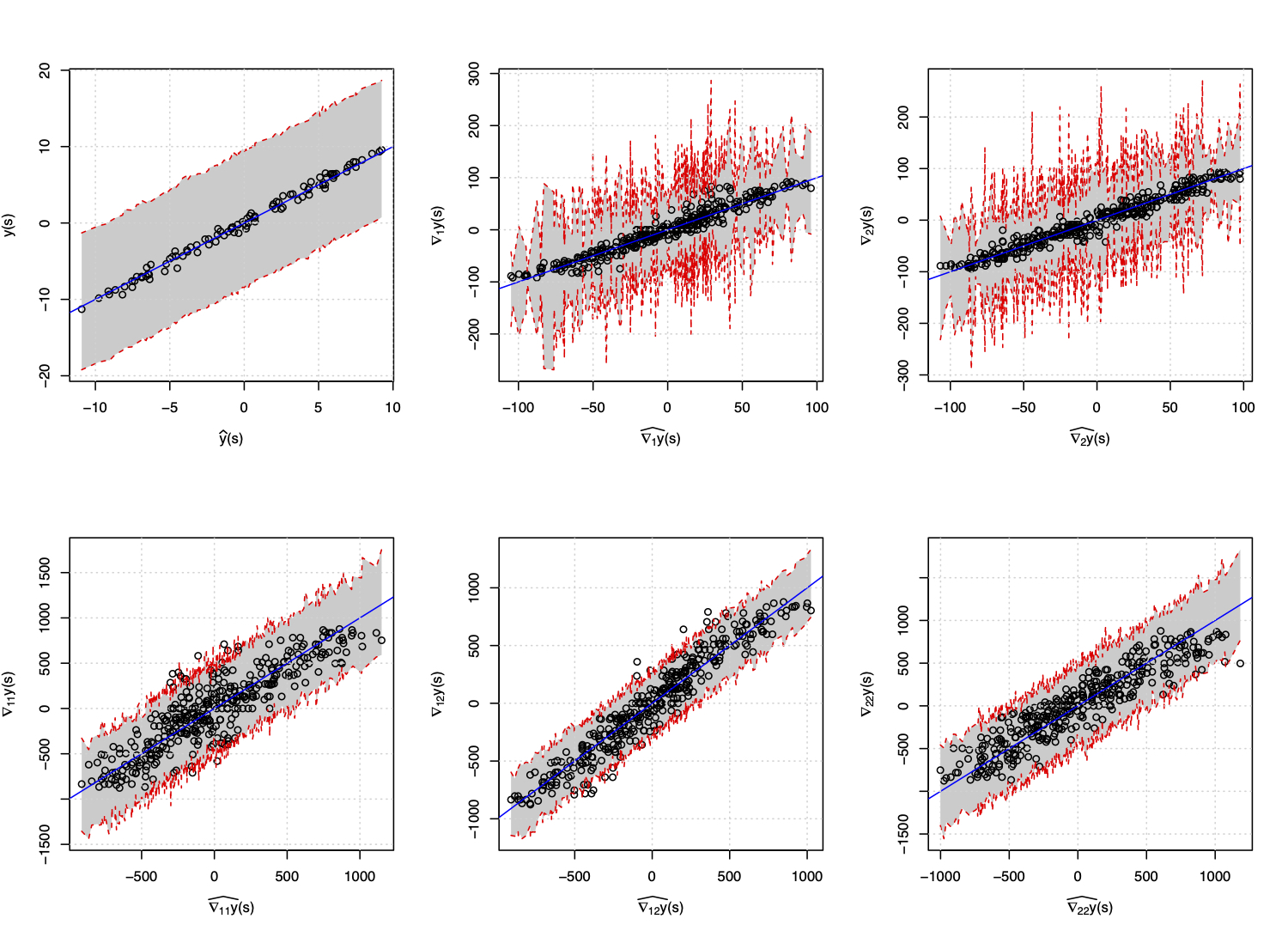}
	\caption{\small Plots showing observed versus fitted values for (a) the response variable $Y(s)$; (b) gradients with respect to x-axis; (c) gradients with respect to $y$-axis; (d) curvature with respect to $x$-axis; (e) mixed curvature over $x$-$y$; (f) curvature for $y\sim N(10[\sin(3\pi s_1)\cdot\cos(3\pi s_2)],1)$ with respect to $y$-axis. The gray shades represent the 95\% HPD regions for each estimate.}\label{fig::ovf-2-t}
\end{figure}


\begin{figure}[H]
	\centering
	\includegraphics[width=1\linewidth , height=0.33\linewidth]{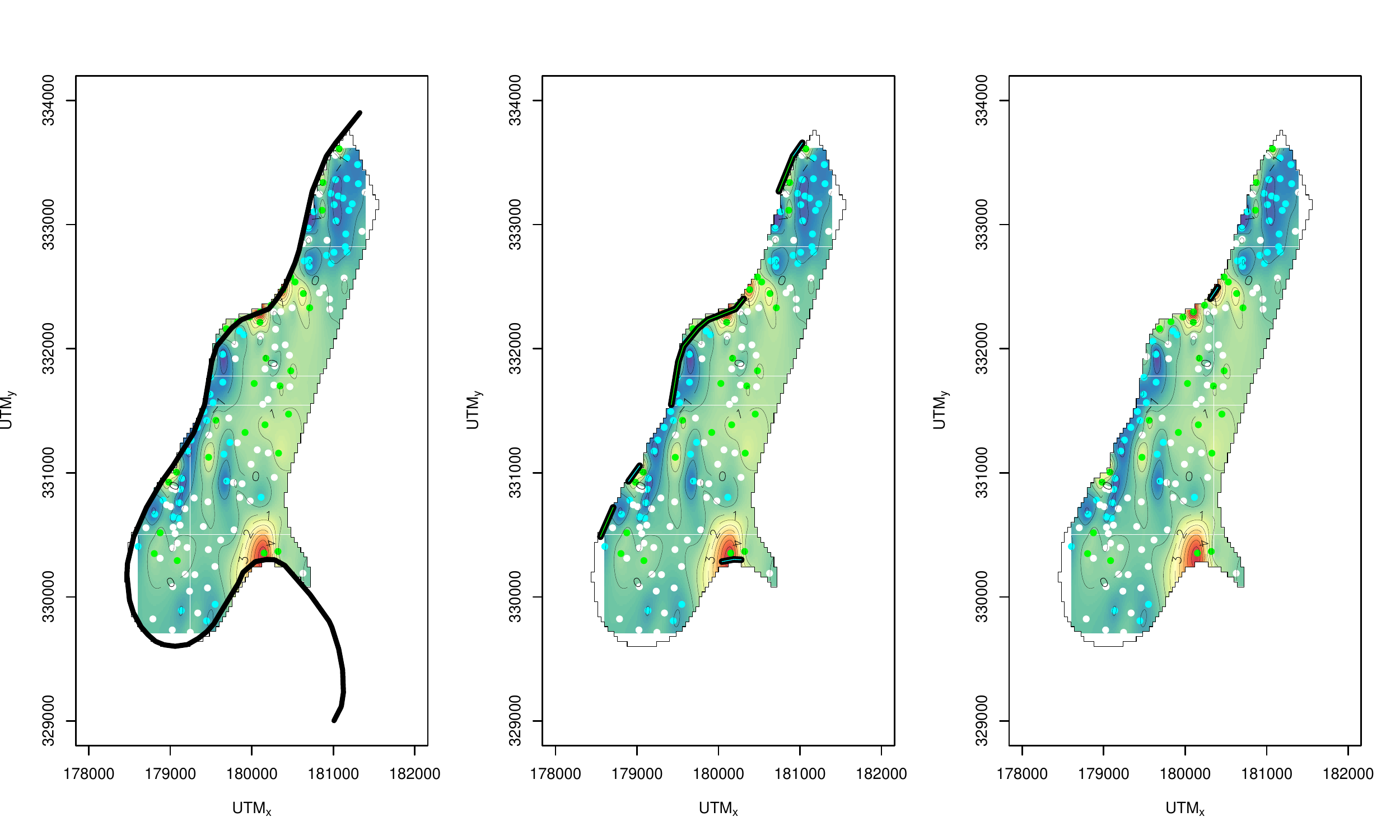}
	\includegraphics[width=1\linewidth , height=0.33\linewidth]{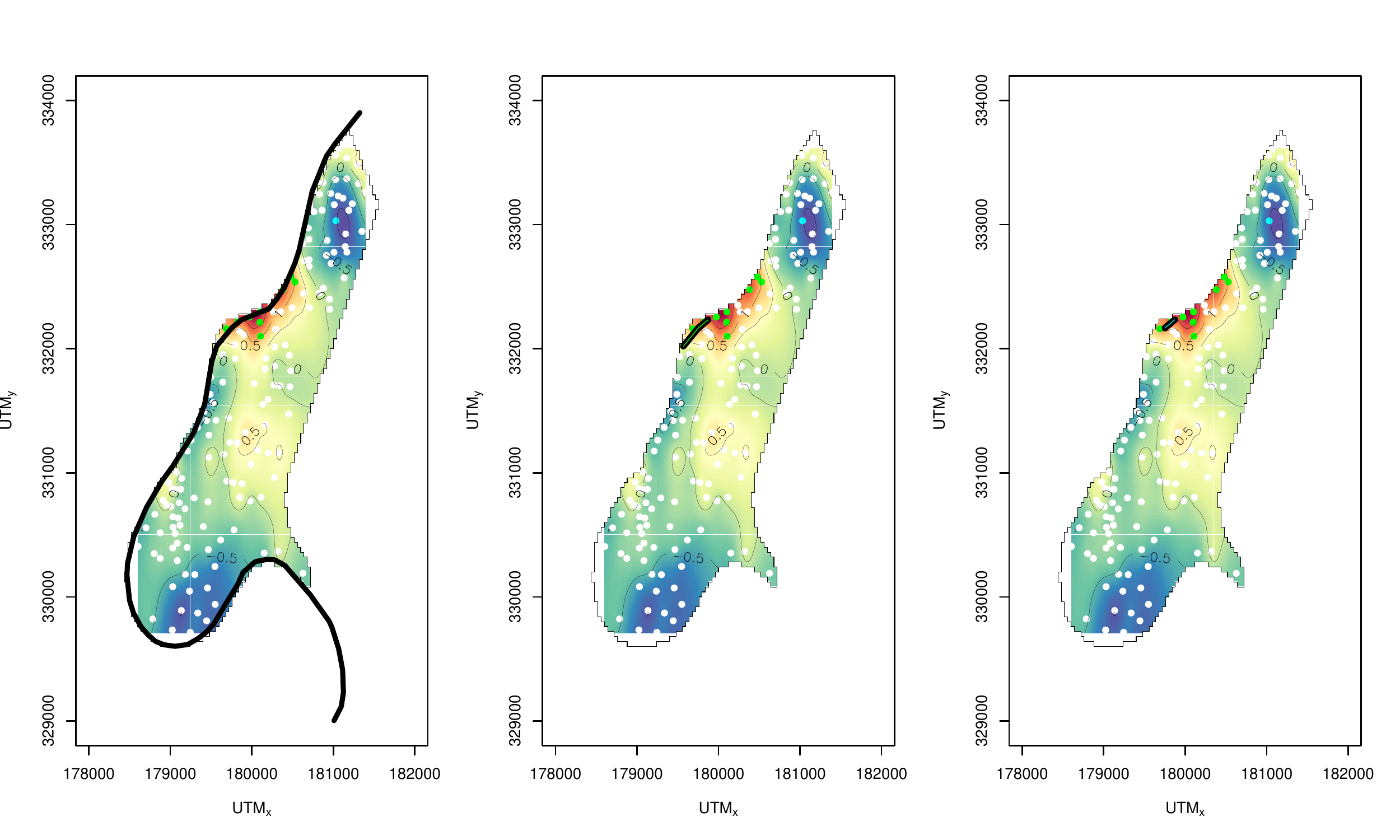}
	\includegraphics[width=1\linewidth , height=0.33\linewidth]{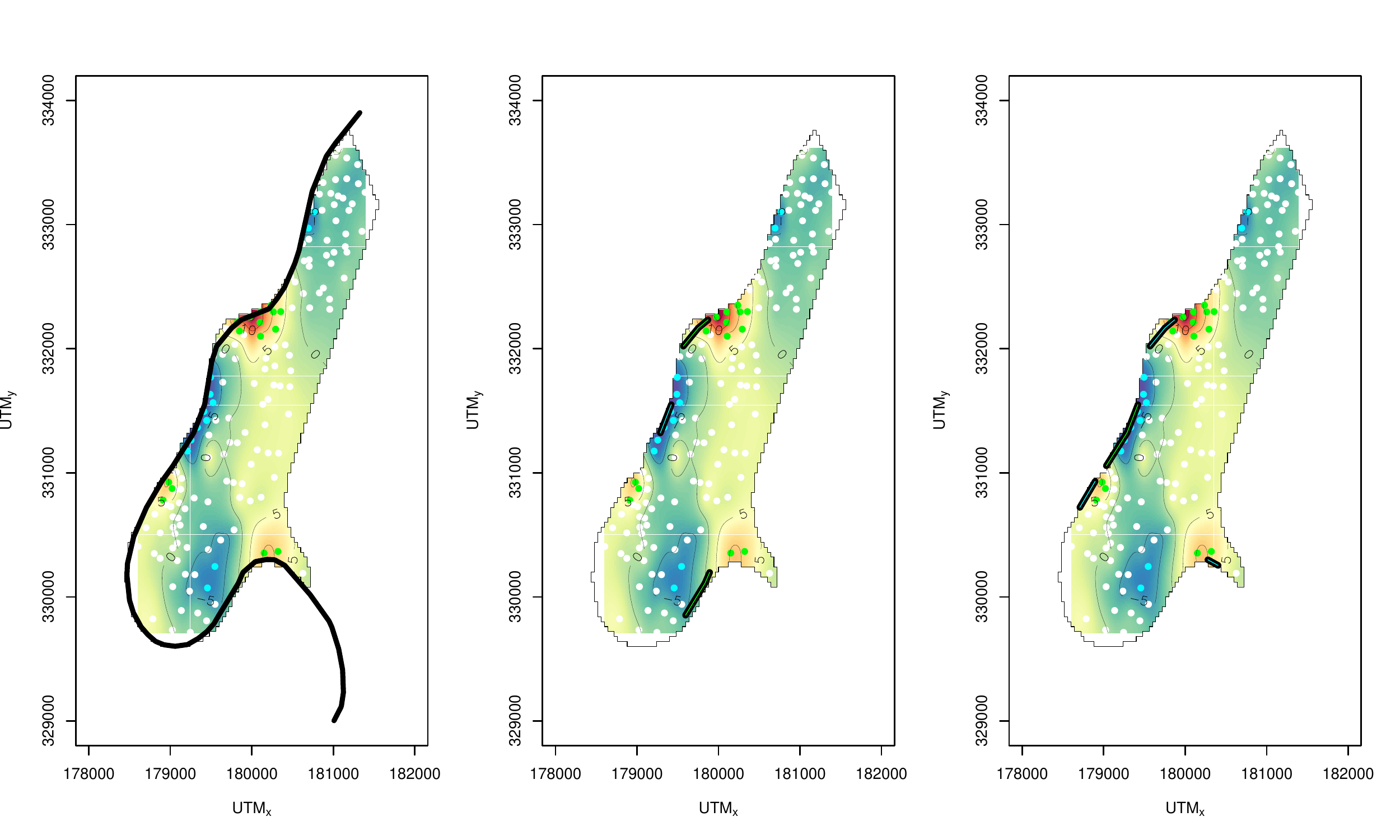}
	\caption{Plots showing results for curvature wombling on the Meuse river (first row) Copper (Cu) (second row) Lead (Pb) (third row) Zinc (Zn).}
	\label{fig::meuse-wombling}
\end{figure}

\section{Further Application}\label{sec::fapp}
\subsection{Temperatures in Northeastern US}
\begin{figure}[H]
	\centering
	\includegraphics[width=1\linewidth , height=0.4\linewidth]{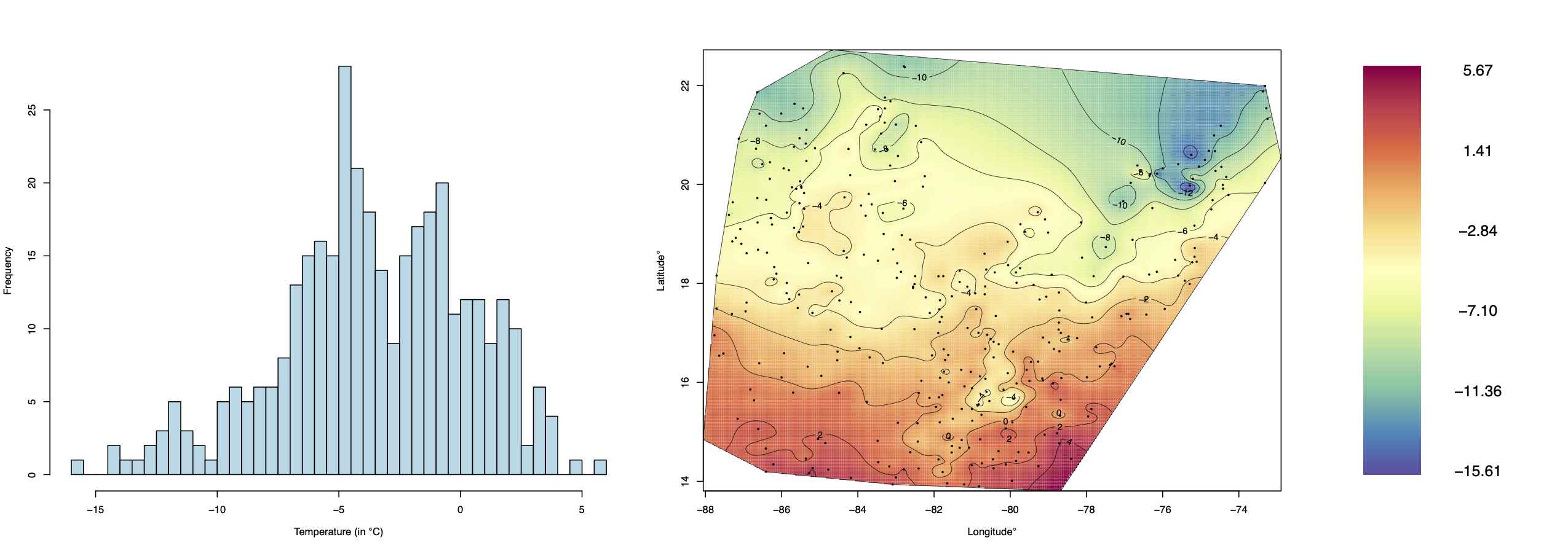}
	\caption{Plots showing (left) probability density of temperatures (in \textdegree C) (right) spatial plot of temperatures in Northeastern US during January 2000.}
	\label{fig::netemp-data}
\end{figure}

\begin{figure}[H]
	\centering
	\includegraphics[width=1\linewidth , height=0.4\linewidth]{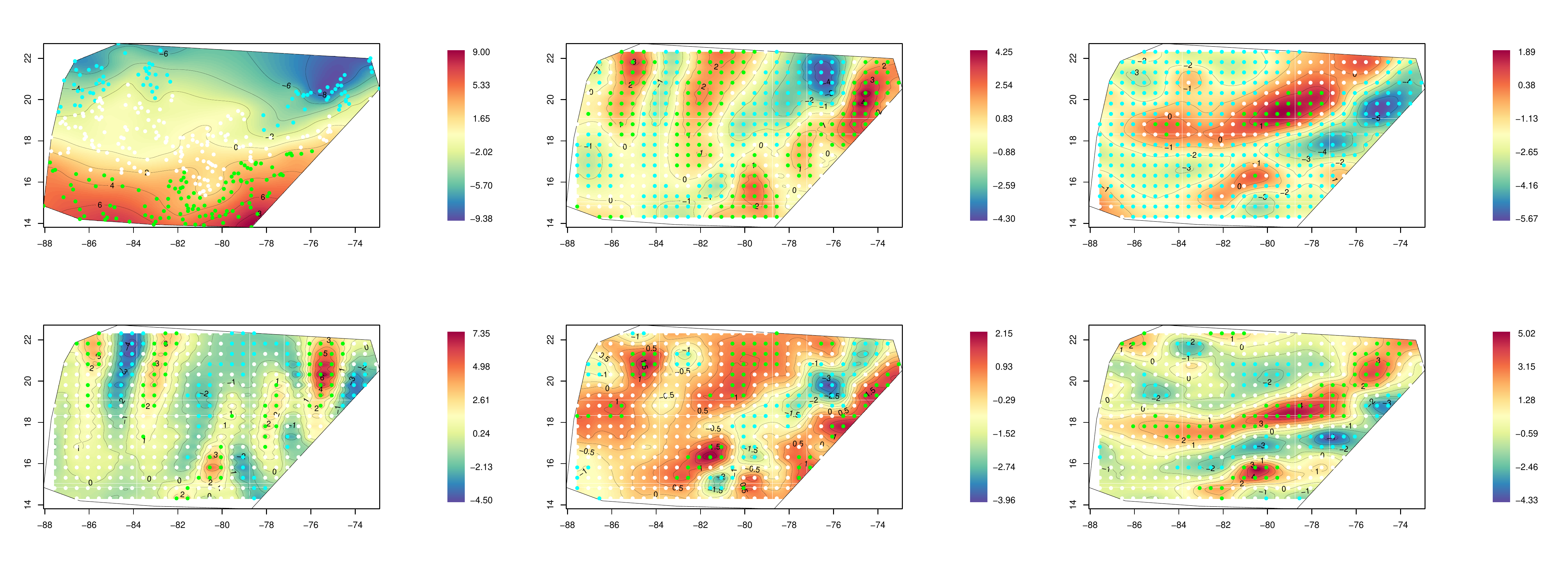}
	\caption{Plots showing (top row) (left) the fitted spatial process (center) the estimated gradient, $\nabla_x$ process along $x$-axis (right) the estimated gradient, $\nabla_y$ process along $y$-axis (bottom row) (left) estimated curvature $\nabla^2_{xx}$ along $x$-axis (center) estimated curvature, $\nabla^2_{xy}$ (right) estimated curvature, $\nabla^2_{yy}$ along $y$-axis.}
	\label{fig::netemp-grad}
\end{figure}

Temperatures are historically known to exhibit spatial variation. We focus on a data set that records monthly temperatures across weather monitoring stations in the Northeastern United States during January, 2000 from the {\tt R}-package {\tt spBayes} \citep{finley2007spbayes}. Temperature gradients and curvature are of interest from an environmental science perspective to track and perform boundary analysis on zones that exhibit significant changes in the surface during a month. Curvature wombling performed on temperature reveals climate zones featuring rapid atmospheric changes. Quantifying such variations in atmospheric conditions is central to statistical modeling in environmental applications. The data consists of temperatures (in degree Celsius) from 356 weather monitoring stations. The probability distribution for temperatures and interpolated spatial plot is shown in Figure~\ref{fig::netemp-data}. We model the data using the hierarchical model outlined in (13) 
in the manuscript. We used the following hyper parameters for the model, $\phi\sim {\rm Unif}\left(\mfrac{3}{max_{\s\in {\cal S}}{||\Delta||}},300\right)$, $\sigma^2\sim IG(2,1)$ (mean 1 variance infinite), $\tau^2\sim IG(2,1)$ (mean 1 variance infinite), $\beta_0\sim N(0,10^6)$, and $\nu=5/2$ for the Mat\'ern kernel. We consider $10^4$ iterations for MCMC chains, with burn-in diagnosed at $5\times10^3$. The posterior estimates from the model fit are shown in Table~\ref{tab::netemp-spatial}.
We fit the model with only an intercept, that allows the spatial process $Z(\s)$ to capture most of the variation in the data. We observe from Table~\ref{tab::netemp-spatial} that $\sigma^2/(\sigma^2+\tau^2)\approx95.86\%$. Significance in process parameter estimates is assessed by checking containment of 0 within the HPD intervals. The fitted spatial process, along with significance is shown in Figure~\ref{fig::netemp-grad} (top row, left). For temperature observations falling in the middle no significant spatial effect is observed, with stations located in the northern and southern regions showing positive and negative spatial effects. This indicates a clear variation in the north-south direction for temperatures in January. The variance of the nugget process $\tau^2$ is small compared to $\sigma^2$. The average estimated temperature is $-3.78~^{\circ}$C $(-5.92,-1.69)$.
\begin{table}[hb]
	\centering
	\caption{Posterior Estimates for Temperatures in Northeastern US. The highest posterior density intervals are shown alongside for respective estimates.}\label{tab::netemp-spatial}
	\resizebox*{\linewidth}{!}{
		\begin{tabular}{l|@{\extracolsep{20pt}}cc@{}}
			\hline
			\hline
			\multirow{2}{*}{Parameters ($\btheta$)}& \multirow{2}{*}{Posterior Estimates ($\widehat{\btheta}$)} & \multirow{2}{*}{Highest Posterior Density intervals (HPD)}\\ 
			&&\\
			\hline
			$\phi$ & 0.38  & (0.27, 0.51) \\ 
			$\sigma^2$ & 21.44 & (10.11, 41.03)\\ 
			$\tau^2$ & 0.92 & (0.76, 1.10) \\ 
			$\beta_0$ & -3.78 & (-5.92, -1.69) \\ 
			\hline
			\hline
		\end{tabular}
	}
\end{table}
Using the posterior estimates for $\btheta$ and the spatial process $\Z$ we perform a gradient and curvature assessment over the estimated posterior surface. We use Algorithm \ref{algo::gradient} outlined in section \ref{sec::algos} for sampling gradients and curvature. This is done over a grid encompassing the spatial domain of reference, ${\cal S}$. A grid ${\cal G}=\{\s_g:\s_g\in {\tt convex-hull}({\cal S})\}$ consisting of 461 points is laid out over the convex-hull of ${\cal S}$. The estimated gradients along $x$ and $y$-axis are shown in the first row for Figure~\ref{fig::netemp-grad}. Elements of the directional curvature process along $x$ and $y$-axis are shown in the second row. Significance is again indicated by checking containment of 0 within HPD intervals for posterior gradients and curvature. Significant directional curvature is indicative of significant rate of change in temperature gradients along the chosen axis. $\nabla^2_{xy}$ quantifies {\em interaction} between rates of change in gradients along $x$ and $y$ axis respectively. A significant $\nabla^2_{xy}(\s_g)$ at the grid location $\s_g$ indicates that the rate of change with respect to axis $x$ was significantly influenced by change along $y$-axis. Regions of local maxima (elliptic points) generally manifest themselves with alternating zones of increasing and decreasing gradients (or curvature) separated by a zone of saddle points (around the location of said maxima). This can be witnessed in the north-eastern region and south-eastern regions where these maxima (minima) occur. 

\begin{figure}[H]
	\centering
	\includegraphics[width=1\linewidth , height=0.2\linewidth]{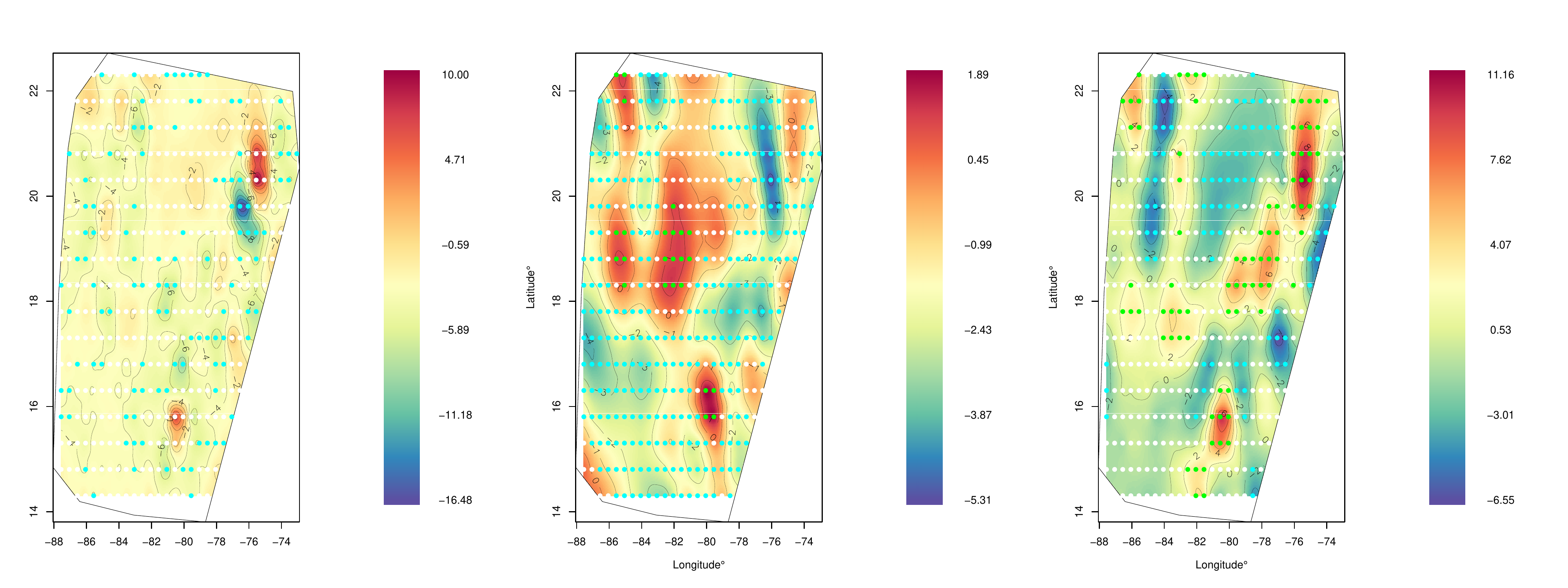}
	\caption{Plots showing surfaces for (left) Gaussian curvature (center) divergence and (right) Laplacian of temperatures in Northeastern US during January 2000.}
	\label{fig::netemp-surfa}
\end{figure}

\begin{figure}[!]
	\centering
	\includegraphics[width=1\linewidth , height=0.3\linewidth]{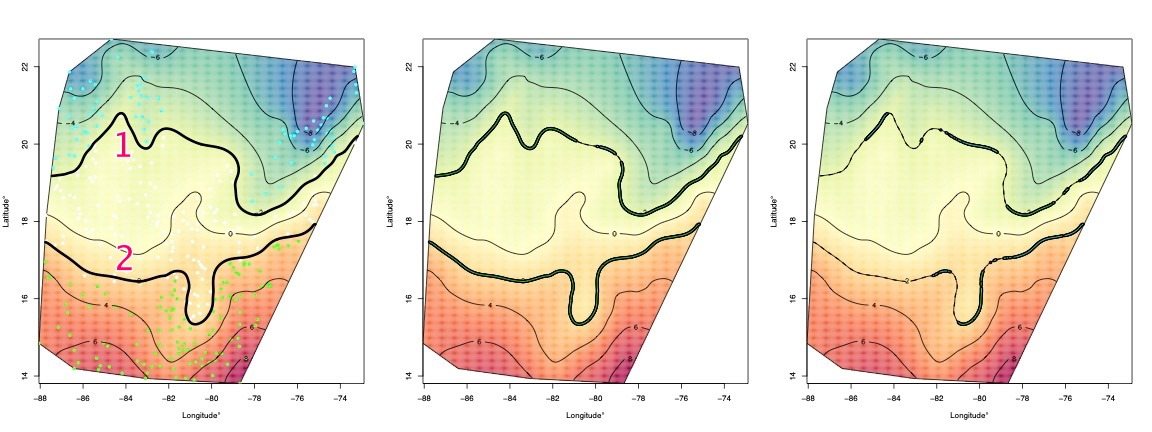}
	\includegraphics[width=1\linewidth , height=0.3\linewidth]{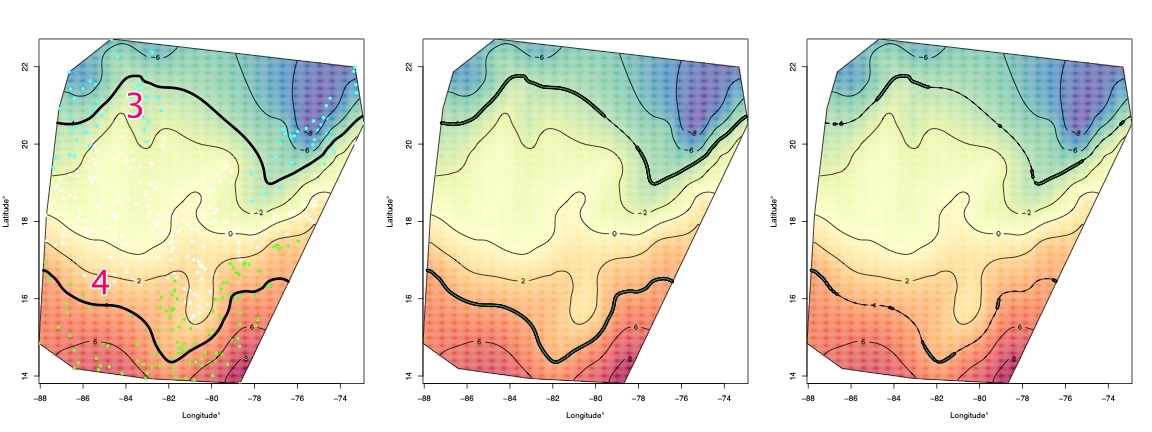}
	\includegraphics[width=1\linewidth , height=0.3\linewidth]{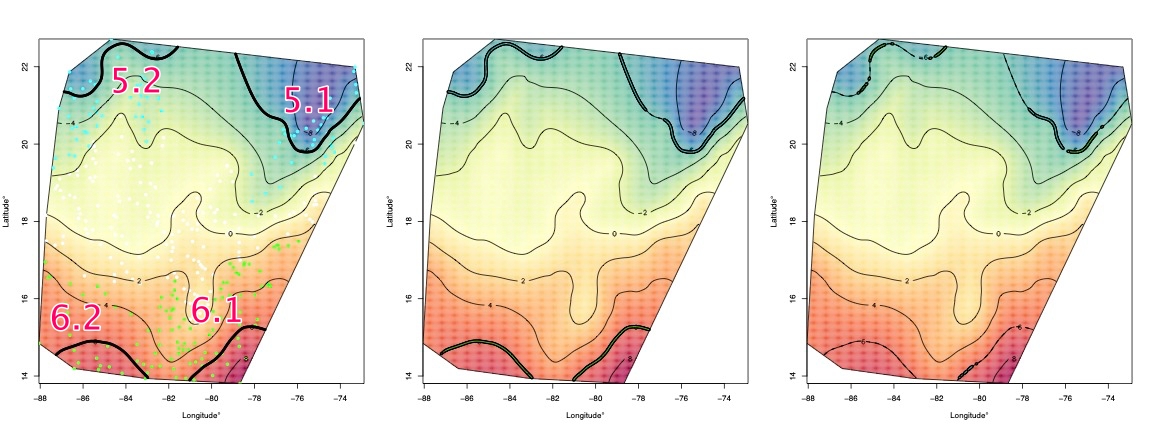}
	\caption{Plots showing curvature wombling on temperatures data. The curves are marked in each row on the figure to the left are to be referenced with Table~\ref{tab::cwomb-netemp} showing average wombling measures for gradient and curvature for respective curves.}
	\label{fig::netemp-womb}
\end{figure}

Posterior surface is effected on the same grid ${\cal G}$ using the posterior estimates for $\nabla^2_{xx}$, $\nabla^2_{xy}$ and $\nabla^2_{yy}$. We leverage these to produce estimated surfaces for the 
Gaussian curvature (determinant) 
(shown in left of Figure~\ref{fig::netemp-surfa}), the divergence operator (shown in center of Figure~\ref{fig::netemp-surfa}) and Laplacian (shown in right of Figure~\ref{fig::netemp-surfa}). Posterior surfaces for the 
Gaussian curvature 
are indicative of locations/presence of maximas and saddle points, divergence surfaces show regions of rapid change in temperatures while the Laplacian shows regions of maximum change in gradients. Gaussian calibration for these estimate attaches significance allowing us to distinguish between contiguous zones housing significant change.
We perform curvature wombling using inference obtained from the posterior analysis of the surface. Figure~\ref{fig::netemp-womb} and Table~\ref{tab::cwomb-netemp} show the results for curvilinear wombling on the resulting posterior estimates of the surface. The curves chosen are shown in plots on the left for Figure~\ref{fig::netemp-womb}. We begin with curves (level sets) ``1" and ``2" that delineate zones of significant (positive in the south and negative in the north) spatial effects, iteratively proceeding to higher (lower) level sets while inspecting them for curvilinear gradients and curvature. Referring to Table~\ref{tab::cwomb-netemp}, we observe that all curves located with respect to curvilinear gradients, observed from significant average gradients. With respect to directional curvature, we refer to significant segments located in the Figure~\ref{fig::netemp-womb} which show enormous heterogeneity, i.e. changes in directional concavity when traversing the curve, with separated contiguous segments indicating significant changes in concavity. For instance traversing curve ``2" in the west-east direction, we observe this clearly. This naturally renders the average directional curvature (shown in Table~\ref{tab::cwomb-netemp}) insignificant when considered along the entirety of curve ``2", which is also the case for other level sets. To be able to detect significance we could only summarize across significant segments (as was done for the Meuse river data).

\begin{table}[!]
	\centering
	\caption{Curvilinear Wombling measures for boundaries in Northeastern US Temperatures, each measure is accompanied by its corresponding HPD interval in brackets below.}\label{tab::cwomb-netemp}
    \resizebox{\linewidth}{!}{
        \begin{tabular}{l|@{\extracolsep{100pt}}cc@{}}
		\hline
		\hline
		\multirow{2}{*}{Curve ($C$)} & \multirow{2}{*}{Average Gradient ($\overline{\Gamma}^{(1)}(C)$)} & \multirow{2}{*}{Average Curvature ($\overline{\Gamma}^{(2)}(C)$)}  \\ 
		&&\\
		\hline
		\multirow{2}{*}{Boundary 1} & 2.44 & -0.38 \\ 
		& (1.90, 2.97) & (-1.48, 0.74)\\
		\multirow{2}{*}{Boundary 2} & 2.70 & -0.40 \\ 
		& (2.24, 3.18) & (-1.67, 0.79)\\\hline
		\multirow{2}{*}{Boundary 3} & 2.69 &  -0.34 \\ 
		& (2.10, 3.28) & (-1.71, 0.84)\\
		\multirow{2}{*}{Boundary 4} & 2.23  & -0.31  \\ 
		& (1.74, 2.72) & (-1.67, 1.11)\\\hline
		\multirow{2}{*}{Boundary 5.1} & -2.96 & 1.13 \\ 
		& (-4.22, -1.73)& (-0.88, 3.16)\\
		\multirow{2}{*}{Boundary 5.2} & -3.24 & -0.23\\ 
		& (-4.22, -2.25) &  (-2.19, 1.69)\\
		\multirow{2}{*}{Boundary 6.1} & 1.63 & -0.23 \\ 
		& (0.70, 2.58)& (-2.96, 2.62)\\
		\multirow{2}{*}{Boundary 6.2} & 2.90  & -0.38  \\ 
		& (1.93, 3.92) & (-3.00, 2.31)\\
		\hline\hline
	\end{tabular}
    }
\end{table}

\bibliographystyle{agsm}
\bibliography{curvaturep}

@book{webster2007geostatistics,
	title={Geostatistics {F}or {E}nvironmental {S}cientists},
	author={Webster, Richard and Oliver, Margaret A},
	year={2007},
	publisher={John Wiley \& Sons}
}

@book{burrough2015principles,
	title={Principles {O}f {G}eographical {I}nformation {S}ystems},
	author={Burrough, Peter A and McDonnell, Rachael A and Lloyd, Christopher D},
	year={2015},
	publisher={Oxford University Press}
}

@book{schabenberger2017statistical,
	title={Statistical {M}ethods {F}or {S}patial {D}ata {A}nalysis},
	author={Schabenberger, Oliver and Gotway, Carol A},
	year={2017},
	publisher={CRC press}
}

@book{plant2018spatial,
	title={Spatial {D}ata {A}nalysis {I}n {E}cology {A}nd {A}griculture {U}sing {R}},
	author={Plant, Richard E},
	year={2018},
	publisher={CRC Press}
}

@book{law2000simulation,
	title={Simulation {M}odeling {A}nd {A}nalysis},
	author={Law, Averill M and Kelton, W David and Kelton, W David},
	volume={3},
	year={2000},
	publisher={McGraw-Hill New York}
}

@book{santner2003design,
	title={The {D}esign {A}nd {A}nalysis {O}f {C}omputer {E}xperiments},
	author={Santner, Thomas J and Williams, Brian J and Notz, William I and Williams, Brain J},
	volume={1},
	year={2003},
	publisher={Springer}
}

@book{jones2014geographical,
	title={Geographical {I}nformation {S}ystems {A}nd {C}omputer {C}artography},
	author={Jones, Chris B},
	year={2014},
	publisher={Routledge}
}

@book{vaughan2018mapping,
	title={Mapping {S}ociety: {T}he {S}patial {D}imensions {O}f {S}ocial {C}artography},
	author={Vaughan, Laura},
	year={2018},
	publisher={UCL Press}
}

@book{winkler2003image,
	title={Image {A}nalysis, {R}andom {F}ields {A}nd {M}arkov {C}hain {M}onte {C}arlo {M}ethods: {A} {M}athematical {I}ntroduction},
	author={Winkler, Gerhard},
	volume={27},
	year={2003},
	publisher={Springer Science \& Business Media}
}

@book{chiu2013stochastic,
	title={Stochastic {G}eometry {A}nd {I}ts {A}pplications},
	author={Chiu, Sung Nok and Stoyan, Dietrich and Kendall, Wilfrid S and Mecke, Joseph},
	year={2013},
	publisher={John Wiley \& Sons}
}

@book{dryden2016statistical,
	title={Statistical {S}hape {A}nalysis: {W}ith {A}pplications {I}n {R}},
	author={Dryden, Ian L and Mardia, Kanti V},
	volume={995},
	year={2016},
	publisher={John Wiley \& Sons}
}

@book{lesage2009introduction,
	title={Introduction {T}o {S}patial {E}conometrics},
	author={LeSage, James and Pace, Robert Kelley},
	year={2009},
	publisher={Chapman and Hall/CRC}
}

@book{elliot2000spatial,
	title={Spatial {E}pidemiology: {M}ethods {A}nd {A}pplications},
	author={Elliot, Paul and Wakefield, Jon C and Best, Nicola G and Briggs, David John and others},
	year={2000},
	publisher={Oxford University Press}
}

@book{waller2004applied,
	title={Applied {S}patial {S}tatistics {F}or {P}ublic {H}ealth {D}ata},
	author={Waller, Lance A and Gotway, Carol A},
	volume={368},
	year={2004},
	publisher={John Wiley \& Sons}
}

@book{lawson2013statistical,
	title={Statistical {M}ethods {I}n {S}patial {E}pidemiology},
	author={Lawson, Andrew B},
	year={2013},
	publisher={John Wiley \& Sons}
}

@book{haining1993spatial,
	title={Spatial {D}ata {A}nalysis {I}n {T}he {S}ocial {A}nd {E}nvironmental {S}ciences},
	author={Haining, Robert},
	year={1993},
	publisher={Cambridge University Press}
}

@book{wise2007gis,
	title={GIS {A}nd {E}vidence-{B}ased {P}olicy {M}aking},
	author={Wise, Stephen and Craglia, Max},
	year={2007},
	publisher={CRC Press}
}

@book{adler1981geometry,
	title={The {G}eometry {O}f {R}andom {F}ields},
	author={Adler, Robert J},
	year={1981},
	publisher={SIAM}
}

@article{kent1989continuity,
	title={Continuity {P}roperties {F}or {R}andom {F}ields},
	author={Kent, John T},
	journal={The Annals of Probability},
	pages={1432--1440},
	year={1989},
	publisher={JSTOR}
}

@article{banerjee2003smoothness,
	title={On {S}moothness {P}roperties {O}f {S}patial {P}rocesses},
	author={Banerjee, S and Gelfand, AE},
	journal={Journal of Multivariate Analysis},
	volume={84},
	number={1},
	pages={85--100},
	year={2003},
	publisher={Elsevier}
}

@article{banerjee2003directional,
	title={Directional {R}ates {O}f {C}hange {U}nder {S}patial {P}rocess {M}odels},
	author={Banerjee, Sudipto and Gelfand, Alan E and Sirmans, CF},
	journal={Journal of the American Statistical Association},
	volume={98},
	number={464},
	pages={946--954},
	year={2003},
	publisher={Taylor \& Francis}
}

@article{mardia1996kriging,
  title={Kriging {A}nd {S}plines {W}ith {D}erivative {I}nformation},
  author={Mardia, KV and Kent, JT and Goodall, CR and Little, JA},
  journal={Biometrika},
  volume={83},
  number={1},
  pages={207--221},
  year={1996},
  publisher={Oxford University Press}
}

@article{majumdar2006gradients,
	title={Gradients {I}n {S}patial {R}esponse {S}urfaces {W}ith {A}pplication {T}o {U}rban {L}and {V}alues},
	author={Majumdar, Anandamayee and Munneke, Henry J and Gelfand, Alan E and Banerjee, Sudipto and Sirmans, CF},
	journal={Journal of Business \& Economic Statistics},
	volume={24},
	number={1},
	pages={77--90},
	year={2006},
	publisher={Taylor \& Francis}
}

@article{terres2015using,
	title={Using {S}patial {G}radient {A}nalysis {T}o {C}larify {S}pecies {D}istributions {W}ith {A}pplication {T}o {S}outh {A}frican {P}rotea},
	author={Terres, Maria A and Gelfand, Alan E},
	journal={Journal of Geographical Systems},
	volume={17},
	number={3},
	pages={227--247},
	year={2015},
	publisher={Springer}
}

@article{wang2016estimating,
	title={Estimating {S}hape {C}onstrained {F}unctions {U}sing {G}aussian {P}rocesses},
	author={Wang, Xiaojing and Berger, James O},
	journal={SIAM/ASA Journal on Uncertainty Quantification},
	volume={4},
	number={1},
	pages={1--25},
	year={2016},
	publisher={SIAM}
}

@article{terres2016spatial,
	title={Spatial {P}rocess {G}radients {A}nd {T}heir {U}se {I}n {S}ensitivity {A}nalysis {F}or {E}nvironmental {P}rocesses},
	author={Terres, Maria A and Gelfand, Alan E},
	journal={Journal of Statistical Planning and Inference},
	volume={168},
	pages={106--119},
	year={2016},
	publisher={Elsevier}
}

@article{wang2018process,
	title={Process {M}odeling {F}or {S}lope {A}nd {A}spect {W}ith {A}pplication {T}o {E}levation {D}ata {M}aps},
	author={Wang, Fangpo and Bhattacharya, Anirban and Gelfand, Alan E},
	journal={Test},
	volume={27},
	number={4},
	pages={749--772},
	year={2018},
	publisher={Springer}
}

@article{banerjee2006bayesian,
	title={Bayesian {W}ombling: {C}urvilinear {G}radient {A}ssessment {U}nder {S}patial {P}rocess {M}odels},
	author={Banerjee, Sudipto and Gelfand, Alan E},
	journal={Journal of the American Statistical Association},
	volume={101},
	number={476},
	pages={1487--1501},
	year={2006},
	publisher={Taylor \& Francis}
}

@article{womble1951differential,
  title={Differential {S}ystematics},
  author={Womble, William H},
  journal={Science},
  volume={114},
  number={2961},
  pages={315--322},
  year={1951},
  publisher={JSTOR}
}

@article{lu2005bayesian,
	title={Bayesian {A}real {W}ombling {F}or {G}eographical {B}oundary {A}nalysis},
	author={Lu, Haolan and Carlin, Bradley P},
	journal={Geographical Analysis},
	volume={37},
	number={3},
	pages={265--285},
	year={2005},
	publisher={Wiley Online Library}
}

@article{fitzpatrick2010ecological,
	title={Ecological {B}oundary {D}etection {U}sing {B}ayesian {A}real {W}ombling},
	author={Fitzpatrick, Matthew C and Preisser, Evan L and Porter, Adam and Elkinton, Joseph and Waller, Lance A and Carlin, Bradley P and Ellison, Aaron M},
	journal={Ecology},
	volume={91},
	number={12},
	pages={3448--3455},
	year={2010},
	publisher={Wiley Online Library}
}

@article{liang2009bayesian,
	title={Bayesian {W}ombling {F}or {S}patial {P}oint {P}rocesses},
	author={Liang, Shengde and Banerjee, Sudipto and Carlin, Bradley P},
	journal={Biometrics},
	volume={65},
	number={4},
	pages={1243--1253},
	year={2009},
	publisher={Wiley Online Library}
}

@article{heaton2014wombling,
	title={Wombling {A}nalysis {O}f {C}hildhood {T}umor {R}ates {I}n {F}lorida},
	author={Heaton, Matthew J},
	journal={Statistics and Public Policy},
	volume={1},
	number={1},
	pages={60--67},
	year={2014},
	publisher={Taylor \& Francis}
}

@article{li2015bayesian,
	title={Bayesian {M}odels {F}or {D}etecting {D}ifference {B}oundaries {I}n {A}real {D}ata},
	author={Li, Pei and Banerjee, Sudipto and Hanson, Timothy A and McBean, Alexander M},
	journal={Statistica Sinica},
	volume={25},
	number={1},
	pages={385},
	year={2015},
	publisher={NIH Public Access}
}

@article{qu2021boundary,
	title={Boundary {D}etection {U}sing {A} {B}ayesian {H}ierarchical {M}odel {F}or {M}ultiscale {S}patial {D}ata},
	author={Qu, Kai and Bradley, Jonathan R and Niu, Xufeng},
	journal={Technometrics},
	volume={63},
	number={1},
	pages={64--76},
	year={2021},
	publisher={Taylor \& Francis}
}

@book{williams2006gaussian,
	title={Gaussian {P}rocesses {F}or {M}achine {L}earning},
	author={Williams, Christopher KI and Rasmussen, Carl Edward},
	volume={2},
	number={3},
	year={2006},
	publisher={MIT press Cambridge, MA}
}

@book{do2016differential,
	title={Differential {G}eometry {O}f {C}urves {A}nd {S}urfaces: {R}evised {A}nd {U}pdated, {S}econd {E}dition},
	author={Do Carmo, Manfredo P},
	year={2016},
	publisher={Courier Dover Publications}
}

@book{kreyszig2019introduction,
	title={Introduction {T}o {D}ifferential {G}eometry {A}nd {R}iemannian {G}eometry},
	author={Kreyszig, Erwin},
	year={2019},
	publisher={University of Toronto Press}
}

@book{matern2013spatial,
	title={Spatial {V}ariation},
	author={Mat{\'e}rn, Bertil},
	volume={36},
	year={2013},
	publisher={Springer Science \& Business Media}
}

@article{magnus1980elimination,
	title={The {E}limination {M}atrix: {S}ome {L}emmas {A}nd {A}pplications},
	author={Magnus, Jan R and Neudecker, H},
	journal={SIAM Journal on Algebraic Discrete Methods},
	volume={1},
	number={4},
	pages={422--449},
	year={1980},
	publisher={SIAM}
}

@misc{abramowitz1988handbook,
	title={Handbook {O}f {M}athematical {F}unctions {W}ith {F}ormulas, {G}raphs, and {M}athematical {T}ables},
	author={Abramowitz, Milton and Stegun, Irene A and Romer, Robert H},
	year={1988},
	publisher={American Association of Physics Teachers}
}

@book{rudin1976principles,
	title={Principles {O}f {M}athematical {A}nalysis},
	author={Rudin, Walter},
	volume={3},
	year={1976},
	publisher={McGraw-hill New York}
}

@book{banerjee2014hierarchical,
  title={Hierarchical {M}odeling {A}nd {A}nalysis {F}or {S}patial {D}ata},
  author={Banerjee, Sudipto and Carlin, Bradley P and Gelfand, Alan E},
  year={2014},
  publisher={CRC press}
}

@book{gallier2000curves,
	title={Curves {A}nd {S}urfaces {I}n {G}eometric {M}odeling: {T}heory {A}nd {A}lgorithms},
	author={Gallier, Jean and Gallier, Jean H},
	year={2000},
	publisher={Morgan Kaufmann}
}

@article{harrison1978hedonic,
	title={Hedonic {H}ousing {P}rices {A}nd {T}he {D}emand {F}or {C}lean {A}ir},
	author={Harrison Jr, David and Rubinfeld, Daniel L},
	journal={Journal of environmental economics and management},
	volume={5},
	number={1},
	pages={81--102},
	year={1978},
	publisher={Elsevier}
}

@article{pebesma2012package,
	title={Package ‘sp’},
	author={Pebesma, Edzer and Bivand, Roger and Pebesma, Maintainer Edzer and RColorBrewer, Suggests and Collate, AAA},
	journal={The Comprehensive R Archive Network},
	year={2012},
	publisher={Citeseer}
}

@article{leenaers1988variability,
	title={Variability {O}f {T}he {M}etal {C}ontent {O}f {F}lood {D}eposits},
	author={Leenaers, H and Schouten, CJ and Rang, MC},
	journal={Environmental Geology and Water Sciences},
	volume={11},
	number={1},
	pages={95--106},
	year={1988},
	publisher={Springer}
}

@article{albering1999human,
	title={Human {H}ealth {R}isk {A}ssessment: {A} {C}ase {S}tudy {I}nvolving {H}eavy {M}etal {S}oil {C}ontamination {A}fter {T}he {F}looding {O}f {T}he {R}iver {M}euse {D}uring {T}he {W}inter {O}f 1993-1994.},
	author={Albering, Harma J and Van Leusen, Sandra M and Moonen, EJ and Hoogewerff, Jurian A and Kleinjans, JC},
	journal={Environmental Health Perspectives},
	volume={107},
	number={1},
	pages={37--43},
	year={1999}
}

@article{greenwood1984unified,
	title={A {U}nified {T}heory {O}f {S}urface {R}oughness},
	author={Greenwood, JA},
	journal={Proceedings of the Royal Society of London. A. Mathematical and Physical Sciences},
	volume={393},
	number={1804},
	pages={133--157},
	year={1984},
	publisher={The Royal Society London}
}

@article{quick2015bayesian,
  title={Bayesian {M}odeling {A}nd {A}nalysis {F}or {G}radients {I}n {S}patiotemporal {P}rocesses},
  author={Quick, Harrison and Banerjee, Sudipto and Carlin, Bradley P},
  journal={Biometrics},
  volume={71},
  number={3},
  pages={575--584},
  year={2015},
  publisher={Wiley Online Library}
}

@article{heaton2019case,
  title={A {C}ase {S}tudy {C}ompetition {A}mong {M}ethods {F}or {A}nalyzing {L}arge {S}patial {D}ata},
  author={Heaton, Matthew J and Datta, Abhirup and Finley, Andrew O and Furrer, Reinhard and Guinness, Joseph and Guhaniyogi, Rajarshi and Gerber, Florian and Gramacy, Robert B and Hammerling, Dorit and Katzfuss, Matthias and others},
  journal={Journal of Agricultural, Biological and Environmental Statistics},
  volume={24},
  number={3},
  pages={398--425},
  year={2019},
  publisher={Springer}
}

@article{hu2019monitoring,
  title={Monitoring {H}ousing {R}ental {P}rices {B}ased {O}n {S}ocial {M}edia: {A}n {I}ntegrated {A}pproach {O}f {M}achine-{L}earning {A}lgorithms {A}nd {H}edonic {M}odeling {T}o {I}nform {E}quitable {H}ousing {P}olicies},
  author={Hu, Lirong and He, Shenjing and Han, Zixuan and Xiao, He and Su, Shiliang and Weng, Min and Cai, Zhongliang},
  journal={Land use policy},
  volume={82},
  pages={657--673},
  year={2019},
  publisher={Elsevier}
}

@InProceedings{gleyze2001wombling,
author="Gleyze, J. F. and Bacro, J. N. and Allard, D.",
editor="Monestiez, Pascal and Allard, Denis and Froidevaux, Roland",
title="Detecting Regions of Abrupt Change: Wombling Procedure and Statistical Significance",
booktitle="geoENV III --- {G}eostatistics {F}or {E}nvironmental {A}pplications",
year="2001",
publisher="Springer Netherlands",
address="Dordrecht",
pages="311--322",
isbn="978-94-010-0810-5"
}

@article{plummer2015package,
  title={Package ‘coda’},
  author={Plummer, Martyn and Best, Nicky and Cowles, Kate and Vines, Karen},
  journal={URL http://cran. r-project. org/web/packages/coda/coda. pdf, accessed January},
  volume={25},
  pages={2015},
  year={2015}
}

@article{chen1999monte,
  title={Monte {C}arlo {E}stimation {O}f {B}ayesian {C}redible {A}nd {HPD} {I}ntervals},
  author={Chen, Ming-Hui and Shao, Qi-Man},
  journal={Journal of Computational and Graphical Statistics},
  volume={8},
  number={1},
  pages={69--92},
  year={1999},
  publisher={Taylor \& Francis}
}

@article{finley2007spbayes,
	title={spBayes: {A}n {R} {P}ackage {F}or {U}nivariate {A}nd {M}ultivariate {H}ierarchical {P}oint-{R}eferenced {S}patial {M}odels},
	author={Finley, Andrew O and Banerjee, Sudipto and Carlin, Bradley P},
	journal={Journal of statistical software},
	volume={19},
	number={4},
	pages={1},
	year={2007},
	publisher={NIH Public Access}
}

@article{morris1993bayesian,
  title={Bayesian {D}esign and {A}nalysis {O}f {C}omputer {E}xperiments: {U}se {O}f {D}erivatives {I}n {S}urface {P}rediction},
  author={Morris, Max D and Mitchell, Toby J and Ylvisaker, Donald},
  journal={Technometrics},
  volume={35},
  number={3},
  pages={243--255},
  year={1993},
  publisher={Taylor \& Francis}
}

@book{stein1999interpolation,
  title={Interpolation {O}f {S}patial {D}ata: {S}ome {T}heory {F}or {K}riging},
  author={Stein, Michael L},
  year={1999},
  publisher={Springer Science \& Business Media}
}

@book{spivak1970comprehensive,
  title={A {C}omprehensive {I}ntroduction {T}o {D}ifferential {G}eometry},
  author={Spivak, M.},
  number={v. 1--5},
  series={A Comprehensive Introduction to Differential Geometry},
  year={1999},
  publisher={Publish or Perish Inc.}
}

@inproceedings{kent2008modelling,
  title={Modelling {S}trategies {F}or {B}ivariate {C}ircular {D}ata},
  author={Kent, John T and Mardia, Kanti V and Taylor, Charles C},
  booktitle={Proceedings of the Leeds Annual Statistical Research Conference, The Art and Science of Statistical Bioinformatics, Leeds University Press, Leeds},
  pages={70--73},
  year={2008}
}

@book{o2006elementary,
  title={Elementary {D}ifferential {G}eometry},
  author={O'Neill, Barrett},
  year={2006},
  publisher={Elsevier}
}

@book{pressley2010elementary,
  title={Elementary {D}ifferential {G}eometry},
  author={Pressley, Andrew N},
  year={2010},
  publisher={Springer Science \& Business Media}
}

@book{gauss1902general,
  title={General {I}nvestigations {O}f {C}urved {S}urfaces {O}f 1827 {A}nd 1825},
  author={Gauss, Carl Friedrich},
  year={1902},
  publisher={Princeton University Library}
}

@article{stevens1981visual,
  title={The {V}isual {I}nterpretation {O}f {S}urface {C}ontours},
  author={Stevens, Kent A},
  journal={Artificial Intelligence},
  volume={17},
  number={1-3},
  pages={47--73},
  year={1981},
  publisher={Elsevier}
}

@article{schnare1976segmentation,
  title={Segmentation {I}n {U}rban {H}ousing {M}arkets},
  author={Schnare, Ann B and Struyk, Raymond J},
  journal={Journal of Urban Economics},
  volume={3},
  number={2},
  pages={146--166},
  year={1976},
  publisher={Elsevier}
}

@article{gao2022wombling,
    author = {Gao, Leiwen and Banerjee, Sudipto and Ritz, Beate},
    title = "{Spatial Difference Boundary Detection for Multiple Outcomes Using Bayesian Disease Mapping}",
    journal = {Biostatistics},
    year = {2022},
    month = {06},
    issn = {1465-4644},
    doi = {10.1093/biostatistics/kxac013},
    url = {https://doi.org/10.1093/biostatistics/kxac013},
    note = {kxac013},
    eprint = {https://academic.oup.com/biostatistics/advance-article-pdf/doi/10.1093/biostatistics/kxac013/43946897/kxac013.pdf},
}

\end{document}